\documentclass[%
pre,%
 twocolumn,
 showpacs,%
 ,secnumarabic%
,nofootinbib
,amssymb, amsmath,nobibnotes, aps]{revtex4-1}
\usepackage{amsmath}
\usepackage{color}
\usepackage{wrapfig}
\usepackage{subfigure}
\usepackage[outercaption,wide]{sidecap}
\usepackage{graphicx}
\usepackage[normalem]{ulem}

\newcommand \bfx{{{\bf{x}}}}
\newcommand \q{{{\theta}}}
\newcommand{\qq}{{\theta\theta}}

\newcommand{\ru}{{\rm u}}
\newcommand{\bU}{{\bar{U}}}
\newcommand{\bW}{{\bar{\rm W}}}
\newcommand{\beq}{\begin{equation}}
\newcommand{\eeq}{\end{equation}}
\newcommand{\ddr}{{\frac{d}{dr}}}
\begin{document}

%\title{A sheet on a sphere \\ Part I:  ``wrinklogami" patterns suppress curvature-induced delamination} 
\title{A sheet on deformable sphere: \\
  ``wrinklogami" patterns suppress curvature-induced delamination} 
%delamination from curved solid foundations}
%of adhesive films} %on curved topographies} 
%\title{''Pro-lamination'' of adhesive films on curved topographies}
% assisted by wrinkles} 
\author{Evan Hohlfeld and Benny Davidovitch}
\address{Department of Physics, University of Massachusetts, Amherst, MA 01003}
\date{\today}

%{\bf 
\begin{abstract}
%The adhesion of stiff films onto curved topographies is required for common operations, such as placing bandages on knuckles or noses, and  underlies a broad range of technologies, from the production of a hemispherical electronic eye \cite{Ko08} to wear-resisting coating of joint implants \cite{Fries02}.  
The adhesion of a stiff film onto a curved substrate often generates elastic stresses in the film that eventually give rise to its delamination. %\cite{Majidi08, Vella09, Hure11}. 
Here we predict that delamination of very thin films can be dramatically suppressed through tiny, smooth deformations of the substrate, dubbed here ``wrinklogami", that barely affect the macroscale topography.       
This ``pro-lamination" effect reflects a surprising capability of smooth wrinkles to suppress compression
in elastic films even when 
%undevelopable 
spherical or other doubly-curved %undevelopable 
topography is imposed,   
in a similar fashion to origami folds that enable construction of curved structures from an unstretchable paper. 
We show that the emergence of a wrinklogami pattern signals a nontrivial isometry of the sheet to its planar, undeformed state, in the doubly asymptotic  limit of small thickness and weak tensile load exerted by the adhesive substrate. We explain how such an ``asymptotic isometry" concept broadens the standard usage of isometries for describing the response of elastic sheets to geomertric constraints and mechanical loads.    
%We show that pro-lamination can be dialed in by numerous mechanisms, suggesting a broad potential usage of this effect in nature and in  technology.
\end{abstract}
\maketitle
\section{Introduction}
\subsection{Background \label{sec:Background}}
The adhesion of stiff films onto curved topographies is required for common operations, such as placing bandages on knuckles or noses, and  underlies a broad range of technologies, from producing
%the production of 
a hemispherical electronic eye \cite{Ko08} to wear-resisting coating of joint implants \cite{Fries02}. When the adhesive film is sufficiently large the elastic stress required to maintain its attachment to the curved substrate increases the energetic cost of adhesion, and  
%becomes too large and the film delaminates from 
the film delaminates.  
%one often observes that it delaminates from the curved substrate 
Despite its obvious importance, studies of 
%the basic mechanisms underlying 
such a geometry-induced delamination, and more broadly, of the 
basic mechanisms by which a curved shape can be imposed on a solid film whose stress-free state is planar,  
appeared only recently. These studies have focused on the deformations of solid films and their consequent delamination from curved substrates that are effectively infinitely rigid 
%, and the consequent delamination of the films  
\cite{Majidi08,Hure11, Hure12}. 
Here and in a subsequent %companion 
paper \cite{Number2}, we seek to provide a general theoretical framework to this fundamental problem. Our purpose is to characterize the various morphological types exhibited by an adhesive film on a curved substrate that is either rigid or deformable, and the relevant dimensionless groups of parameters that govern the laminated state of the film and its ultimate delamination from the substrate. 
%A particular object of interest here is the effect of a finite stiffness of the substrate on the morphology of adhesive films and on their delamination. 
For simplicity, and to discuss the problem in a context that is close to some recent experiments, we choose to focus our work %papers 
on the behavior of thin films attached to a spherically-shaped substrate.   

%This Letter seeks to elucidate the crucial and subtle role that finite stiffness of a curved substrate has on delamination mechanics - THIS SENETNCE SHOULD BE REVISED AND EXPANDED  

%Zi - referee? 

%and a pattern of blisters is subsequently formed, where the film remians in conact with the substrate 
%Assuming for simplicity 
Considering a spherical substrate, 
one may notice that delamination of an adhesive film  
%The delamination of elastic film from a spherically-shaped substrate  
has a similar origin to the unavoidable distortion of 
distances in planar maps of earth.
% famously encapsulated by Gauss' \emph{Throerama Egregium} CITATION?. This theorem states that any deformation of flat surface in which both principle curvatures are nonzero must incur in-plane strains, and therefore stress.
Assuming for instance a direct projection of a circular film of radius $W$ on a large rigid sphere of radius $R \gg W$, one may estimate the average strain in the film as $(W/R)^2$, the approximate percentage by which longitudes are elongated (see Fig.~1a). The elastic energy cost (per area) is %of this strain is 
%$u_{el} \sim (E_f t) (W/R)^2$
$U\sim (E_f t) (W/R)^4$, where $E_f$ is the Young modulus and $t$ the thickness of the film, and delamination 
occurs when this energy exceeds the areal adhesion energy density % (surface) energy density
$\Gamma$ \cite{Vella09}. 
Noting that the fraction $\phi$ of the sphere covered by the film is proportional to $(W/R)^2$, 
we obtain the maximal laminated fraction of a rigid sphere \cite{Majidi08}:   
\begin{equation}
\phi_{rig} \sim \sqrt{\Gamma/E_f t} \ . \label{max-phi-rigid}
\end{equation} 
This scaling law has been confirmed in recent experiments that used a glass ball as a substrate \cite{Hure11}. 
However, in another experiment, in which the rigid substrate was replaced by a liquid drop, delamination was not observed even when the coverage fraction $\phi$ was apparently much larger than $\phi_{rig}$ \cite{King12}. Instead, the radial profile of the drop %and the attached film 
was gradually flattened beneath the attached film, deviating substantially from the original spherical shape of the drop (Fig.~1c). 
Furthermore, the drop-film system developed a periodic pattern of radial wrinkles of tiny wavelength and amplitude, indicating the relaxation of compression along latitudes near the perimeter of the film (Fig.~1b).    

The observations of a distinctive behavior of adhesive films on rigid and liquid substrates motivates our theoretical study of adhesion on a deformable, curved solid substrate. The simplest model system that enables us to study this problem is the adhesion of a thin elastic film on a spherically-shaped Winkler foundation of radius $R$ and stiffness $K$,  
%
%: A sphere of radius $R$ with energetic penalty that is proportional to a stiffness parameter $K$ and to the square of the deformation of its spherical shape. The Winkler foundation 
which may be thought of as a ball of $N$ harmonic springs, each with rest length $R$ and spring constant $4\pi R^2 K/N$ (Fig.~1d). The simplified nature of the Winkler's respone allows us to carry out a quantitative analyis, from which we extract the scaling laws that govern the adhesion and delamination of a thin solid film on real, isotropic, spherically-shaped solid substrate, of radius $R$ and Young modulus $E_s$. 

%%%%%%%%%%%%%%%%%%%%%%%%%
\begin{figure}
\hspace{-0 cm}
\includegraphics[width=8cm]{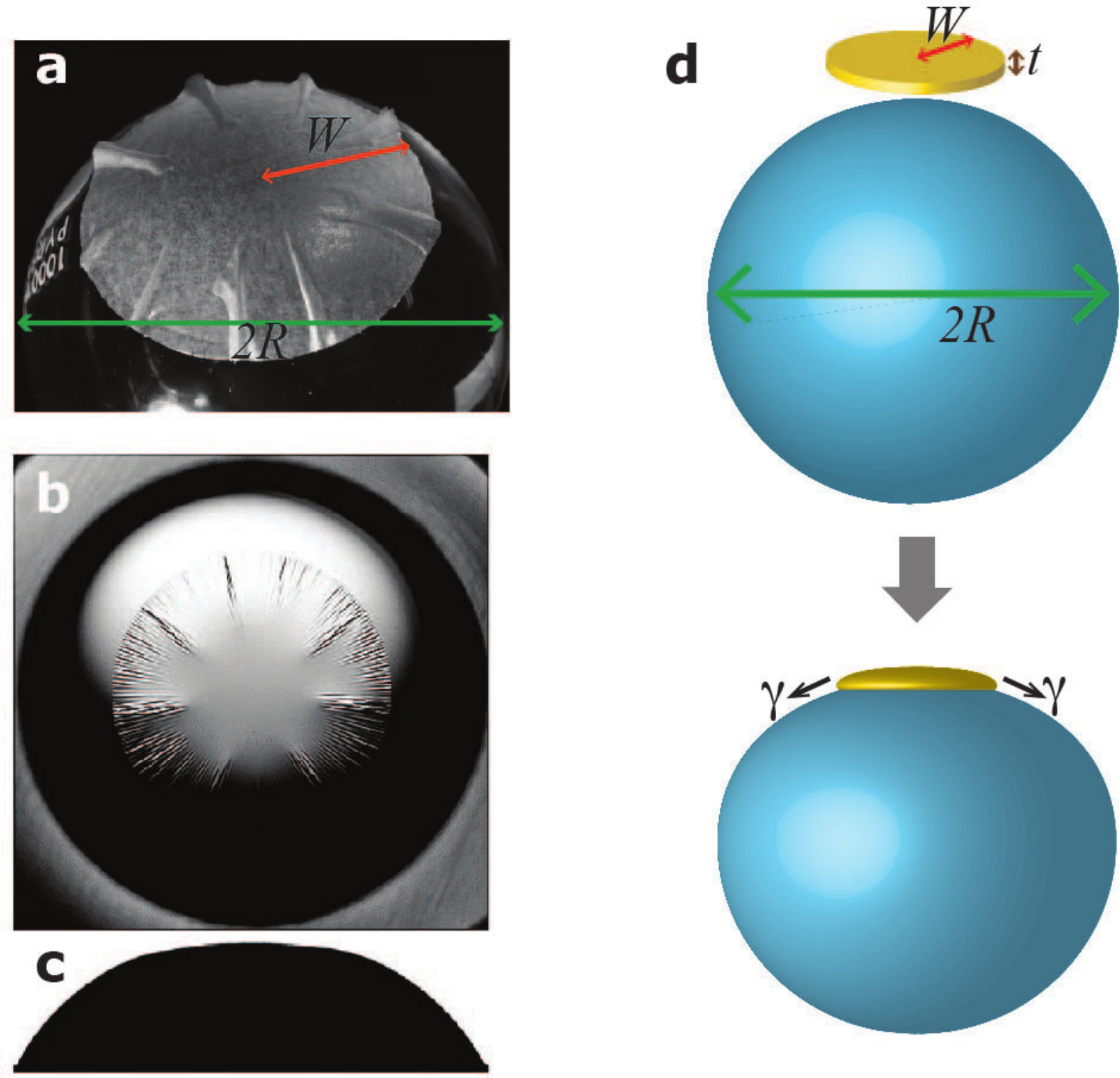}
%\includegraphics[width=8cm]{Fig1new.pdf}
%\includegraphics[width=3.75in]{MF1.pdf}
%\includegraphics[width=\columnwidth]{MF1.pdf}
%\begin{wrapfigure}{r}{3.0in}
%\includegraphics[width=\paperwidth] {MF1.pdf}
\caption{\label{fig:fig1} {\bf (a)} A thin adhesive film delaminates from a glass ball of radius $R$ when its area  ($\sim W^2$) exceeds a critical fraction $\phi_{rig}$ (Eq.~\ref{max-phi-rigid}) of the surface area of the ball ($\sim R^2$). Before delamination, longitudes of the film acquire an average strain $\sim (W/R)^2$.  {\bf (b-c)} Top and side views of an ultrathin PS film floating on a curved liquid surface (see \cite{King12}). 
%(reproduced from \cite{King12}). 
The film remains attached to the surface, and their joint deformation consists of flattening of the liquid portion beneath the film (c), and a periodic array of radial wrinkles (b). Courtesy of H. King and N. Menon.{\bf (d)} Schematic figure of our model system: A thin disk (yellow) of Young's modulus $E_f$, thickness $t$, and radius $W$ is attached to a spherical substrate (blue) of radius $R \gg W$, by the substrate-vaopr surface tension $\gamma$ that pulls on the film's edge. (For simplicity, we assume $\gamma \approx \Gamma$, where $\Gamma$ is the adhesion energy, see Sec.~\ref{sec:surfaceenergy-1}). The resistance of the substrate to deformation of its spherical shape is modeled through a Winkler's stiffness $K$ (a ball of $N$ springs, each with an effective constant $4\pi R^2 K/N$). The generalization to the case of an isotropic solid with Young's modulus $E_s$ is discussed in Sec.~\ref{sec:fromWinkler}.}        
%\end{wrapfigure}
\end{figure}

%%%%%%%%%%%%%%%%%%%%%%%%%%%%

\subsection{Main results \label{sec:MainResults}}
\subsubsection*
%\paragraph*
{Prolamination}
A central predicition of our study is that,  
%CONSIDER TWO TYPES = $K AND E_s$
%EXPAND ON THE SURPRISE - WE TWO (RATHER THAN ONE) CHARACTERSTIC VALUES OF THE STIFFNESS PARAMETER. REIGID - ABOVE K1, SOFT BELOW K2. THUS, WE PREDICT THAT IN ..   
%We predict that 
in addition to the two aforementioned classes of adhesion on rigid and soft substrates, there exists a
%is yet another, 
novel type of adhesion
%We predict 
%the existence of a novel type of adhesion, 
where the film delaminates at a coverage fraction $\phi_{def} \gg \phi_{rig}$ without any macroscale deformation of the spherically-shaped substrate. 
We call this delamination suppression phenomenon {\emph{``prolamination"}}, and argue that 
%Furthermore,
as the film becomes thinner, it %this behavior 
is expected to prevail at a broad range of physical parameters, denoted below as ``regime III''.  
%%%
%\begin{figure*}
\begin{figure*}
\hspace{-0 cm}
\includegraphics[width=15cm]{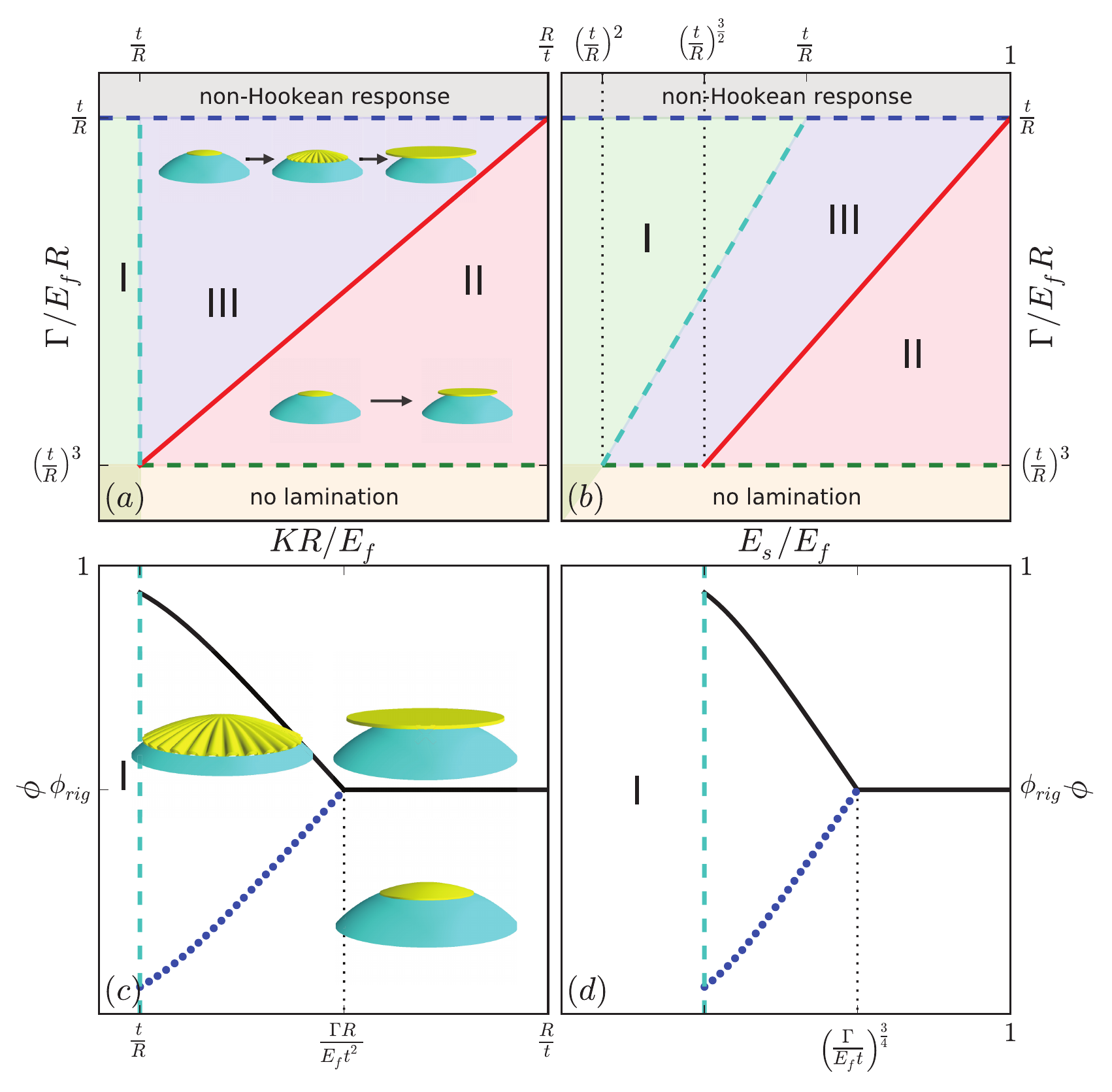}
%\includegraphics[width=3.75in]{MF1.pdf}
%\includegraphics[width=\columnwidth]{MF1.pdf}
%\begin{wrapfigure}{r}{3.0in}
%\includegraphics[width=\paperwidth] {MF1.pdf}
\caption{\label{fig:fig1}{\bf (a)} Phase diagram for the adhesion of thin films on curved substrates. 
%A lamination phase diagram for a film of thickness $t$ and modulus $E_f$ on a Winkler substrate with stiffness $K$ and curvature $1/R$. The adhesive energy scale is $\Gamma$. 
Regime I (green): Soft substrates are deformed significantly, %by the film, 
suppressing elastic strain and thus delamination at all values of the coverage fraction $\phi$.
 %$\phi=W^2/R^2$. 
Regime II (red): Rigid substrates are undeformed, and the film is severely strained. 
For a given point in regime II we illustrate the delamination process upon increasing $\phi$.
%the coverage fraction 
%$\phi$ approaches $\phi_{rig}$ (Eq.~\ref{max-phi-rigid}). 
 %      
%We illustrate the process of increasing the coverage fraction $\phi$ at a given point in regime II. The film delaminates when $\phi$ approaches $\phi_{rig}$ (see Eq. 1.)
 % for $\phi<\phi_{rig}$, the critical coverage for delamination. See inset schematic illustration. 
Regime III (blue): 
For substrates with intermediate stiffness,  
%Intermediate stiffness substrates 
both their macro-scale deformation and strain in the laminated film are suppressed by forming 
fine wrinkles for
%a fine wrinkle pattern at coverages 
$\phi_{wr}<\phi<\phi_{def}$. Delamination occurs at $\phi=\phi_{def}\gg\phi_{rig}$ (Eqs.~\ref{eq:regime-III-A-def},~\ref{eq:regime-III-B-def}). In the ``no lamination'' regime (amber),  the energetic cost of %macroscale film 
bending precludes any adhesion. The boundaries between regimes II/III and III/I are described, respectively, by the relations:
$K_{rig} \sim  \Gamma/t^2$ (Eq.~\ref{eq:Krig}), and $K_{soft} \sim (E_ft)/R^2$ (Eq.~\ref{eq:Ksoft}). Logarithmic scales are used for the axes, making the phase boundaries appear as straight lines. % .. (K_rig)_ and Eq. .. (K_{soft}). 
%  with  intercepts marked on the  axes.  
{\bf (b)} An analogous phase diagram %to (a) 
for an elastic substrate with modulus $E_s$. The lines that separate regimes II/III and III/I, are, respectively, $E_{s,rig} \sim  {\gamma^{3/4}E_f^{1/4}} / {t^{3/4}}$ (Eq.~\ref{eq:Esrig}), and  
$E_{s,soft} \sim  \gamma^{1/2}E_f^{1/2} t^{1/2} / {R}$ 
(Eq.~\ref{eq:Essoft}).
{\bf (c)} Varying the substrate stiffness $K$, we plot the predicted evolution of the system for a fixed adhesion energy $\Gamma$ and film's thickness $t$, as the coverage fraction $\phi$ increases. The blue dotted curve is described by Eqs.~
(\ref{eq:regime-III-A-wr},\ref{eq:regime-III-B-wr}), and the solid black curve is described by Eqs.~(\ref{eq:regime-III-A-def},\ref{eq:regime-III-B-def}).
 %The configuration of a film with thickness $t$ and adhesion energy $\Gamma$.
 %with schematic inset illustrations. 
Delamination occurs above the black line, wrinkling occurs between the blue dotted line and black line, and large substrate deformations occur to the left of the dashed cyan line.
% in the region labeled I. The axes are logarithmically  scaled. 
{\bf (d)} An analogous diagram to (c) for an elastic substrate with modulus $E_s$.}
%\end{wrapfigure}
\end{figure*}
%%
%For a structureless, spherically-shaped solid substrate,   
%this prediction is borne out by
For an isotropic, spherically-shaped solid substrate, this prediction is borne out by 
identifying two distinct characteristic values of the substrate's Young modulus, $E_s^{soft}$ and $E_s^{rig}$, which mark the transitions between three different regimes of adhesion. The explicit dependence of $E_s^{soft}$ and $E_s^{rig}$ on the physical parameters (film's thickness and modulus, substrate radius, and strength of adhesion) is described in Fig.~\ref{fig:fig1}b [see also Eqs.~(\ref{eq:Esrig},\ref{eq:Essoft})].

\underline{Regime I:} $E_s < E_s^{soft}$, where delamination is totally suppressed and the substrate is substantially deformed beneath the film, similarly to a liquid drop.

\underline{Regime II:} $E_s > E_s^{rig}$, where the substrate is effectively rigid, and delamination occurs at $\phi = \phi_{rig}$ [Eq.~\eqref{max-phi-rigid}].  

%and delamination occurs as $\phi \sim (W/R)^2$ approaches $\phi_{rig}$, Eq.~(\ref{max-phi-rigid}).   

\underline{Regime III:} $ E_s^{soft} < E_s < E_s^{rig}$. In this intermediate regime, the film %remains attached to the substrate at $\phi>\phi_{rig}$, and delamination occurs 
delaminates from the substrate at a coverage fraction $\phi_{def}$
%\begin{equation}
%\phi_{def} \sim \phi_{rig} \frac{\Gamma^{1/2} E_f^{1/6}}{E_s^{2/3} t^{1/2}} \ ,  
%\end{equation}      
which may significantly exceed $\phi_{rig}$ if the film is sufficiently thin (see Fig.~\ref{fig:fig1}d). 

%As is shown in Fig.~2, 
The {\emph{prolamination}} effect in parameter regime III is enabled by wrinkles around the original shape of the substrate, whose amplitude decreases with the film's thickness $t$.  
In contrast to adhesion on a liquid drop (or regime I above), the substrate retains its spherical shape and does not deform beneath the film, except for those tiny wrinkles. 
%This novel behavior is the main focus of our study.
%the main focus of our Letter.  
Notably, we find that the ratio $E_s^{soft}/E_s^{rig}$ becomes indefinitely  small as the film's thickness diminishes, %gets thinner 
so that regime III is prominent for the adhesion of ultrathin films.
%so we note  the particular relevance of Regime III for the adhesion of ultrathin films. 
For instance, for a polystyrene film ($E_f \approx 3 \ GPa$) of thickness $t=100 \ nm$ attached to spherical substrate of radius $R = 1 \  cm$ with adhesion energy $\Gamma =70 \ dyn/cm$, we find that the system is within the intermediate regime III if the sphere is made of material similar in stiffness to polydimethylsiloxane ($E_s \approx 2 \ MPa$). For a graphene sheet (estimating $E_f \approx 1 \ TPa$ and $t\approx 0.1 \ nm$ \cite{Yakobson96} and assuming similar values of $R$ and $\Gamma$), where wrinkling and delamination are often observed \cite{Einstein}, we find that this parameter regime is obtained for substrates whose stiffness can be as large as a few $G Pa$ ({\emph{e.g.}} low- and high-density polyethylene).
 
How does the wrinkle pattern %in regime III 
suppress delamination of the film ? %an adhesive film from a curved substrate ? 
As we explained above, %(Subsec.~\ref{sec:Background}), 
curvature-induced delamination stems from the 
%energetic cost of the 
strain, $\sim (W/R)^2$, acquired by a flat film of size $W$ upon placing it on a sphere of radius $R$. 
The energetic cost of strain is proportional to the stretching modulus $ \sim E_f t$.  
Recalling the Gauss' \emph{Theorama Egregium}, which links the strain ({\emph{i.e.}} deviation from a flat metric of a manifold) to the Gaussian curvature, it is tempting to think that such a geometry-induced strain in the film is inevitable, and so is the elastic energy cost associated with it. However, we find that   
%At first, one's intitution may be that the existsnce of such a strain is unavoidable, in accord with Gauss theorem Egregium,    
the wrinkle pattern acts precisely to diminish almost entirely this strain, 
%and enables the film to be attached to the substrate at a much lower energetic cost,  %Actually, the formation of wrinkles allows the film (and the attached substrate) avoid strain almost fin a way that the energetic cost is determined by the 
such that the energetic cost becomes governed by the bending modulus of the film, $B \sim E_f t^3$, 
%energy of the film through, 
and the deformation energy of the substrate. For a sufficiently thin film, this reduction in the energetic cost makes the laminated state more favorable in comparison to the loss of adhesion due to delamination.
This reduction in the energetic cost, %of the laminated state of the film, 
and the fact that it requires merely tiny perturbation of the substrate rather than macro-scale deformation of its shape, underlies the prolamination effect.   

%\vspace{0.3cm} 
\subsubsection*
%\paragraph*
{Asymptotic isometry and the ``wrinklogami" pattern}
Beyond its importance for the adhesion of films, the wrinkle pattern that enables the prolamination effect constitutes a novel type of {\emph{asymptotic isometry}} of elastic  films. % on which a curved shape is imposed. %thin films. 
The common usage of ``isometry" in elasticity theory
%The standard notion of an ``isometry" % of a surface
%, often used in studies of thin elastic bodies, 
refers to a perfectly strainless mapping of a surface (the film's midplane) 
%(here, the film's midplane) 
to some configuration in 3D space. This familiar mathematical concept %has been used to describe 
is particularly useful for describing %underlies 
the ``stress focusing" of thin solid sheets under geometric confinement (e.g. sheet confined by a ring \cite{Pomeau97,Cerda98} or a box \cite{Witten02}), whereby the shape attains a piecewise developable shape, namely, a state which is strainless everywhere except in narrow ridges and vertices that become lines and points as the sheet's thickness $t \to 0$ \cite{Pogorelov,Witten07,Pomeau97,Cerda98,AddaBedia-PRL13}. %In this paper we
The ``asymptotic isometry equation" that we introduce in this paper (sec.~\ref{sec:asym-iso}) generalizes this concept, showing that a film may approach an isometry even when its boundary is subjected to a weak tensile load,  
% weak tensile load, 
rather than to a purely geometric confinement. %, is exerted on its boundaries. 
This asymptotic isometry equation characterizes the energy %Such an 
%The asymptotic isometry we consider here  %, attained in our problem by a wrinkle pattern, amounts to 
of a family of states, %mappings of the film's midplane, 
parametrized by two parameters: the film's thickness $t$, and the tension $\gamma$ exerted on the film's edge, %. by the adhesive substrate, 
that approach an isometry %perfect isometry  ({\emph{i.e.}} a strainless mapping of the midplane to some surface in 3D) 
in the limit $(t \to 0,\gamma \to 0)$. 
%, where all other parameters in the problem are held fixed. Second, we show that the states that characterize this may not exhibit any stress focusing, but, similarly to other examples of wrinkle patterns, are homogenous supression of the stress everhwhere in the film.

In addition to highlighting the relevance of isometric maps to sheets subjected to geometric confinement and weak tensile loads, our analysis 
broadens the type of physically admissible states through which the strain can be eliminated, showing that such states are not %they do not 
%shows  
%Furthermore, we argue that the 
%that asymptotically isometric states of the film are not 
necessarily (piecewise) developable, stress focusing patterns. Instead, we show that asymptotically isometric states could be associated with wrinkle patterns that emanate from a nearly homogenous collapse of both 
%. We find that an asymptotic isometry of the forced film may be attained by a wrinkling-type shape, in which both 
compressive and tensile parts of the stress throughout the film. %vanish homogenously 
%throughout the film. %This type of wrinkle pattern, which 
We think that such a homogenous, simultaneous collapse of both compressive and tesnile stress is rather surprising, and deserves its own label, 
%of behvaior surprising, and therefore call it by a special name, ``wrinklogami", 
since the common emergence of wrinkle patterns is associated with the collapse of compression only, whereas the tensile stress along the wrinkle direction is retained.    
%We call this type of asymptotically isometric pattern ``wrinklogami", in order to highlight its difference from common examples of wrinkle patterns, which allow collapse of compression but retain the tensile stress in the sheet.   
%generalizes the common usage of wrinkle patterns, which are often considered to eliminate only the compressive component, but not tensile of the stress. 
%Inspired by the recent idea of ``buckligami'' patterns \cite{Shim-Reis12}, 
We propose the word  ``wrinklogami'' (which is different though from ``buckligami" \cite{Shim-Reis12}) to highlight the conceptual similarity to origami artistry that creates curved structures from unstrectchable sheets through designed networks of sharp (inelastic) folds \cite{Huffman, Dias12}.

\subsection{Outline}

We start in Sec.~\ref{sec:Model} with a detailed description of the physical constants, the energies, and the various forces in our model system; those pertain to the elastic film, the deformable substrate, and the adhesion between them. We conclude this section by listing the dimensionless groups of parameters that govern the problem: the geometrical and mechanical strains, their ratio (called confinement), the bendability, and the deformability parameter. In Sec.~\ref{sec:deformability} we discuss the unwrinkled, axisymmetric state of the system, where a reduction of the strain in the laminated film is associated with large deformation (flattening) of the substrate. This discussion highlights the physical meaning of the deformability parameter, and enables us to identify a range of parameters where the laminated state is only slightly deformed from a spherical shape. This parameter range includes regimes II and III in Fig.~2, and is the main focus of the rest of the current paper. In Sec.~\ref{sec:Wrinkledstate} we analyze the effect of the wrinkle pattern on the  laminated state of a film attached to such a slightly deformable spherical substrate. We start by explaining the ``far from threshold" approach, used in recent studies of radially-stretched sheets, and employ it to evaluate the energy gain due to the formation of wrinkles. In contrast to previous studies, the focus of our analysis here 
%, which makes it distinguished from previous studies, 
is on the parameter regime of large confinement and large bendability, corresponding to the limit ($t \to 0, \gamma \to 0$) at which the wrinkle pattern becomes an asymptotic isometry of the laminated film. In Sec.~\ref{sec:Prolamination}   
%\ref{sec:CompareEnergy1},\ref{sec:CompareEnergy2} 
we construct the phase diagrams in Fig.~2a and 2b, by
comparing the energies of the unwrinkled and wrinkled states with the energy cost of delamination. We start %in Sec.~\ref{sec:CompareEnergy1} 
with the Winkler substrate of stiffness $K$, assumed throughout our study, and then ``translate" the relevant scaling laws to an isotropic elastic substrate with Young's modulus $E_s$. Finally, in Sec.~\ref{sec:Asym-iso} we discuss the concept of asymptotic isometry, its applicability for other systems, and the framework it provides for studying morphological transitions between wrinkles and deformation patterns that are governed by stress-focusing. 

In a subsequent paper \cite{Number2} we plan to elaborate on the physics in parameter regime I in Fig.~2, which corresponds to an adhesive film on a curved, highly deformable substrate.   

\section{Model \label{sec:Model}}
%%%%%%%%%%%%%%%%%%%%%%%%%%%%%%%%%%%%%%%%%%%%%%%%%%%%%%%
The laminated state in our model system consists of surface energy $U_{\rm sur}$, as well as the energies $U_{\rm Win}$, $U_{\rm strain}$, and $U_{\rm bend}$, which are associated, respectively, 
with the deformation of the spherical substrate (modeled by Winkler's response), and the straining and bending of the laminated film. These energies are expressed as functionals of the displacement field, and the associated strain and curvature of the film. We start this section by describing these fields.

\subsection{Displacement, strain, stress, and normal force \label{sec:DisplacementEtc}}
%ASSUME NU=0 ?? 
We consider a circular elastic film of radius $W$ attached to a spherical substrate of radius $R$, as depcited in Fig.~1d. For our current study it is sufficient to address the case $W\ll R$, hence we will
%of film whose radius $W$ is much smaller than the radius $R$ of the spherical substrate,allowing us to 
use the common theory, often attributed to F\"oppl and von K\'arm\'an (FvK), which assumes the amplitude and slope of the deformed film are small everywhere. (Our theory can be developed with the aid of covariant derivatives in the full geometrically-nonlinear framework, but this complication will not be required in the current study). Additionaly, we assume the strains are small, which allows us to use a Hookean stress-strain relation (namely, linear material response). In this simplified framework 
%of {\emph{weak nonlinear geometry}}, 
the two tangential directions can be taken to be the radial $\boldsymbol{\hat{r}}$ and azimuthal $\boldsymbol{\hat{\q}}$ at the plane of the undeformed film, and the normal direction $\boldsymbol{\hat{n}} $ to be along the perpendicular $\boldsymbol{\hat{z}}$ to that plane.       
%
%and the normal direction can be taken to be fixed($\mathbf{\hat x}$, $\mathbf{\hat y}$, and $\mathbf{\hat z}$, correspondingly). In polar coordinates, where 
The displacement field is then expressed as: % \cite{notations-u}:
\begin{equation}
\mathbf{u}(r,\q) = \ru_r(r,\q)\mathbf{\hat r} +  \ru_\q(r,\q)\boldsymbol{\hat\q} +  \zeta(r,\q) \boldsymbol{\hat z} \ . 
\label{displacement}
\end{equation}
\subsubsection*{Strain and stress}
The strain tensor $\boldsymbol\varepsilon$ is given by: 
\begin{subequations} \label{eq:strains}
\begin{gather}
\varepsilon_{rr} = \partial_r \ru_r + \tfrac{1}{2} (\partial_r\zeta)^2\ , \label{eq:strain-radial-1} \\
\varepsilon_\qq = \tfrac{1}{r} \partial_\q \ru_\q + \tfrac1r \ru_r + \tfrac{1}{2r^2} (\partial_\q \zeta)^2 \ ,\label{eq:hoopstrain}\\
\varepsilon_{r\theta} = \epsilon_{\theta r} = \tfrac12 \left( \tfrac1r \partial_\q \ru_r +  \partial_r \ru_\q + \tfrac1r \partial_r\zeta\partial_\q \zeta\right)\ ,  \label{eq:strain-shear-1}
\end{gather}
\end{subequations}
and the stress in the film is given by the Hookean relationship \cite{LL86,Timoshenko,MansfieldBook}: 
%so that this relationship is linear (Hookean response). 
%Denoting the strain tensor $\boldsymbol\epsilon$, and using a polar coordinate system {\emph{with its origin at the center of the film}}, which is the natural choice for our study, this Hookean relationship becomes \cite{LL86}: 
\begin{subequations} \label{eq:stresses}
\begin{gather}
\sigma_{rr} = \frac{Y}{1-\Lambda^2} \left(\varepsilon_{rr} + \Lambda \varepsilon_\qq \right)\ \label{eq:stresses-radial} , \\
\sigma_{\qq} = \frac{Y}{1-\Lambda^2} \left( \varepsilon_\qq + \Lambda \varepsilon_{rr} \right)\ , \\
\sigma_{r\theta} = \frac{Y}{1+\Lambda} \varepsilon_{r\theta} \ , 
\end{gather}
\end{subequations}
where $Y=E_ft$ is the stretching modulus and $\Lambda$ the Poisson ratio of the sheet \cite{Comment-Disp-Strain-Stress}. %Again, we emphasize that Eqs.~(\ref{eq:strains}) describe the geometric (strain-displacement) connection only for $|\nabla \ru| \ll 1$ (which neccesarily implies $|\epsilon_{ij}| \ll 1$), where the Hookean (stress-strain) response only requires $|\epsilon_{ij}| \ll 1$ (which may be satisfied even in situation where the gradient $|\nabla \ru|$ is not small and the geometric connection, Eq.~(\ref{eq:strains}), is not valid.   
\subsubsection*{Curvature}
Since we assume the shape $\zeta(r,\theta)$ to be characterized by small slopes, the various components of the curvature tesor $\kappa_{ij}$ can be approximated as: 
\begin{equation}
\kappa_{rr} = \partial^2_{rr}\zeta \ \ ; \ \ \kappa_{\theta\theta}  = \tfrac{1}{r} \partial_r \zeta + \tfrac{1}{r^2} \partial^2_{\theta\theta} \zeta \ \ ; \ \ \kappa_{r\theta} = 2\partial^2_{r\theta} \zeta 
\label{eq:disp-curvature}
\end{equation} 
\subsubsection*{Winkler's restoring force}
As we mentioned already, the simplest model for a curved solid substrate is a Winkler's sphere of radius $R$ and stiffness $K$, which exerts a restroing force $F_{\rm Win}(\bfx) = -K [r(\bfx) - R]$ for deviations  of its spherical shape (where $\bfx$ is a point on the surface of the substrate, and $r(\bfx)$ is the distance of this point from the center of the spherical substrate). 
Since $W\ll R$, the Winkler's restoring force on the film and the attached substrate can be approximated as
\begin{equation}
\mathbf{F}_{\rm Win} (r,\theta) \approx -K  [\zeta(r,\q) - r^2/2R] \boldsymbol{\hat z} 
\label{approximateWinkler}
\end{equation}

%
%
%, than as isotropic solid with Young's modulus.  modulus $E_s$, we prefer to consider a Winkler-type substrate whose deformation energy (per area) for deviations $\delta\zeta_{sph}(\bfx)$ from a spherical shape at a point $\bfx$ on its surface is: $U_{Win} = ({K}/{2})\delta\zeta_{sph}^2$, where $K$ is a stiffness parameter. A Winkler substrate exertes a force $\mathbf{F}_{Win}$ in the direction normal to its surface whose magnitude is \begin{equation} F_{Win}(\bfx) = -K [r(\bfx) - R] \ ,  \label{WinklerForce} \end{equation} where $\bfx$ is a point on the surface of the substrate, $r(\bfx)$ is the distance from the center of the spherical substrate (assumed to be fixed in space), and $R$ is the radius of the undeformed substrate. 
The Winkler model is advantageous, both conceptually and computationally,  due to the local nature of the substrate response \cite{footnote1}. We will assume a fixed $K$ throughout the analysis, and will explain in Sec.~\ref{sec:Prolamination} how the results can be used to describe the behavior of an elastic substrate with Young modulus $E_s$, by identifying an appropriate ``effective stiffness". 
%Finally, for $W\ll R$ the Winkler restoring force (Eq.~\ref{WinklerForce}) on the film and the attached substrate can be approximated as \begin{equation} \mathbf{F} (r,\theta) \approx -K  [\zeta(r,\q) - r^2/2R] \boldsymbol{\hat z} \label{approximateWinkler}\end{equation}

\subsection{Energies}
%%%%%%%%%%%%%%%%%%%%%%%%%
%%%%%%%%%%%%%%%%%%%%%%%%%
Here we discuss the energy of the system and the boundary conditions implied on deformations of the laminated film at its perimeter, $r=W$, by the surrounding surface of the substrate.  
%We evaluate the energtic cost of the laminated state of the film on the substrate, with respect to the delaminated state. %, namely, we set: $U_{delam}= 0$. 
We will denote actual energies as $\bU_{(\cdot)}$, energy densities (per area) as $U_{(\cdot)} = \bU_{(\cdot)}/W^2$, and the normalized (dimensionless) energies as $u_{(\cdot)}$, where:
\begin{equation} 
u_{(\cdot)} \equiv \bU_{(\cdot)}/(E_ft)W^2 \ , 
\label{eq:notation-normalized-energy}
\end{equation}
(and similarly for work terms, which will be denoted by the normal typeface ${\rm W}$ and ${\rm w}$). Note that the normal typeface $\ru$ stands for the displacement field, whereas the italic $u$ stands for energy normalized by $(E_f t) \cdot W^2$. For clarity, we will derive first the actual energies, $\bU_{(\cdot)}$, and only later introduce their normalized versions, $u_{(\cdot)}$. 
%%%%%%%%%%%%%%%%%%%%%%%%%%%%
%%%%%%%%%%%%%%%%%%%%%%%%%%%%%%%%%%%%
\subsubsection{Adhesion energy \label{sec:adhesion-energy}} %Energy of delamination}
Before discussing the energetic cost of an elastic film laminated on a spherical substrate, let us consider two ``ideal" configurations: a delaminated state, where the film is completely detached from the substrate (except, perhaps, at a few isolated points or lines), and an unstretchable film ({\emph{i.e.}} with $E_f =\infty$) attached to a planar substrate. It is useful to define the adhesion energy $\bU_{\rm ad}$ as the energetic difference between these two ideal states:  
% laminated film, let us address the delaminated state. For simplicity, we will address only the ``ideal" delamination, where the film is completely deached from the spherical substrate (except, perhaps, at a few isolated points or lines).  
%The energy cost of such a state, with respect to the surface energy of a laminated state. 
\begin{equation} 
\bU_{\rm ad} = \Gamma A \equiv (\gamma_{\tiny fil,vap} - \gamma_{\tiny fil, subst} + \gamma )A   
\label{eq:Udelam-1}
\end{equation}
%
%2\gamma_{\tiny fil,vap} A  + 4 \pi R^2 \gamma_{\tiny subst,vap} \ , 
%\end{equation} 
where $A=\pi W^2$ is the area of the film, $\gamma_{fil, vap} \ , \ \gamma_{fil, subst}$, are the surface tensions (energies per molecular area) of the film-vaopr and film-substrate interfaces, respectively, and $\gamma$ is the surface tension of the substrate-vaopr interface.    
%where $\gamma_{a,b}$ is the surface tension {\emph{energy per molecular area}} between two materials of types $a,b$ (here $vap, substs,fil$ stand for \emph{vapor, substrate} and \emph{film}, respectively). 
%The above expression for $\bU_{ad}$ requires two clarifications. First, it is the energetic difference between the completely delaminated state and a hypothetical state, in which an unstretchable film ({\emph{i.e.}} with $E_f =\infty$) is attached on a planar substrate. 

Two notes are in order here. First, the actual energy of the laminated state consists of additional contributions, due to the strain and curvature of the film, which are the subject of the subsequent subsections.   
Second, it is possible that as delamination occurs, the film retains partial contact with the substrate, as is shown in Fig.~1a. Nevertheless, as long as some finite portion of the film detaches from the substrate, the energetic cost of delamination can be estimated through Eq.~(\ref{eq:Udelam-1}) multiplied by an appropriate numerical pre-factor.         

%Strictly speaking, the above energy is the cost of surface energy of the fully delaminated state (i.e. where the film detaches everywhere from the sheet, execpt perhaps at a few isolated points or lines). The actual demaination mode may be somewhat deifferent, retaining a finite portion .. but nevertheless its good.      

%\subsubsection{Surface energy \label{sec:surfaceenergy}}
\subsubsection{Work of adhesive substrate on elastic film \label{sec:surfaceenergy-1}} 
%In the delaminated state, the area of the film is $A = \pi W^2$, and the surface energy of the system is: \begin{equation} \bU_{delam} = 2\gamma_{\tiny fil,vap} A  + 4 \pi R^2 \gamma_{\tiny subst,vap} \ , \end{equation} where $\gamma_{a,b}$ is the surface tension {\emph{energy per molecular area}} between two materials of types $a,b$ (here $vap, substs,fil$ stand for \emph{vapor, substrate} and \emph{film}, respectively). 
%It is important to realize that the total surface area of the system is not conserved quantity. 
%In the delaminated state, it is $4\pi R^2 + 2 A$, whereas in the laminated state, 

Eq.~(\ref{eq:Udelam-1}) defines the energy $\bU_{\rm ad}$ by considering  the lamintaed state of an unstretchable film on a planar substrate, where the area of the film is $A = \pi W^2$. When studying a solid film with finite stretching modulus $Y$ attached to a spherical substrate of radius $R$, we must consider also the induced changes in surface area, which give rise to additional contributions to surface energy: $(\gamma_{fil,subst} + \gamma_{fil,vap})  \  dA_{fil}$ and 
$- \  \gamma  \ dA_{sph} $. Here, $dA_{fil}$ is the 
modification to the area %(from the original $A$) 
of the deformed film, and $dA_{sph}$ is the modification to the area removed from the substrate-vapor surface due to the laminated film. The explicit expressions for $dA_{fil}$ and $dA_{sph}$ are: 
%from whose evaluation requires its actual strain field: 
\begin{subequations}
\label{eq:dA}
\begin{gather}
dA_{film} =  \int dS \ \varepsilon_{ii} \label{dA1} \ ,  \\
dA_{sph} =  2\pi W [\ru_r(W) + W^3/8R^2] \ ,  \label{dA2}
\end{gather}
\end{subequations}
%whereas $A + dA_{sph}$ is the portion of the substrate-vapor surface which is removed and replaced with the (possibly wrinkled) film whose edge is at $r= W+\ru_r(W)$ \cite{Comment-BC-1}:        
%
%In the laminated state the area of the film is $A+dA_{film}$, and the area of the substrate-vapor interface becomes $4\pi R^2 - (A +dA_{sph})$, where: 
%covered by the film is $A+dA$ where $dA$ can be expressed using the displacement field (\ref{displacement}) as:   
%\begin{subequations} \label{eq:dA} \begin{gather}dA_{film} =  \int dS \ \epsilon_{ii} \label{dA1}   \\ dA_{sph} =  2\pi W [\ru_r(W) + W^3/8R^2] \ ,  \label{dA2} \end{gather} \end{subequations}
where in Eq.~(\ref{dA1}) $\varepsilon_{ii}$ is the trace of the strain tensor (Eq.~\ref{eq:strains}) and the integral is over the  range $r<W$.  %(of the undeformed film), 
In Eq.~(\ref{dA2}), we obtained the term $W^3/8R^2$ by considering the radial displacement $u_r(W)$ of the film's edge for an area-preserving, axisymmteric projection of the film onto a rigid sphere \cite{Comment-BC-1}. 
%Note the difference between $dA_{film}$ and $dA_{sph}$. The former is the change in area (from the original $A$) of the film, whose evaluation requires its actual strain field. In contrast, $dA_{sph}$ is the portion of the substrate-vapor surface which is removed and replaced with the (possibly wrinkled) film whose edge is at $r= W+\ru_r(W)$ \cite{Comment-BC-1}. 
% %
%(where we used $W/R \ll 1$). 
Note that, generally, $dA_{film}  \neq dA_{sph}$; namely, 
%The fact that $dA_{film}  \neq dA_{sph}$ reflects the fact that 
the total surface area (of substrate-vapor, substrate-film, and film-vaopr) is not necessarily conserved. 
%Namely, in the laminated state, the portion of the spherical substrate covered by the film and the portion in contact with vapor do not sum up to $4\pi R^2$. 
An equality ($dA_{fil} = dA_{sph}$), which for a planar substrate is trivially satisfied, would have been achieved only if the film was axisymmetrically stretched on the sphere, such that the radial displacement at its edge is $\ru_r(W) = -W^3/8R^2$ \cite{Comment-BC-1}. 

A central outcome of the forthcoming analysis 
%in the forthcoming sections 
is the strong deviation of the laminated film from %such 
the highly energetic axisymmetric state,  %(if the substrate is not too rigid) 
which is enabled by the formation of wrinkles such that
%due to its high elastic energy cost, such that 
$|dA_{sph}/dA_{fil}| \gg 1$. %\cite{Comment-Asymptotic-Areas-Ratio}.
%The exact asymptotic meaning of this inequality requires appropriate limits of the deformability and von-K\'arm\'an numbers, which will be discussed later, see Eq.~(\ref{scaling-11}). 
The physical meaning of this result is that the laminated film can remain almost unstretched ($dA_{fil} \approx 0$) by
forming a larger substrate-vapor contact area,  
%``sacrifying" some contact  
%
%having less (than the maximally possible) 
%on the expense of less 
%contact area with the substrate,
such that $dA_{sph}  < 0$.      
%
%Notice that the total surface area of the system ({\emph{i.e.}} film-vaopr, substrate-vapor and film-substrate) is not a conserved quantity (namely, $dA_{film} \neq dA_{sph}$), hence the total surface area of the laminated state is different than the delaminated state). 
%The different forms of the mathematical expressions for $dA_{film}$ (Eq.~\ref{dA1}) and $dA_{sph}$ (Eq.~\ref{dA2}) indicates on  
Therefore, we find that the surface energy of the laminated state can be approximated through the expression:   
\begin{equation}
\bW_{\rm surf} = %(\gamma_{fil,subst} + \gamma_{fil,vap})  \  dA_{film}     \  
-  \  \gamma  \ dA_{sph}   = - 2\pi \gamma W [\ru_r(W) + W^3/8R^2] \ , 
\label{eq:define-Wsurf} \ ,  
\end{equation}
which is simply the tensile work exerted by the adhesive spherical substrate on the edge of the laminated film. 
Since the tangential force exerted on the perimeter of the film, $r=W$, is simply the derivative of this work with resepct to the radial displacement of the edge $\ru_r(W)$, we obtain the 
%is 
%$-(\delta \bU/\delta A_{sph} + \delta \bU/\delta A_{film})$
%$-\delta U/\delta u_r(W)$
%, this energy gives rise to a 
boundary condition: % \cite{Comment-BC-2}:
\begin{equation}
\sigma_{rr}(W) = \gamma 
\label{BC1} \ . 
\end{equation}    

Since the three surface tensions $\gamma_{subst,fil}$,$\gamma_{subst,vap}$ and $\gamma$ are separate physical constants,
% the substrate-vapor surface tension $\gamma$ and 
the adhesion energy $\Gamma$ (Eq.~\ref{eq:Udelam-1}) and the tensile boundary force (Eq.~\ref{BC1}), are two independent quantities. For simplicity, we assume in the current study $\gamma \approx \Gamma$,  %(often called ``capillary adhesion''),
% \cite{CapillaryAdhesion}), 
but our results (up to numerical factors that do not affect the scaling rules) are valid for any finite ratio of $\gamma/\Gamma$.

\subsubsection{Substrate deformation  \label{sec:Winkler-1}}
%%%%%%%%%%%%%%%%%%%%%%%%%%%%%%%%%%%%
The energy associated with the local, Winkler-type restoring force, Eq.~(\ref{approximateWinkler}), is: 
\begin{equation}
\bU_{\rm Win} = \tfrac{K}{2} \int d^2x \  [r(\bfx) - R]^2 \ , \label{WinklerEnergy}
\end{equation} 
where the integration is over the whole surface area of the substrate. This energy, 
which penalizes for deviations from the favorable spherical shape of the substrate, can be written as: 
\begin{gather}
\bU_{\rm Win}= \tfrac{K}{2} \int_0^{2\pi} \! d\q \int_0^W \! r dr  \  (\zeta-r^2/2R)^2  
\ + \  \bU_{\rm men} 
%
%\nonumber \\
%\bU_{subst}^{(2)} = \tfrac{K}{2} \int_{outside} d^2x \  [r(\bfx) - R]^2   \ ,
\label{WinklerEnergy2}
\end{gather}  
where the integral is now only over the substrate surface beneath the laminated film, and, similarly to Eq.~(\ref{approximateWinkler})  
%we expressed $r(\bfx)$ through the polar coordinates (see Eq.~\ref{displacement}), and 
we used $W/R \ll 1$ to express the leading order (in $W/R$) of this term.  
%and on the rest of the substrate surface. In the first line, we expressed $r(\bfx)$ through the polar coordinates (see Eq.~\ref{displacement}), and used the fact $W/R \ll 1$ to express the leading order (in $W/R$) of this term. 
The second term $\bU_{\rm men}$ in Eq.~(\ref{WinklerEnergy2}) corresponds to the ``meniscus'' -- deformation of the substrate's surface that decays away from the film's edge at $r=W$. In Appendix~\ref{sec:intro} we show that any effects of the energy $\bU_{\rm men}$ can be safely neglected in our analysis.
%%%%%%%%%%%%%%%%%%%%%%%%%%%%%%%%%%%
\subsubsection{Bending energy \label{sec:bendingenergy}} 
The bending energy %per area of the deformed film 
of the curved film is proportional to the square of the principal curvatures. 
%is $\tfrac{B}{2}(Tr\boldsymbol\kappa)^2$. 
For $W/R \ll 1$ and with our polar coordinates this energy becomes: %gives rise to an energy: 
\begin{gather}
\bU_{\rm bend} = \frac{B}{2} \int_0^{2\pi} \! d\theta \int_{0}^{W} \! r dr  \ \Big( (\partial_{rr}\zeta)^2 \nonumber \\
+ (r^{-2}\partial_{\q\q} \zeta)^2 + 2r^{-2} \partial_{rr}\zeta\partial_{\q\q}\zeta \Big) \label{bendingenergy} \ , 
\end{gather}     
where $B = Et^3/(12(1-\Lambda^2)$ is the bending modulus of the film. 
The minimal bending energy for a film of area $\sim W^2$ on a substrate with curvature $R^{-1}$ is 
$\bU_{\rm bend} \sim (B/2) W^2/R^2$, implying that  
the adhesion $\Gamma$ must exceed a critical value $B/R^2 \sim E_f t^3/R^2$ in order for the film to be laminated. % at any $W>0$. 
This lower bound %on $\Gamma$ 
underlies the horizontal dashed green line in the phase diagram, Figs.~2a,2b.  In this study we address the parameter regime $\Gamma \gg E_f t^3/R^2$ (or $\delta_m \gg \tilde{t}^2$ in the dimensionless parameters, to be defined below), where the bending cost associated with the substrate curvature $R^{-1}$ can be neglected 
\cite{Comment-Lower-Bound-Bending}.
If the film is sufficiently thin (and the substrate is not too rigid) the wrinkle pattern governs the bending energy 
%allowed to develop wrinkles the bending energy becomes important 
due to the high curvature of the small wavelength undulations in the azimuthal direction. Hence, Eq.~(\ref{bendingenergy}) can be approximated by:  
\begin{equation}
\bU_{\rm bend}  = \frac{B}{2} \int_0^{2\pi} d\theta \int_{0}^{W} r \ dr  \ (r^{-2}\partial_{\q\q} \zeta)^2  \label{bendingenergy1} \ .  
\end{equation}
%Similarly, the bending force in Eq.~(\ref{eq:normal}) (which is obtained from Eq.~(\ref{bendingenergy}) upon variation with respect to $\zeta$), is approximated as: % in the high bendabilty regime as: 
%\begin{equation}
%B \Delta^2 \zeta \approx B \frac{1}{r^4} \partial_\q^4 \zeta  \ . 
%\label{bendingforce1}
%\end{equation} 
%Similarly, the bending force in Eq.~(\ref{eq:normal}) (which is obtained from Eq.~(\ref{bendingenergy}) upon variation with respect to $\zeta$), is approximated as \begin{equation} B \Delta^2 \zeta \approx B \frac{1}{r^4} \partial_\q^4 \zeta  \ . \label{bendingforce1} \end{equation} The above estimates of the bending energy and force obviously vanish in the axisymmetric state (where $\partial_\q\zeta = 0$), but have a strong effect in the wrinkled state, as we will discuss in Sec.\ref{sec:wrinkling}. 
%%%%%%%%%%%%%%%%%%%%%%%%%%%%%%%%%%%%
\subsubsection{Straining energy} % (stresses)}
Imposing a spherical shape on the film 
%through the normal force $F_n$, Eq.~(\ref{eq:normal}), as well as the tension $\gamma$ exerted at its boundary $r=W$, Eq.~(\ref{BC1}), 
will generally give rise to elastic stresses $\sigma_{ij}$ in the plane of the film (where $i,j = r,\q$) and to corresponding strains $\varepsilon_{ij}$ (Eq.~\ref{eq:stresses}).
%which are found by solving Eqs.~(\ref{eq:div},\ref{eq:normal}). 
The energy associated with these stresses is: 
\begin{equation}
\bU_{\rm strain} = \frac{1}{2} \int_0^{2\pi} d\theta \int_{0}^{W} r \ dr \sigma_{ij} \varepsilon_{ij}  % -  \gamma dA_{sph} %u_r(W,\q) 
\ .  \label{general-elas-energy}
\end{equation}  
%where $dA_{sph}$ is given by Eq.~(\ref{dA2}), and we incorporated here the increment in the surface energy through the work done on the film at its boundary $r=W$.   

%%%%%%%%%%%%%%%%%%%%%%%%%%%%%%%%%%%%%%%%%%%%%%
\subsection{The F\"oppl--von K\'arm\'an equations \label{sec:FvK}}

At mechanical equilibrium, the %state of the 
system can be described by the F\"oppl--von K\'arm\'an (FvK) equations of elastic sheets:  
\begin{subequations}
%\begin{gather}
\label{eq:div}
\begin{gather}
B \Delta^2 \zeta - \sigma_{rr}\partial_r^2\zeta - \tfrac2r \sigma_{r\q}\left(\partial_r -\tfrac1r\right) \partial_\q \zeta \nonumber \\
- \tfrac{1}{r^2} \sigma_\qq \left(\partial_\q^2 \zeta + r\partial_r \zeta\right) = -K  [\zeta(r,\q) - r^2/2R] \ , \label{eq:normal} \\ 
%\begin{align}
%\text{force balance in $\mathbf{\hat r}$:}& \ 
\partial_r \sigma_{rr} + \tfrac1r \left(\partial_\q \sigma_{r\q}
+ \sigma_{rr} - \sigma_\qq\right) =  0  \label{eq:divr}\ ,\\
%\text{force balance in $\boldsymbol{\hat \q}$:}& \ \ 
\partial_r \sigma_{r\theta} + \tfrac1r \left(\partial_\q \sigma_\qq + 2\sigma_{r\theta}\right) = 0  \ ,  \label{eq:divq} 
%\end{align}
%and Eq.~(\ref{FvK2gen}) becomes: 
%\begin{equation} 
%\end{equation}
%\end{align}
\end{gather}
\end{subequations}
where the Laplacian $\Delta \equiv \partial_r^2 + \frac1r\partial_r + \frac{1}{r^2}\partial_\q^2$, and the stress-strain relations (Eqs.~\ref{eq:stresses}) with the geometric strain-displacement relation  (Eqs.~\ref{eq:strains}) are used to transform Eqs.~(\ref{eq:div}) to a nonlinear set of partial differential equations for the three components of the displacement field $\mathbf{u} (r,\q)$. Eq.~(\ref{eq:normal}) is the $1^{st}$ FvK equation and expresses force balance in the normal direction; Eqs.~(\ref{eq:divr},\ref{eq:divq}) form the $2^{nd}$ FvK equation, and express force balance on each infinitesimal piece of the film in the two directions locally tangent to the sheet \cite{LL86,Timoshenko,MansfieldBook}.   
%respectively, the $2^{nd}$ and $1^{st}$ FvK equations. The $1^{st}$ and $2^{nd}$ FvK equations, which 
These equations can be obtained as the Euler-Lagrange equations of the energy functional $\bU_{\rm strain} + \bU_{\rm bend} + \bU_{\rm Win}$. %and express force balance on each infinitesimal piece of the film in the normal direction and the two directions locally tangent to a reference plane of the  film %as well as in the normal direction 

At the parameter regime that we focus on here, which is relevant for very thin films, the bending energy due to radial curvature and the force associated with it ($= \ B (\partial_r^2 + \frac1r\partial_r)\zeta)$) are always negligible. %\cite{Comment-Radia-Bending-Negligible}.  
%Minimization of the substrate deformation energy $\bU_{Win}$ gives rise to the normal force $F_N =-K  [\zeta(r,\q) - r^2/2R] $ in the RHS of Eq.~(\ref{eq:normal}), and 
 %Furthermore, at the parameter regime that we focus on here, the bending energy due to radial curvature and the force associated with it ($B (\equiv \partial_r^2 + \frac1r\partial_r)\zeta)$) are always negligible \cite{Comment-Radia-Bending-Negligible}. 
In the following sections we will show that this fact allows us to analyze the FvK equations (\ref{eq:div}) both for the axisymmetric (unwrinkled) state and for the fully wrinkled state of the system, by using the %only two 
BCs at $r=W$: 
\begin{subequations}
\label{BC-TOT-W}
\begin{gather}
\sigma_{rr}(W) \!=\! \gamma  \ , 
\label{BC-TOT-W-1}
\end{gather}
\begin{gather}  
\zeta(W) \!=\! \zeta_{sph}(W)  \ \ ; \ \   \frac{\partial}{\partial r}\zeta(W\!,\!\q) \!=\! \zeta_{sph}'(W)   \ , 
\label{BC-TOT-W-2}
\end{gather}
\end{subequations}    
where $\zeta_{sph}(r) \approx -r^2/2R$ is the shape of the undeformed sphere. The BC~(\ref{BC-TOT-W-1}) follows from Eq.~(\ref{BC1}), and the 
second BC~(\ref{BC-TOT-W-2}) stems from the 
negligibility of normal force and torque 
%meniscus effect (see Sec.? and Appendix) implies that the torque %bending moment 
exerted on the film's edge by the deformed substrate at $r>W$. 
In Appendix~\ref{sec:intro} we explain why BC~(\ref{BC-TOT-W-2}) is valid for a slightly deformable substrate, which is the main focus of our study, as well as for a highly deformable substrate, which we plan to discuss elsewhere. %be the subject of another paper \cite{Number2}.     

Two other BCs at $r=0$ are 
\begin{equation}
\ru_r(0) = 0 \ ; \ \zeta_r(0) = 0 \ ,
\label{BC-TOT-0}
\end{equation}
which imply integrity of the film and a non-diverging stress at its center.

 %the adhesion areal energy $\Gamma$ of the delaminated state.  
%where $U_S = (Y/2) (\text{strain})^2$ and $U_B = (B/2)  (\text{curvature})^2$ are, respectively, the stretching and bending areal energies, and $Y \sim  E_ft, B \sim E_f t^3$ are the respective modulii of the film. The detailed mathematical expressions for the strain and curvature are given in {\emph{SI}}.    
\subsection{Dimensionless groups \label{sec:dimensionlessgroups}}
\subsubsection*{From dimensional parameters to dimensionless groups}
%%%%%%%%%%%%%%%%%%%%%%%%%%%%%%%%%%%%%%%%%%%
The physical (dimensional) parameters of our model are the substrate-vapor surface tension ($\gamma$), the adhesion energy per area ($\Gamma$), the stiffness ($K$) and radius ($R$) of the substrate, and the Young modulus ($E_f$), thickness ($t$), and radius ($W$) of the film.
%\footnote{For simplicity, we assume here $\Gamma \approx \gamma$, where $\gamma$ is the surface tension of the substrate. In {\emph{SI}} we address the general case $\Gamma \neq \gamma$.}. 
All energies considered, denoted hence by lowercase $u_{(\cdot)}$, are normalized by 
the stretching modulus $(E_f t)$ and the area $W^2$ of the film, such that the model can be characterized by five dimensionless groups: 
\begin{gather}
% \phi \sim 
\delta_g=  (W/R)^2 \ \ ; \ \ 
%\delta_{m} =   \Gamma/ (E_f t)  \ \ 
\delta_{m} =   \gamma/ (E_f t)  \ \ 
; \ \  \tilde{t} = t/R  \label{nondim1} \\  
\tilde{K}^{-1}  = (E_f t)/KR^2 %\ \ ; \ \ \tilde{t} = t/R  
\label{nondim2}  \ ,  \\
\gamma/\Gamma  \approx 1 \label{nondim3}
%\epsilon_{\Gamma}^{-1}  = {\Gamma W^2}/{(E_f t^3)} 
\end{gather}  
%to which we 
We call the first two numbers, respectively, the {\emph{geometrical}} and {\emph{mechanical}} strains ($\delta_{g},\delta_{m}$) that are imposed on the film, and will elaborate on their physical meaning in the next paragraph. 
%, and the {\emph{deformability}} parameter $\tilde{K}^{-1}$.  
The number $\tilde{t}$ is a useful dimensionless measure of the sheet's thickness,  %,
which simplifies the apperance of various formulas. %comparison to  
%we find to be more convinient to use here rather than 
%the standard von K\'arm\'an number $vK = (W/t)^2 = \phi/\tilde{t}^2$ (that characterizes the aspect ratio of the film). 
The fourth dimensionless group, $\tilde{K}^{-1}$, which we call the {\emph{deformability}} parameter, is the only one that depends on the substrate stiffness $K$ \cite{Comment-deformability}. In the following section we will highlight the basic difference between a high deformability regime ($\tilde{K}^{-1} \gg 1$),
and the low-deformability regime ($\tilde{K}^{-1} \ll 1$), on which we will focus in this paper. 
%, we call the {\emph{deformability}} parameter $\tilde{K}^{-1}$. 
Finally, the ratio between the substrate-vapor surface tension and the adhesion energy is another independent dimensionless group, but in the current study we will assume for simplicity that it is identically one \cite{footnote2}. 

The division of the four primary dimensionless groups in Eqs.~(\ref{nondim1},\ref{nondim2}) reflects an important feature of our model. Although $\delta_g,\delta_m$ and $\tilde{t}$ are independent parameters, we will show below that the morphology of the laminated state 
depends only the deformability $\tilde{K}^{-1}$ and two independent combinations of the triplet ($\delta_g,\delta_m,\tilde{t}$), that we call the effective {\emph{confinement}} and {\emph{bendability}} parameters. In order to understand their meaning, let us discuss first the strain parameters.

\subsubsection*{Geometrical strain, mechanical strain, and confinement}
The mechanical strain $\delta_{m}$ is defined as the tensile strain at the film's perimeter $r=W$, which exists also in the lamination of a flat substrate. Such a tensile load tends to induce uniform tension $\sim \gamma $, %\Gamma$, 
and hence expansion of both longitudes and latitudes 
 %=E_f t \delta_{m}$ 
across the film. In contrast, the geometrical strain $\delta_{g}$ imposed on the film is associated with the projection of a flat shape onto a sphere, and therefore induces confinement of latitudes.
% \cite{footnote3}.
This basic difference between the imposed geometrical strain and mechanical strain is a cornerstone of our study. We will show that, in contrast to the mechanical strain, the geometrical strain does not necessarily imply stretching or compressing the film; instead - the film may ``trade in" such a highly-energetic straining with wrinkles whose energetic cost is associated with bending the sheet and deforming the substrate. This difference underlies the asymptotic isometry that may be attained by the film, and the pro-lamination phenomenon.  
%Note that $\delta_g$ is proportional to the fraction $\phi$ of the sphere covered by the film. 
% \footnote{In the F\"oppl-von K\'arm\'an framework the $\sigma \propto E_f t\epsilon,$ where the strain $\epsilon_{ij}=\frac{1}{2}\left(\nabla_iu_j+\nabla_ju_i + \nabla_i \zeta\nabla_j\zeta\right)$. When the film is a nearly spherical cap, $\zeta = r^2/2R$, which gives the scaling of the geometric strain.}.
In order to elucidate the conflicting nature of the geometrical and mechanical strains it is useful to introduce their ratio: 
\begin{equation} \alpha = \delta_{g}/\delta_{m}  = (E_f t) W^2 / \gamma R^2  \ , %\sim \phi/\delta_m \label{confinement} \ , 
\label{gener-alpha}
\end{equation} 
which was called the {\emph{confinement}} parameter in \cite{PNAS11,King12}. The parameter $\alpha$ describes the degree of %azimuthal 
confinement of latitudes imposed on the film by stretching it over a spherical substrate.  

As we mentioned in the beginning of this section, it will be sufficient for us to consider small values of both geometrical and mechanical strains, $\delta_g,\delta_m \ll 1$, such that %Under such condition, 
the coverage fraction $\phi$ of the sphere laminated by the film satisfies $\phi \sim (W/R)^2 \sim \delta_g\ll1$, and the stress is proportional to the strain as we discussed in Secs.~\ref{sec:DisplacementEtc},\ref{sec:FvK}. 
%[as is implied by the quadratic form of the energies in Eq.~\eqref{energies}]. 
The condition $\delta_m \ll 1$ underlies the upper bound in the phase diagram, Fig.~1a: If $\Gamma >(E_ft)$, %or more precisely, $\Gamma > \gamma^2/(E_f t)$, 
the adhesion is too strong 
%(alternatively, the film is too thin), 
and the large strain may generate non-Hookean response. (The lower boundary in Fig.~1a is explained in Sec.~\ref{sec:bendingenergy} and in \cite{Comment-Lower-Bound-Bending}). 
%which we explained already in Sec.~\ref{sec:bendingenergy}, stems from the bending response of the film. If $\Gamma < E_ft^3/R^2$ the adhesion is too weak,  and the energetic cost of bending prohibits lamination for any $\phi>0$.) 

\subsubsection*{The bendability parameter}

In addition to the defomability parameter $\tilde{K}^{-1}$ and the confinement $\alpha$, another crucial ratio for understanding the system is the effective {\emph{bendability}}, denoted as $\epsilon^{-1}$, which expresses the ratio between the imposed strain and the bending resistance of the film. This ratio is conveniently expressed as the product of the von K\'arm\'an number $vK = (W/t)^2$, which characterizes the aspect ratio of the film, and the characteristic imposed strain $=\max[\delta_g,\delta_m]$:  
%
% and can be expressed as the prodocut of the $vK$ number and the characteristic strain imposed on the film: 
\begin{equation}
\epsilon^{-1} \sim vK \ \cdot  \ (\small{\text{imposed strain}})  = %\left\{ 
\begin{array}{ll}
\epsilon_m^{-1}  = \frac{\gamma W^2}{B}  \ \ & \ \ (\alpha < 1) \\ \\
\epsilon_g^{-1}   = \frac{W^4}{t^2R^2}  \ \ & \ \ (\alpha > 1)  , 
\end{array}
\label{gener-bendability}
\end{equation}
where we refer to $\epsilon_m^{-1}$ and $\epsilon_g^{-1}$, respectively,  as the {\emph{mechanical and geometrical bendailities}}, in analogous manner to the mechanical and geometrical strains, $\delta_m$ and $\delta_g$, introduced above. Note that the definition (\ref{gener-bendability}) of the bendability parameter generalizes previous usage of this term  %that had been used in previous works
\cite{King12,PNAS11,PRE12, PNAS13,Toga13,Peinura13} 
which addressed only %which assumed a 
the ``mechanical regime", where $\epsilon =\epsilon_m$. %corresponding to situations where $\epsilon \sim \epsilon_m$. 

\subsubsection*{Three relevant dimensionless groups and asymptotic limits}

%     
%
%and the inverse von K\'arm\'an number $\tilde{t}$. 
%

%Our exposition will provide the 
%physical meaning %morphological meaning and physical role 
%of these numbers.
%{\emph{(tensional) bendability}} $\epsilon_{\Gamma}^{-1}$. 
%Our findings are valid for an arbitrary ratio $W/R$ between the radii of the film and substrate, but it will suffice to 

Recent works on radially-stretched elastic films described the characteristic morphologies in these systems through two dimensionless parameters: confinement and bendability. The study of an elastic film laminated on a solid substrate requires us to consider three dimensionless groups: the generalized versions of the confinement and bendability parameters, Eqs.~(\ref{gener-alpha},\ref{gener-bendability}), and the substrate deformability $\tilde{K}^{-1}$ (Eq.~\ref{nondim2}).     
% 
%By introducing the generalized versions of these parameters, Eqs.~(\ref{gener-alpha},\ref{gener-bendability}) we are able to describe the laminated state of the film in our system 
%through a set of three dimensionless parameters, which includes the substrate deformability $\tilde{K}^{-1}$ (Eq.~\ref{nondim2}) in addition to the confinement (\ref{gener-alpha}) and the bendability (\ref{gener-bendability}). 

In terms of the effective confinement, bendability, and deformability parameters, the core of the current paper is %the majority of this paper is 
analysis of the parameter regime $\tilde{K}^{-1} \ll 1  \  ,  \ \alpha \gg 1 \ ,  \ \epsilon^{-1} \gg 1$. The assumption of low deformability ($\tilde{K}^{-1} \ll 1$) allows us to simplify calculations since the shape of the laminated shape is close to the original, spherical shape of the substrate. This assumption,which will be used intensively in the next sections, will be relaxed in a subsequent paper \cite{Number2}, where we plan to address also the high-deformability regime. The focus on high confinement and bendability values is the essential contribution of the present work. The asymptotic limit $\alpha \to \infty$, which in dimensional form reads $\gamma \ll  (E_f t) (W/R)^2$, means that the mechanical strain is negligible in comparison to the geometric strain; the asymptotic limit $\epsilon \  (=\epsilon_g)  \to 0$, which in dimensional form reads $t \ll W^2/R$, 
means that the imposed geometric strain can be eliminated by wrinkles. Thus, when the tension imposed on the film is weak enough and the film's thickness is sufficiently small, we have to study the doubly asymptotic limit,$(\alpha,\epsilon^{-1}) \to \infty$ at a fixed value of $\tilde{K}$. 
%and $(\alpha,\epsilon^{-1}) \to \infty$.
This singular limit, on which we will expand further in Sec.~\ref{sec:wrinklogami}, underlies the concept of asymptotic isometry that we study here.  

%In dimensional form, the doubly asymptotic limit $\epsilon^{-1},\alpha \to \infty$ correspond to the parameter regime $t \ll W^4/R^2, \gamma \ll  (E_f t) (W/R)^2$; namely, where both film's thickness $t$ and the tension $\gamma$ exerted on it by the substrate, are sufficiently small. As we noted already in the introduction (Sec.~\ref{sec:MainResults}), in this limit the wrinkled state becomes a ``wrinklogami" pattern -- a smooth, low-energy deformation of the confined sheet which we describe as {\emph{asymptotic isometry}}. This concept will be further elucidated in Sec.~\ref{sec:wrinklogami}.

An additional complexity stems from the nontrivial dependence of those effective parameters on the four ``pristine" dimensionless parameters, Eqs.~(\ref{nondim1},\ref{nondim2}).      
%With the aid our  behvaior of our model is essentially described by the three parameters. Namely, at each parameter range, only three parameters govern the physics. This is just a "minimal" extension beyond the confinement-bendability problems discussed earlier. However, an additional level of complexity originates from the intricate dependence of $\alpha$ and furthermore $\epsilon$ on the ``pristine" control parameters, requires us to consider the problem through a 4d parameter space. 
Therefore, Fig.~1a (or any other planar plot of the phase diagram) must be understood as a projection of this 4d parameter space onto a specific planar section. In Table 1 we summarize the dimensionless parameters of our model, and specifiy the parameter regimes that are the focus of the current study.    

%\begin{multicols}[1]
\begin{table*}[t]
\centering
\caption{``Pristine" dimensionless groups of the model}
\begin{tabular}{@{\vrule height 10.5pt depth4pt  width0pt}lrcccc}
%table text
      & Group & Definition & Focus of current study  & Comments \\ \hline\hline
%      &Bendability  & $\epsilon = B/\gamma W^2 $  & $\epsilon \ll 1$ & high bendability (Sec.~\ref{sec:bendingenergy})    \\ \hline 
& normalized thickness  & $ \tilde{t} = t/R $  & $ \tilde{t} \ll \phi^2 $ &
compression collapses by wrinkling  (Sec.~\ref{sec:Principles-FT})    \\ \hline 
      &mechanical  strain & $\delta_m = \gamma/E_ft $  & $ \tilde{t}^2 \ll \delta_m \ll 1$ &  Hookean response (Sec.~\ref{sec:adhesion-energy}) \\ 
& \mbox{} & \mbox{} & \mbox{} & adhesion sufficiently strong  (Sec.~\ref{sec:bendingenergy}) \\ \hline
      &geometric strain & $\delta_g = (W/R)^2 $ & $\delta_g \sim \phi \ll 1$ & small slopes, FvK equations are valid (Sec.~\ref{sec:FvK}) \\
      &laminated fraction of sphere & $\phi \sim \delta_g $ & \mbox{} & \mbox{} \\ 
 \hline
      &deformability & $\tilde{K}^{-1} = (E_ft)/(KR^2)$ \cite{Comment-deformability}  &$ \tilde{K}^{-1} \ll 1$ & low substrate deformability (Sec.~\ref{sec:deformability-parameter})  \\ \hline
      &tensile ratio & $\gamma/\Gamma$ & $\approx 1$ & simplification (Sec.~\ref{sec:dimensionlessgroups})  \\  \\ \hline  \\ \hline \hline
\end{tabular}
\caption{Effective dimensionless parameters}
%(dependent on ``pristine" parameters of Table I)}
\begin{tabular}{@{\vrule height 10.5pt depth4pt  width0pt}lrcccc}
      & Group & Definition & Focus of current study & Physical meaning \\ \hline\hline
      &confinement& $\alpha = \delta_g/\delta_m$ & $\alpha \gg 1$ & asymptotic isometry (Sec.~\ref{sec:Principles-FT})   \\ 
      &mechanical bendability& $\epsilon_m^{-1} = \tilde{t}^{-2}\delta_m \delta_g$ & \mbox{}  & \mbox{}   \\
       &geometrical bendability& $\epsilon_g^{-1} = \tilde{t}^{-2} \delta_g^2$ & $\epsilon_g^{-1} \gg \epsilon_m^{-1} \gg 1 $ & asymptotic isometry (Secs.~\ref{sec:Principles-FT})     \\
%      &effective bendability& $\epsilon^{-1} = \max[\epsilon_g^{-1},\epsilon_m^{-1}]$ & $\epsilon^{-1} \gg 1$ & asymptotic isometry (Secs.~\ref{sec:wrinkling?})   \\ 
&deformability& $\tilde{K}^{-1}$ (see above) & $\tilde{K}^{-1} \ll 1$ & low substrate deformability  \\ 
\hline
\end{tabular}
\label{table1}
\end{table*} 
%\end{multicols}

% 
%$W/R \ll 1 $ (such that $\phi xerted strain $\delta_m \ll 1$. 
%We focus on a small values of $W/R$, allowing us to express 
%the coverage fraction $\phi \sim (W/R)^2 \sim \delta_g\ll1$. We also assume the small strains, i.e. $\delta_m$, so that stress is proportional to  strain. This gives the top dashed line in Fig.~1. 
%In these limits the, the F\"oppl-von K\'arm\'an elastic energy density is appropriate for the adherent film. This energy density has the scaling form
%\begin{multline}
%U_{def}=\frac{E_ft^3}{2}\text{(curvature)}^2 + \frac{E_ft}{2}\text{(strain)}^2 \\+ \frac{K}{2}\text{(displacement)}^2\label{eq:FvK-energy}
%\end{multline}
%We note that when adhesion is weak, the bending energy in \eqref{eq:FvK-energy} can drive delamination. Estimating curvature as $1/R$,  bending energy becomes comparable to adhesion when $E_ft^3/R^2\sim \Gamma$, so lamination evidently requires $\Gamma/E_f R\gg (t/R)^3,$ which gives the bottom dashed line in Fig.~1.
%When $W/R\ll1$, the equations of in-plane and out-of-plane force balance are the well known F\"oppl-von K\'arm\'an equations
%\begin{subequations}
%\begin{gather}
%\nabla_j \sigma_{ij}=0\label{eq:FVK-1}\\
%B\Delta^2 \zeta - \sigma_{ij}\nabla_i\nabla_j \zeta=-K\delta \zeta\label{eq:FVK-2}
%\end{gather}
%\label{eq:FVK}
%\end{subequations}
%where the bending modulus $B\sim E_f t^3$ and the effect of the substrate is given by the right hand side of \eqref{eq:FVK-2}. CITATION 
\section{The axisymmetric state
\label{sec:deformability}}
%%%%%%%%%%%%%%%%%%%%%%%%%%
%\subsection{The axisymmetric state}
%%%%%%%%%%%%%%%%%%%%%%%%%
We start 
%analyzing our model system 
with the laminated, %by studying the laminated 
axisymmetric state of the system, whose energy underlies the delamination from a highly rigid substrate (Eq.~\ref{max-phi-rigid}) \cite{Majidi08,Hure12}. The analysis, which is focused on the high bendability limit,  will allow us to elucidate the role of the geometric and mechanical strains, $\delta_g$ and $\delta_m$, %their ratio (which we call the ``confinement'' parameter), 
the confinement parameter $\alpha$, and the deformability parameter $\tilde{K}^{-1}$.    

The axisymmetric state is characterized by radial and normal displacements of the form $\ru_r(r), \zeta(r)$. The only components of the strain tensor are: 
\begin{equation}
\varepsilon_{rr} = \ru_r' + \frac{1}{2} (\zeta')^2 \ \  ; \ \ \varepsilon_{\q\q} = \ru_r/r  \ , 
\label{geom-axi}
\end{equation}
and the corresponding components of the stress tensor are $\sigma_{rr}(r),\sigma_{\q\q}(r)$, as determined from Eq.~(\ref{eq:stresses}). The FvK equations~(\ref{eq:div}) %,\ref{eq:normal}) 
thus transform into a coupled set of ODE's for the functions $\ru_r(r),\zeta(r)$:   
\begin{subequations}
\label{eq:div-axi}
\begin{align}
%\text{force balance in $\mathbf{\hat r}$:}& \ 
\partial_r \sigma_{rr} + \tfrac1r \left(
\sigma_{rr} - \sigma_\qq\right) =  0  \label{eq:divr-axi}\ ,\\
 \sigma_{rr}\partial_r^2\zeta 
%- \tfrac2r \sigma_{r\q}\left(\partial_r -\tfrac1r\right) \partial_\q \zeta
 + \tfrac{1}{r} \sigma_\qq \left(\partial_r \zeta\right) = K  [\zeta - r^2/2R] \ ,  \label{eq:normal-axi}
%\text{force balance in $\boldsymbol{\hat \q}$:}& \ \ \partial_r \sigma_{r\theta} + \tfrac1r \left(\partial_\q \sigma_\qq + 2\sigma_{r\theta}\right) = 0  \ ,  \label{eq:divq}
\end{align}\end{subequations}
where we neglected the bending force due to the radial curvature $R^{-1}$ %(since we address the parameter regime $\delta_m \gg \tilde{t}^2$, 
(see Sec.~\ref{sec:bendingenergy}).
%, and hence for the axisymmetric state (where $\partial_\theta = 0$), 
These equations, together with Eqs.~(\ref{eq:stresses},\ref{geom-axi}), are $2^{nd}$ order in $\ru_r$ and $\zeta$, and therefore the BCs (\ref{BC-TOT-W},\ref{BC-TOT-0}) suffice to find the axisymmetric state.  
Notably, solutions to these FvK equations  
%solutions to Eqs.~(\ref{eq:div-axi}) (supplemented by Eqs.~(\ref{eq:stresses},\ref{geom-axi}) and the above BCs) 
are determined by two dimensionless groups only: The confinement parameter $\alpha = \delta_g/\delta_m$, %which we associate below with the degree of azimuthal confinement imposed on the film, 
and the deformability parameter $\tilde{K}^{-1}$. We denote these solutions by the superscript $^{\rm axi}$, for instance the stress components are denoted as: 
% these solutions: 
$\sigma^{\rm axi}_{rr}(r;\tilde{K},\alpha),\sigma^{\rm axi}_{\q\q}(r;\tilde{K},\alpha)$. 
We address first the case of an infinitely rigid substrate and then turn to discuss deformable substrates.    

%\subsection{Rigid substrate: strain and confinement}
\subsection{Strain and confinement \label{sec:StrainConfinement}}
%{\emph{Strain and confinement:}} 
%Let us first discuss the strains $\delta_g,\delta_m$.
%[, and the related stress field $\sigma$ which is governed by Eq. \eqref{FVK-1}.] 

%Note that $\alpha$ can take any positive value.
%The relevance of this ratio is reflected 

\begin{figure}
\includegraphics[width=89mm]{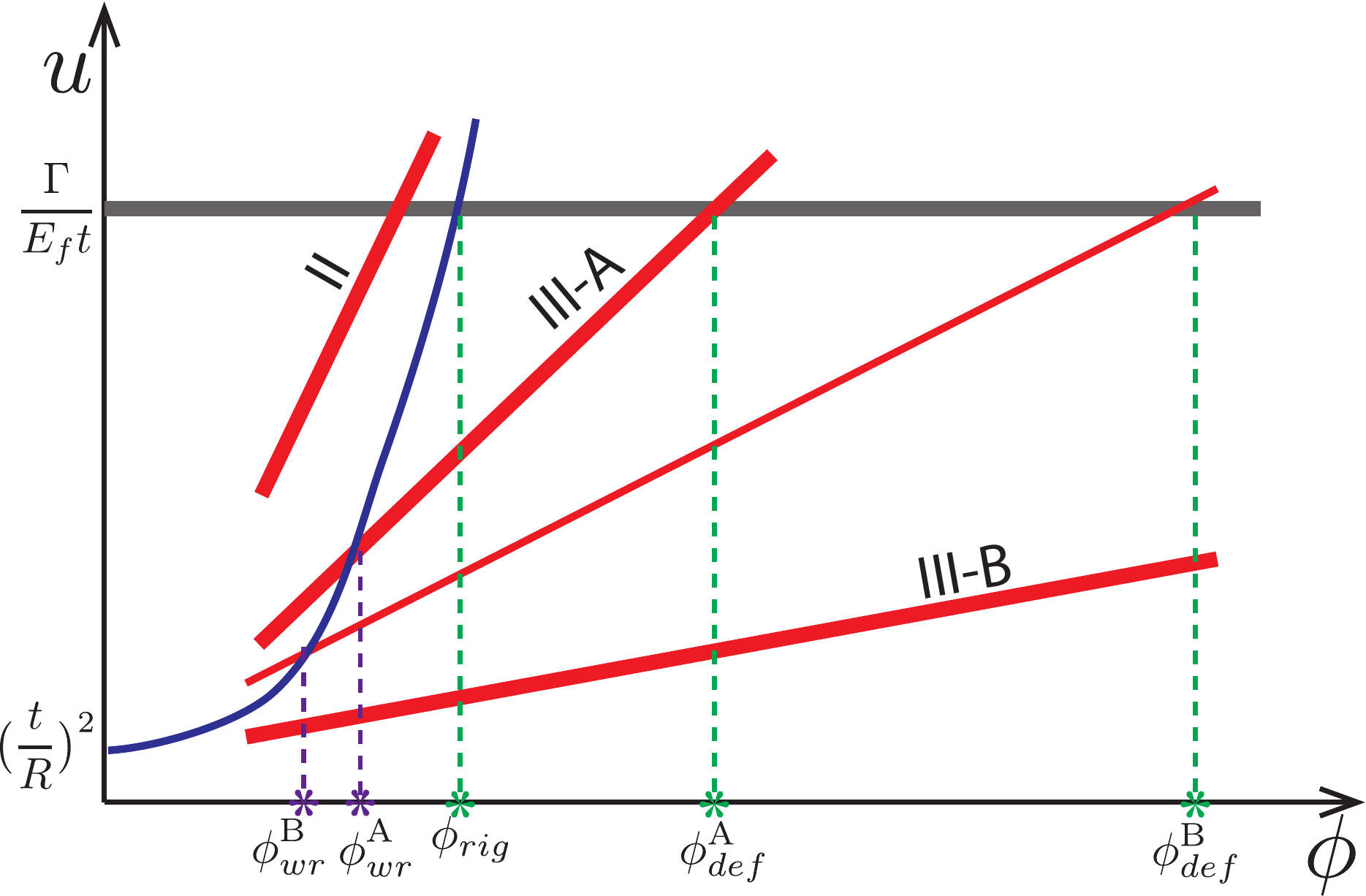}
\caption{\label{fig:fig2} A schematic diagram comparing the various normalized energies upon increasing the coverage fraction $\phi$: the energy of axisymmetric configuration, $u^{\rm axi}$ (blue line); the energetic cost of adhesion, $\Gamma/E_f t$ (black line); and the two components of the wrinkle energy, $u^{\rm wr} = u^{\rm dom} +u^{\rm  sub}$ (Eq.~\ref{uwr}), where $u^{\rm dom} \sim w_{\rm surf}$ is the 
tensile work done by the adhesive substrate pulling on the film (thin red line), which overrides the  
energy stored in the compression-free stress field (see Eqs.~\ref{elasenergy-fft-1},\ref{eq:wrinkle-energy}), 
and $u^{\rm sub}$ is the energetic cost of bending and substrate deformation associated with the formation of wrinkles (Eq.~(\ref{usub2}), thick red line). 
%The energy of the wrinkled configuration $u_{wr}$ is the sum of $u_{dom}$ (thin red line) and $u_{sub}$ (thick red lines).
 %$u_{wr}=u_{dom}+u_{sub}$.  
A wrinkled state emerges at values of $\phi > \phi_{wr} $, where the axisymmetric state is compressive near the edge and the combination of bending and substrate deformation cost is sufficiently small.   
The energy component $u^{\rm sub}$ is plotted 
 %as labeled dotted blue lines 
for parameter regimes II, III-A, and III-B.
% , and the energy $u_{sub}$ is plotted as the light blue line. 
 In regime II, $u^{\rm wr}\approx u^{\rm sub} \gg u^{\rm axi}$ for all  coverages $\phi$, and delamination occurs at $\phi=\phi_{rig}$  when $u^{\rm axi}$ exceeds $\Gamma/E_f t$. In regime III-A $u^{\rm wr}\approx u^{\rm sub}$, wrinkling sets in at $\phi=\phi_{wr}^A$, %when $u_{axi}>u_{sub}$,
 and delamination occurs at $\phi=\phi_{def}^A$. In regime III-B, $u^{\rm wr}\approx u^{\rm dom}$ (thin red line), wrinkles emerge at $\phi_{wr}^B$, and delamination is suppressed until 
a large coverage fraction $\phi_{def}^B \sim O(1)$ is reached. 
% for all  $\phi\lesssim 1$. 
The nonzero intercept of the energy $u^{\rm axi}$ 
%and $u_{sub}$ 
reflects the (small) bending energy of a film with macroscale curvature $1/R$.}
\end{figure}

An infinitely rigid, undeformable substrate %({\emph{i.e.}} $K=\infty$ and hence $\tilde{K}^{-1}=0$) 
must keep its spherical shape, hence $\zeta(r)%^{\rm axi}(r;\alpha) 
= \zeta_{\rm sph} \approx -r^2/2R$. In this case, Eqs.~(\ref{eq:div-axi}) can be solved analytically \cite{Majidi08} and we find:
\begin{gather}
%\sigma^{\rm axi}_{\theta \theta} (r)/\gamma=\frac{\alpha}{16}\Big[ 1- 3 \Big(\frac{r}{W} \Big)^2 \Big] + 1 \nonumber 
\sigma^{\rm axi}_{\theta \theta} (r;\tilde{K}\!=\!\infty,\alpha)\ =\ \gamma\{{\alpha}[ 1- 3 ({r}/{W})^2 ]/16 + 1 \}\nonumber 
\\  %\label{stressaxi1} \\
\sigma^{\rm axi}_{rr} (r;\tilde{K} =\infty,\alpha) \ = \ \gamma \{{\alpha}[ 1-  ({r}/{W})^2]/16  + 1 \} \label{stressaxi2} ,
%\sigma^{\rm axi}_{rr} (r)/\gamma =\frac{\alpha}{16}\Big[ 1-  \Big(\frac{r}{W} \Big)^2 \Big]  + 1 \label{stressaxi2} ,
\end{gather}
Thus, up to a constant factor ($\gamma$), the stresses are completely determined by the confinement parameter $\alpha = \delta_g/\delta_m$. Obviously, for a substrate of infinite stiffness, the substrate deformation energy vanishes, and the energy of the axisymmetrically laminated film is dominated by the part $\bU_{\rm strain}$, Eq.~(\ref{general-elas-energy}), whose normalized version $\bU_{\rm strain}/(E_ft)W^2$ becomes:  
\begin{equation} 
u^{\rm axi}  = %u_{axi} = 
\pi \delta_m^2 [-1+\Lambda + \tfrac{1}{384} \alpha^2]  \ . \label{elasenergy-axi}
\end{equation}
These exact expressions for the stress and energy provide us an insight into the nature of the axisymmetric state, which is useful also for the case of a deformable substrate.     
%In the axisymmetrically laminated state of the system (see Fig.~1 of {\emph{SI}}). 
%For the  the shape is characterized by a profile $\zeta^{axi}(r;\alpha)$, and the stress components are radial, $\sigma^{axi}_{rr}(r;\alpha)$, and azimuthal (hoop), $\sigma^{axi}_{\q\q}(r;\alpha)$. 
When the confinement $\alpha$ is small,  %below some threshold value, 
the tension exerted by the substrate on the film 
%adhesion 
is sufficiently strong (alternatively, curvature is sufficiently weak), and both stress components are tensile everywhere. In this range, the normalized elastic energy 
can be estimated as $u^{\rm axi} \sim \delta_{m}^2$, independent of the coverage fraction $\phi$. In contrast, when $\alpha \gg 1$, the strong geometric strain induces azimuthal compression ($\sigma_{\q\q} <0$) near the perimeter, at a zone that extends as $\alpha$ increases. At large confinement ($\alpha \gg 1$) the energy of the axisymmetric state becomes dominated by the geometric strain $u^{\rm axi} \sim \delta_g^2 \sim \phi^2$. We thus obtain the asymptotic scaling rules: 
\begin{equation}
u^{\rm axi}(\phi) \sim  \delta_{m}^2 \ \  \text{for} \ \alpha \ll 1  \ \ ; \ \ 
u^{\rm axi} (\phi)\sim  \phi^2 \ \ \text{for} \ \alpha \gg 1  
\label{axi-trans}
\end{equation} 
The function $u^{\rm axi}(\phi)$ 
%in the regime large confinement 
is depicted in Fig.~\ref{fig:fig2} (blue line). 
Interestingly, as the confinement $\alpha$ becomes large, we find from Eq.~(\ref{geom-axi}) that $\varepsilon_{\q\q} + \varepsilon_{rr} \approx 0$, such that area of the sheet is unchanged (Eq.~\ref{dA1}). 
%namely $dA_{\rm fil}$ approches zero as $\alpha \to \infty$. 
However, one should note that this invariance of the area is comprised of significant stretching of radials and shrinking latitudes on the sheet, such that $\varepsilon_{rr}$ and $\varepsilon_{\theta\theta}$ are both proportional to the geometric strain $\delta_g \sim (W/R)^2$. 
The corresponding stress profiles (Eq.~\ref{stressaxi2}) are plotted in Fig.~4 (black curves). From the expression for $\sigma^{\rm axi}_{\q\q}$ one immediately obtains that the critical value at which a compressive zone emerges is $\alpha^*[\tilde{K} =\infty] = 8$.   

% and compared with the other energies in the problem.  
As the coverage fraction $\phi$ increases, the (normalized) adhesion energy $u_{\rm ad}  = \Gamma/(E_ft)$ becomes comparable to $u^{\rm axi}(\phi)$, 
%at the large confinement regime, 
and for  $\phi \!>\! \phi_{rig}$ %(Eq.~\ref{max-phi-rigid}) %= \sqrt{\Gamma/E_f t}$ 
delamination becomes energetically favorable in comparison to
the axisymmetrically laminated state. % (Eq.~\ref{max-phi-rigid}). 
This analysis shows that 
%Since 
delamination is expected %to occur 
at high confinement, $\alpha \gg 1$; hence, we focus our analysis in this paper on this asymptotic parameter regime.   
%of the film.   
%%%%%%%%%%%%%%%%%%%%%%%%%%%%

%For $\alpha <8$ both stress components are tensile ($>0$) everywhere, whereas for $\alpha >8$ the hoop stress becomes compressive ($\sigma^{\rm axi}_{\q\q}(r) <0$) at an annulus near the perimeter of the film. 

\begin{figure*}
\centering
%\subfigure{\label{lame-a}\includegraphics{FFT-plots}}
\subfigure{\label{lame-aa}
\includegraphics[width=15cm]
{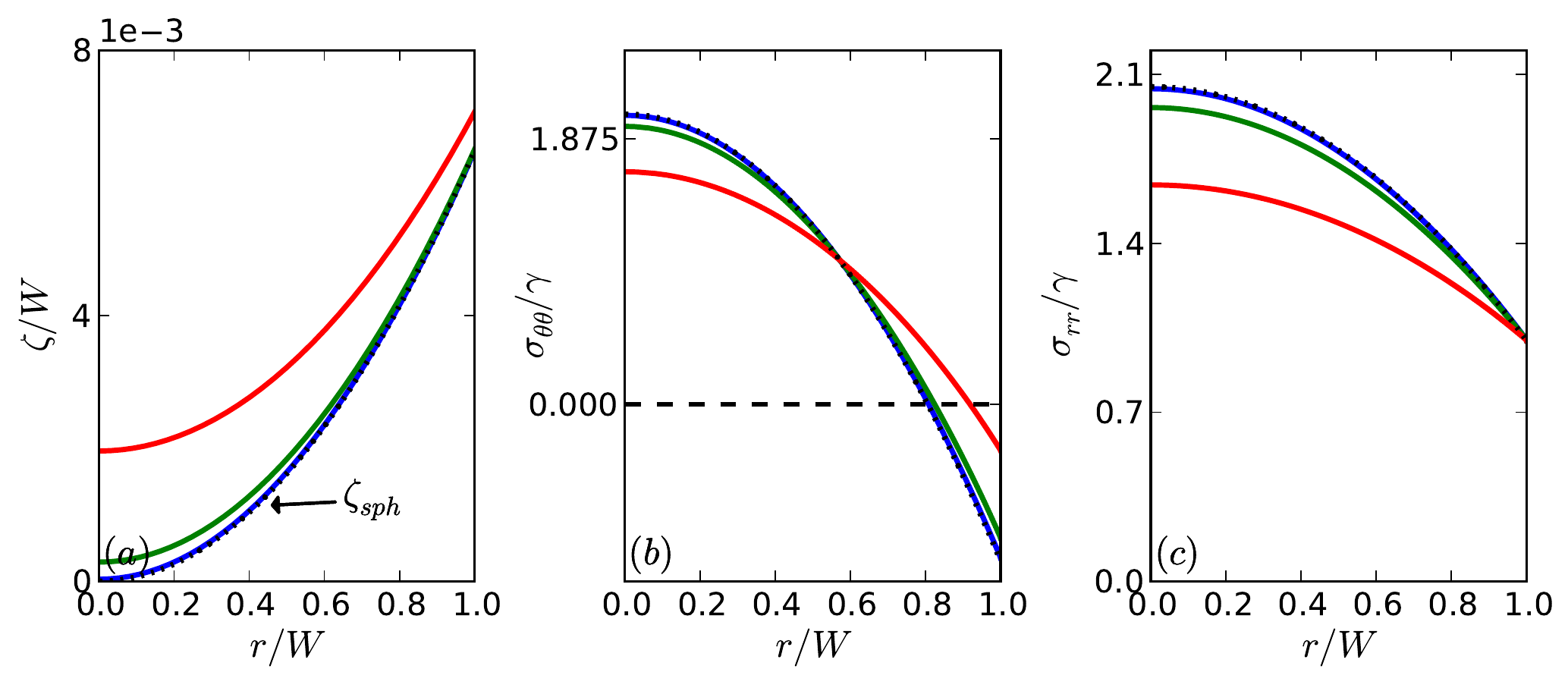}}
%\subfigure{\label{lame-b}}
%\subfigure{\label{lame-c}}
\caption{\label{fig:axi-symmetric-1}
Plots of the shape $\zeta(r)$ and the stresses $\sigma_{rr}(r),\sigma_{\q\q}(r)$ in the unwrinkled, axisymmetric configuration for a fixed 
%(large) 
value of the confinement $\alpha\approx 17$ and a few representative values of $\tilde{K}$:  $\tilde{K}=1$ (red),   $\tilde{K}=10$ (green), $\tilde{K}=100$ (blue), and $\tilde{K}=\infty$ (dotted black lines) in which case the %(non-oscillatory) 
vertical displacement of the plate is exactly $\zeta_{sph}$ and the stress is given by the analytic solution, Eqs.~(\ref{stressaxi2}).}
%0},\ref{eq-new2},\ref{eq:lengh-wr}).}  
%and $\tilde{K}$ close to $1/\alpha$. }
%Plots of the shape $\zeta(r)$ and the stresses $\sigma_{rr}(r),\sigma_{\q\q}(r)$ for a fixed (large) value of the confinement $\alpha$ and a few representative values of $\tilde{K} < 1$.}
\end{figure*}

%{\emph{Deformability:}}  
\subsection{The deformability parameter \label{sec:deformability-parameter}}
What happens when the substrate is not infinitely rigid? Since the profile $\zeta^{\rm axi}(r)$ is allowed to deviate from the ideal spherical shape, analytic solution to the FvK Eqs.~(\ref{eq:div-axi}) is not available, and 
we resort to numerical analysis (using an integration method similar to \cite{HohlfeldThesis}).  
%we employ instead a standard numerical package to integrate the equations. 
A few representative plots of the stresses and profiles are shown in Fig.~\ref{fig:axi-symmetric-1}. %This figure shows that 
The characteristic behavior of the stress field remains valid also for finite values of $\tilde{K}$, whereby both radial and hoop components are tensile for sufficiently small confinement, and a hoop compression emerges above a ctirical value $\alpha^*(\tilde{K})$. The critical value $\alpha^{*}(\tilde{K})$ is defined by the implicit equation:
\begin{equation} 
\sigma^{\rm axi}_{\q\q}[r=W; \tilde{K},\alpha^{*}(\tilde{K})] = 0
\label{eq:critical}
\end{equation}
For a nearly rigid substrate, $\tilde{K} \to \infty$, our numerical solution shows that $\alpha^*(\tilde{K}) \to 8$, in agreement with the above analytic result for the infinitely rigid substrate. In this limit, the substrate's spherical profile is barely deformed. 
However, as $\tilde{K}$ decreases, $\alpha^*(\tilde{K})$ increases.
% and appears to diverge as   $\alpha^*(\tilde{K}) \to 0$. 
This trend is accompanied by the significant deformation of the substrate beneath the film, as can be observed in experiments of a film floating on a liquid drop (Fig.~1c). 

In order to understand the effect of the deformability parameter $\tilde{K}$, it is useful to consider the enery $u^{\rm axi}$ of the axisymmetric state. Intuitively, if the susbstrate is not infinitely rigid, the strain, and hence the elastic energy $u^{\rm axi}$ may be reduced by flattening the substrate beneath the film such that the effective radius of curvature there becomes $R_{\rm eff} >R$. Such a mechanism is clearly instrumental for suppressing delamination of films from a liquid drop \cite{Py07}. 
How soft must a substrate be in order that such a mechanism be operative? In the high confinement limit ($\alpha\gg1$) we estimate the strain of an axisymmetric state 
%with 
%in an unwrinkled film with 
%curvature $1/R_f$ 
by $\sim (W/R_{\rm eff})^2$ and the stretching energy is $u_{\rm strain} \sim (E_f t) (W/R_{\rm eff})^4$, favoring large $R_{\rm eff}$. The displacement %$\delta\zeta$ 
from the original spherical shape is estimated as $\delta\zeta_{\rm sph} \sim  W^2 (1/R_{\rm eff} - 1/R)$, and the resulting substrate energy is $u_{\rm Win} \sim K \delta \zeta_{\rm sph}^2$. Comparing $u_{\rm strain}$ and $u_{\rm Win}$, 
we find that $R_{\rm eff} \approx R$ if the dimensionless deformability parameter $\tilde{K}^{-1} \ll 1$, hence we conclude that in this parameter range lamination yields only small deviations from the original spherical shape of the substrate. In contrast, regime I, which we define as \cite{Comment-deformability}: 
\begin{equation}
\text{regime I:} \ \ \tilde{K} = KR^2/(E_ft) \  \ll 1  \ , \label{regimeI}
\end{equation}
is characterized by large distortion of the substrate. Accordingly, we expect that the scaling behavior of the energy $u^{\rm axi}$ is correctly decribed by Eq.~(\ref{axi-trans}) for $\tilde{K} \gg 1$, where $R_{\rm eff} \approx R$ and the substrate deformation energy is negligible. However, the energy $u^{\rm axi}$ in regime I, of highly-deformable substrate,  %($\tilde{K}$) 
is significantly lower than the estimate~(\ref{axi-trans}), such that delamination could be avoided. % \cite{Comment-High-Def}. 
%even without considering the further reduction of the energy through wrinkles is not )     
In the rest of this paper we will focus on the low deformability regime, $\tilde{K} \gg 1$. The interesting physics of regime I will be addressed elsewhere \cite{Number2}.

\section{The wrinkled state \label{sec:Wrinkledstate}}
%{\emph{A ``wrinklogami" pattern:}} 
The wrinkling instability has been shown recently to hinder delamination of a uniaxially-compressed film floating on a planar liquid surface \cite{Wagner-Vella11}.
Here we show that wrinkling should emerge even for a film attached to a curved, nearly rigid substrate. Furthermore, in the next section we will show that the suppression of the elastic energy %of the laminated state 
enabled by the formation of wrinkles has dramatic consequences on the delamination mechanism, that could not be addressed by the 1D geometry of \cite{Wagner-Vella11}.

%*** Note that $\epsilon = \epsilon_g$ ***
%We thus focus in this section on wrinkling analysis of our model system in the parameter regime of low substrate deformability and high confinement: $\tilde{K}^{-1} \ll 1, \alpha \gg 1$. While the forthcoming analysis is similar in spirit to recent studies of radial wrinkle phenomena, there are numerous crucial differences. We will emphasize here the central %a few important 
%aspects that distinguish wrinkling in the parameter regime $\tilde{K} ,\alpha \gg 1$ from previous studies of wrinkling phenomena. %{\emph{tensional}} \cite{PNAS11,PNAS12} and {\emph{substrate-dominated}} \cite{PNAS13} wrinkling phenomena that did not address the joint effect of stiff curved substrate and large azimuthal confinement on the wrinkle pattern. 

Recent studies of wrinkle patterns in thin films under tensile loads in 2D set-ups have employed the ``far-from-threshold" (FT) method -- a singular perturbation of FvK equations around a compression-free state of the film. The small parameter in the FT expansion is the inverse of the bendability parameter $\epsilon$, Eq.~(\ref{gener-bendability}), which is the only dimensionless group that depends on the bending modulus. We will implement the FT method, highlighting a
%A 
unique aspect of the current work -- a {\emph{doubly-asymptotic}} analysis of the wrinkle pattern, which involves both limits of high-bendability ($\epsilon^{-1} \to \infty$) and large confinement $\alpha \to \infty$. Considering Fig.~5, which depcits a wrinkled laminated state, one may associate the first limit ($\epsilon^{-1} \to \infty$) as giving rise to divergence in the number of wrinkles and correspondingly to vanishing of their amplitude, and the second limit ($\alpha \to \infty$) with a maximal extension of their length, such that they occupy almost the whole sheet.  
\begin{figure}[b]  
\includegraphics[width=.5\columnwidth]{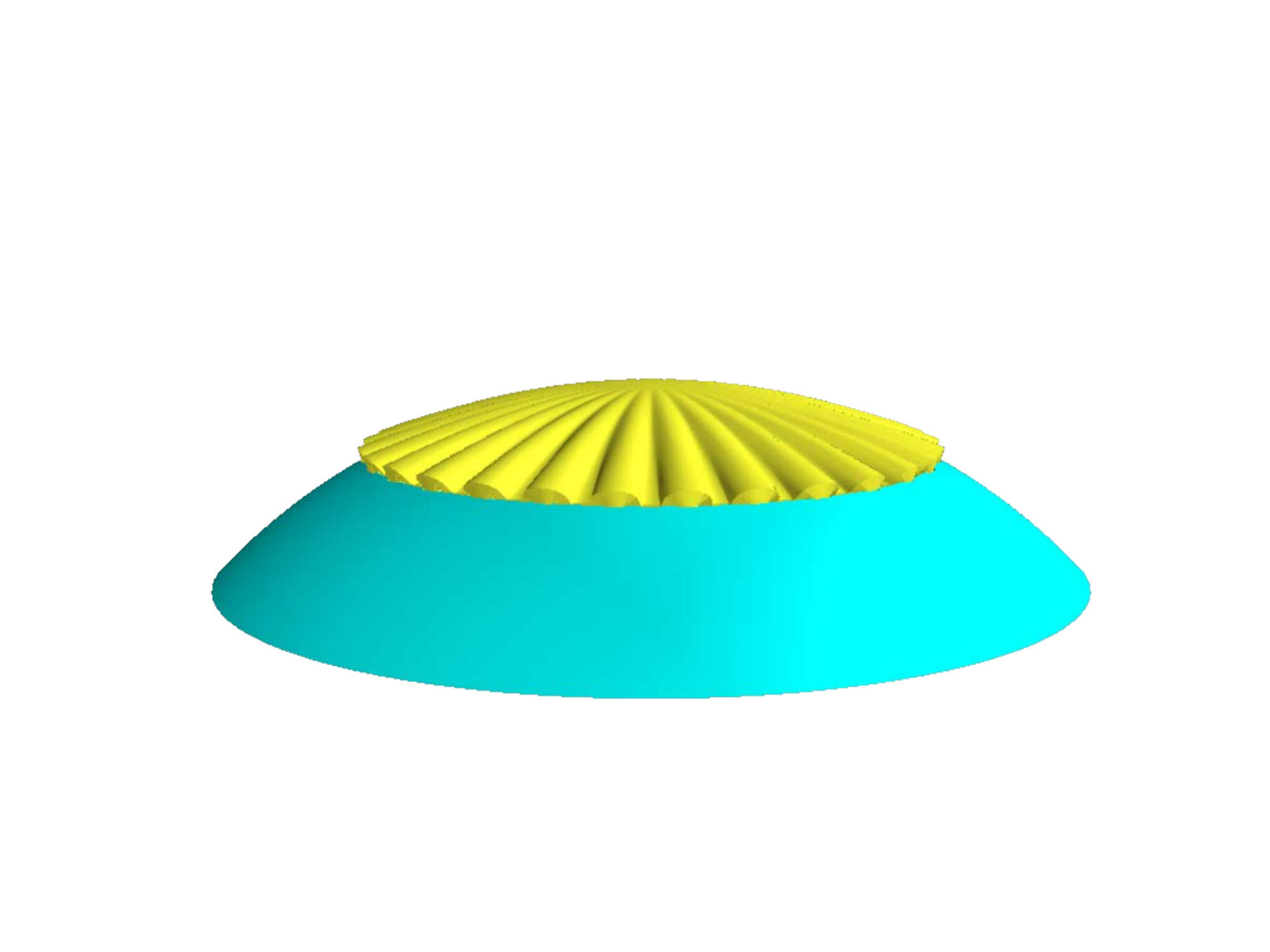}
 \caption{\label{fig:axi-symmetric} 
The wrinkled shape of a laminated film (yellow) on a spherical substrate (blue). The out-of-plane undulations of the film and the attached substrate are accompanied by in-plane oscillations of the perimeter of the film, which are of the same periodicity but with amplitude reduced by a factor of $W/R$. The amplitude of the wrinkles has been exaggerated to highlight their shape. }
\end{figure}

%\begin{figure}[b]
%\centering
%\subfigure{\label{lame-a}\includegraphics[width=.5\columnwidth]{SF3.pdf}}
%\caption{\label{fig:axi-symmetric} 
%The wrinkled shape of a laminated film (yellow) on a spherical substrate (blue). The out-of-plane undulations of the film and the attached substrate are accompanied by in-plane oscillations of the perimeter of the film, which are of the same periodicity but with amplitude reduced by a factor of $W/R$. The amplitude of the wrinkles has been exaggerated to highlight their shape. }
%\end{figure}

This type of asymptotic analysis, which corresponds to the parameter regime at which the wrinkle pattern becomes ``wrinklogami", namely, asymptotically isometric to the planar state of the film,    
distinguishes our work from those recent studies that addressed only the limit  of high-bendability ($\epsilon^{-1} \to \infty$) \cite{PNAS11,PRE12,PNAS13}. Understanding the doubly-asymptotic limit is essential for our study, since we found above (Sec.~\ref{sec:StrainConfinement}) that it is the regime of large $\alpha$ %where surface tension is sufficiently low such that 
where delamination may become energetically favorable. 

\subsection{Principles of the far-from-threshold expansion  \label{sec:Principles-FT}} 
Similarly to the classical Euler buckling of compressed rods, radial wrinkles relax the hoop confinement and enable the laminated film attain a compression-free state whose energy $u^{\rm wr}$ is lower than the elastic energy $u^{\rm axi}$ of the compressed, axisymmetric ({\emph{i.e.}} unwrinkled) state.
Obviously, wrinkles require additional energy, due to the bending resistance of the film and the stiffness of the substrate. 
%\sout{(which are proportional, respectively, to $\epsilon_\Gamma, \tilde{K}$).}  
We evaluate the wrinkle energy $u^{\rm wr}$ by assuming the most basic wrinkled film-substrate shape \cite{footnote4}:
%\footnote{We consider here the most elementary mechanisms for wrinkling [simply periodic shape, Eq. \eqref{wrinkleform}] and delamination (such that the whole film detaches). The actual instabilities may be more complex (e.g. wrinkling cascades, and a periodic pattern of blisters, respectively). Nevertheless, we do not expect this to affect the scaling laws  [Eqs. (\ref{phi-wr}-\ref{eq:elastic-soft-rig})] that are presented in Fig. 1.}
\begin{subequations}
\label{disp-wrink}
\begin{equation}
\zeta (r,\theta) \approx \zeta_{\rm sph}(r) + f(r)\cos(m\theta) \ , \label{wrinkleform} 
\end{equation}
where $\zeta_{\rm sph}(r) \approx -r^2/R$ is the spherical profile, and the wrinkle amplitude $f(r)$ decreases as the sheet becomes thinner, or, more accurately, as the bendability $\epsilon^{-1}$ increases. 
Implicit in Eq.~(\ref{wrinkleform}) is our consideration of the low-deformability regime, $\tilde{K}^{-1} \ll 1$, where the non-oscillating part of the shape is approximated by the undeformed shape of the sphere.
%, such that deviations from $\zeta_{sph}(r)$ %this shape
%of the shape $\zeta(r,\q)$ from $\zeta_{sph}(r)$ are assumed to be 
%are proportional to the asymptotically vanishing wrinkle amplitude $f(r)$.
As we will see below, the vertical displacement (\ref{wrinkleform}) must be accompanied by radial and azimuthal displacements: 
%\begin{subequations}
%\label{disp-wrink}
\begin{align}
\ru_r(r,\theta) &= \ru_r^{(0)}(r) + \ru_r^{(m)}(r)\cos(m\theta) \ , \label{disp-wrink-1} \\
\ru_\theta(r,\theta) &=  \ru_\theta^{(2m)}(r)\sin(2m\theta)   \label{disp-wrink-2} \ . %\\
%\zeta(r,\theta) &= \zeta_{sph}(r) + f(r)\cos(m\theta) \ .  \label{disp-wrink-3} %\label{eq:anzats}
\end{align}
\end{subequations}  
The displacement field, represented in Eqs.~(\ref{disp-wrink}) by the functions $f(r), \ru_r^{(0)}(r) , \ru_r^{(m)}(r), \ru_\theta^{(2m)}$, is found as a singular, far-from-threshold (FT) expansion of FvK equations (\ref{eq:div}) \cite{PNAS11,King12}. In this approach,      
%%%%%%%%%Employing the ``far-from-threshold" formalism which was introduced recently for radial stretching problems \cite{PNAS11,King12},  
%we assume that the wrinkled state is completely determined by the deformability $\tilde{K}^{-1}$, confinement $\alpha$, and bendability $\epsilon^{-1}$ parameters, and 
we expand the FvK equations around the singular limit of infinite bendability ($\epsilon \to 0$). Namely, the energy $u^{\rm wr}$ of the wrinkled state is assumed to have the form: 
%decomposed into two parts:
\begin{equation} 
u^{\rm wr} = u^{\rm dom} + u^{\rm sub}, \, \, \label{uwr}
\end{equation}   
where 
\begin{equation}
{u^{\rm sub}}/{u^{\rm dom}}  \approx g(\tilde{K},\alpha) \cdot \epsilon^{\beta}  \ \ \ \text{for} \ \  \epsilon \to 0 \ , 
\label{energy-ratio}
\end{equation}
with $g(\tilde{K},\alpha)$ some unknown function and $\beta >0$.
%where 
Here, $u^{\rm dom}$ is the ``dominant'' energy stored in the asymptotic, compression-free stress field, which consists of the straining energy $u_{\rm strain}$ (Eq.~\ref{general-elas-energy}) and the work $w_{\rm surf}$ of the tensile load at the edge of the film (Eq.~\ref{eq:define-Wsurf}); this energy term depends only on the ``macroscopic" parameters $\tilde{K}^{-1}$ and $\alpha$, and bares no explicit dependence on the bendability $\epsilon^{-1}$ or any  
%
%does not depend explicitly on the 
small-scale features of the wrinkle pattern.  
%but does depend on the adhesive tension $\Gamma$ exerted at the edge , 
In contrast, the energy $u^{\rm sub}$ is the ``sub-dominant'' energy which depends on the wrinkle number $m$ and amplitude $f(r)$ and is
determined by a balance of bending and substrate-stiffness. 

It is crucial to emphasize a simple yet somewhat confusing point:
The FT relation~(\ref{energy-ratio}) means that if the confinement $\alpha$ (as well as $\tilde{K}$) is held fixed, 
%all other parameters except $\epsilon$ ({\emph{i.e.}} $\tilde{K}$ and $\alpha$) are held fixed, 
and the bendability $\epsilon^{-1}$ increases indefinitely, then the energy $u^{\rm sub}$ becomes smaller than $u^{\rm dom}$ for sufficiently small $\epsilon$. This basic feature, which was noted in previous FT studies of wrinkle patterns \cite{PNAS11}, motivated the usage of the notations ``dom" and ``sub" for the respective energy terms.      
%However, the asymptotic analysis addressed in this section involves the simoultaneous limits $\alpha \to \infty$ and $ \epsilon^{-1} \to \infty$ of both confinement and bendability parameters.
However, when analyzing the doubly asymptotic limit $(\alpha \to \infty$ and $ \epsilon^{-1} \to \infty)$, of both confinement and bendability parameters, we must pay special attention to the pre-factor $g(\tilde{K},\alpha)$ in Eq.~(\ref{energy-ratio}). We will show that $g(\tilde{K},\alpha)$ vanishes for large $\alpha$. As a conclusion, 
%while the energetic expressions computed below do satisfy the FT limit~(\ref{energy-ratio}), we will 
we must take into consideration the counter-intuitive possibility that
%encounter situations where 
the %sub-dominant 
energy term $u^{\rm sub}$ may actually be larger than $u^{\rm dom}$. %\cite{Comment-switch-dom-sub}. 

The above paragraph highlights a potential source of confusion in our analysis, since it requires the implementation of the FT formalism -- an expansion in the inverse bendability $\epsilon$ -- to situations where another independent parameter (the confinement $\alpha$) becomes asymptotically large. We thus need to clarify the meaning of adjectives such as ``finite", ``diverging", and ``vanishing". Unless specifically stated otherwise, we will use the symbol ``$\to$" 
to denote the limit $\epsilon \to 0$ (for fixed values of the parameters $\tilde{K}$ and $\alpha$), and will attribute the above adjectives to the
asymptotic behavior in this limit. % $\epsilon \to 0$ (for fixed values of the parameters $\tilde{K}$ and $\alpha$).
For instance, we shall refer to the energy $u^{\rm dom}$ as ``finite" since it approaches $\epsilon$-independent limit as $\epsilon \to 0$, and to $u^{\rm sub}$ as ``vanishing" since it scales $\sim \epsilon^{\beta}$ as $\epsilon \to 0$, but this terminology does not mean
%wording should not be interpreted as implying 
that $u^{\rm dom} >u^{\rm sub}$ at a given pair of values of $\epsilon$ and $\alpha$.   
Another important example pertains to the amplitude of wrinkles and their number. The amplitude $f(r)$ is vanshing and the number $m$ is diverging, whereas their product $m\cdot f(r)$ approaches a finite ({\emph{i.e.}} $\epsilon$-independent) limit, which is necessary to     
%Nevertheless, 
%in order to enable the 
collapse the compression in the azimuthal direction. 
Other, more obvious examples of finite objects, are the axisymmetric component of the radial displacement $\ru_r^{(0)}(r)$, the fixed slope $\ddr \zeta_{\rm sph}$, and the radial strain $\varepsilon_{rr}$.

The basic structure of this far-from-threshold expansion appeared already in \cite{PNAS11}, which considered radial wrinkles in a planar (Lam\'e) set-up, and its singular nature was elaborated in \cite{PRE12}. However, the displacement field~(\ref{disp-wrink}) differs from that study by the existence of an axisymmetric contribution $\zeta_{\rm sph}$ to the out-of-plane displacement field, and by the related, harmonic contribution $\ru^{(m)}(r)\cos(m\q)$ to the radial displacement. 
%Note also, that implicit in Eq.~(\ref{eq:wrinkle-form}) is our consideration of the low-deformability regime, $\tilde{K}^{-1} \ll 1$, where the non-oscillating part of the shape is approximated by the undeformed shape $\zeta_{sph}(r)$, such that deviations of the shape $\zeta(r,\q)$ from $\zeta_{sph}(r)$ are assumed to be proportional to the asymptotically vanishing wrinkle amplitude $f(r)$.
%\footnote{A natural question is whether the assumed single-mode shape, Eq.~(\ref{eq:wrinkle-form}), does characterize the minimal energy state of a laminated film. While it is possible that a more complicated pattern, such as multi-mode wrinkle shape (where $f(r)\cos(m\theta) \to \sum_m f_m(r) \cos(m \theta)$) may lower the energy further, our predicted scaling laws remain valid. Details of this analysis will appear elsewhere.}.      
%
%As the names suggest, one may expect that $|u_{dom}| \gg u_{sub}$. This inequality does indeed characterize previous studies of {\emph{tensional}} wrinkle phenomena.  In those studies \cite{PNAS11,PNAS12} (that did not address the effect of stiff substrate and large confinement), the compression-free stress field that underlies the wrinkle pattern contains tensional components whose elastic energy constitues a finite fraction of the energy $u_{axi}$ stored in the axisymmetric (unwrinkled) state of the system. In contrast, the current study shows that in the low deformability, large confinement regime ($\tilde{K}^{-1} \ll 1, \alpha \gg 1$): $$ |u_{dom}| \ll u_{sub}  \ \ \Rightarrow u_{wr} \sim u_{sub} \ .  $$
In the following subsections we derive the dominant and sub-dominant energies, and highlight the unique aspects of the wrinkled state in this problem.  

\subsection{The compression-free stress field \label{sec:Compfree}}% and the energy $u_{dom}$}

%The computation of the compression-free stress field underlying radial wrinkle patterns in axisymmetric set-ups has been discussed in detail recently \cite{PNAS11,PNAS12}. 
For the specific system we address here, of a circular film attached to a spherical substrate, the compression-free field and its associated energy $u^{\rm dom}$ have been calculated analytically in the limit $\tilde{K} \to \infty$, of an infinitely rigid substrate \cite{PNAS13}. Therefore, in this subsection we will briefly describe this result, and will refer the reader to \cite{PNAS13} for a detailed calculation. 
%stress field, which we denote as $\sigma^{\rm FFT}$, and its energy $u^{\rm dom}$. 
%The detialed calculation of \cite{PNAS13} is reviewed in appenix ...   

Since very thin films cannot support compression, the stress field that underlies the wrinkle pattern (often called the ``membrane'' limit \cite{MansfieldBook}) is assumed to satisfy $\sigma_{ii} \geq 0$ in the high bendability limit $\epsilon \to 0$, where $i$ labels the two principal directions of the stress tensor \cite{Stein61,Pipkin86,Steigman90}. This principle is also known as ``tension field theory'' \cite{Stein61, MansfieldBook,Steigman90} 
or ``relaxed energy'' \cite{Pipkin86}.   
For our laminated, axially-loaded film, this condition is naturally realized by solving the force balance Eqs.~(\ref{eq:div-axi}) in two distinct zones: an inner one ($0<r<L$) and an outer one $(L<r<W)$, separated at some radius $r=L$, where appropriate matching conditions are invoked \cite{Comment-FT}. 

In the inner disk, ($0<r<L$), both radial and hoop stresses are purely tensile, and are described by the corresponding stress $\sigma^{\rm axi}(r;\tilde{K},\alpha)$ of the axisymmetric state, %(Eq.~\ref{stressaxi2}), 
%\noindent
%$\bullet$ At $0<r<L$, both diagonal stresses are purely tensile: $\sigma_{rr}>0,\sigma_{\q\q} >0$. There, the film remains unwrinkled and its state is described by the axisymmetric stress field $\sigma^{\rm axi}(r;\tilde{K},\alpha)$ 
upon replacing:
\begin{equation} W \to L  \ \  ; \  \ \gamma \to \sigma_{rr} (L)    \  \ ;\   \ 
\alpha \to \overline{\alpha} = \frac{(E_ft)}{\sigma_{rr}(L)} \cdot (\frac{L}{R})^2  \ . 
\label{subst1}
\end{equation}
%The effective confinement parameter exerted on this tensile part of the film becomes, correspondingly: 
%\begin{equation}\alpha \to \overline{\alpha} = \frac{Y}{\sigma_{rr}(L)} \cdot (\frac{L}{R})^2  \ . 
%\end{equation}
In particular, in the limit $\tilde{K} \to \infty$, the axisymmetric stress field is given analytically by Eq.~(\ref{stressaxi2}), such that: 
\begin{subequations}
\label{stressaxi20}
\begin{gather}
%\sigma^{\rm axi}_{\theta \theta} (r)/\gamma=\frac{\alpha}{16}\Big[ 1- 3 \Big(\frac{r}{W} \Big)^2 \Big] + 1 \nonumber 
(0<r<L): \ \ \frac{\sigma_{\theta \theta} (r)}{\sigma_{rr}(L)}=\tfrac{1}{16}{\overline{\alpha}}[ 1- 3 (\tfrac{r}{L})^2 ] + 1 
\label{stressaxi2a}
\\  %\label{stressaxi1} \\
(0<r<L): \ \ \frac{\sigma_{rr} (r)}{\sigma_{rr}(L)} =\tfrac{1}{16}{\overline{\alpha}}[ 1-  (\tfrac{r}{L})^2] + 1 \label{stressaxi2b} \ .
%\sigma^{\rm axi}_{rr} (r)/\gamma =\frac{\alpha}{16}\Big[ 1-  \Big(\frac{r}{W} \Big)^2 \Big]  + 1 \label{stressaxi2} ,
\end{gather}
\end{subequations}

In the outer annulus, $L\!<\!r\!<\!W$, where wrinkles emerge, the radial stress is still tensile ($\sigma_{rr} >0$) and finite, whereas both $\sigma_{\q\q}$ and $\sigma_{r\q}$ are negligible, namely vanish as $\epsilon \to 0$.   
%(The exact meaning of ``negligible" will be clarified in the next subsection).
The radial stress is thus immediately obtained from Eq.~(\ref{eq:div-axi}) by 
%substituting $\sigma_{\q\q} = 0$ and 
using the BC $\sigma_{rr}(W) = \gamma$, obtaining: 
 \begin{equation}
(L<r<W): \ \ \sigma_{rr} \to \gamma W/r  \ \ ; \ \ \sigma_{\q\q} \to 0  \ . \  
\label{eq-new2}
\end{equation}    

The length $L$ and the stress $\sigma_{rr}(L)$ are determined by requiring continuity of $\sigma_{rr}$ and $\sigma_{\q\q}$ at the borderline $r=L$. This implies that the effective confinement $\overline{\alpha}$ (Eq.~\ref{subst1}) felt by the inner disk is just at the critical value $\alpha^*(\tilde{K})$. We thus obtain:   
% between the wrinkled ($r \to L^{+}$) and unwrinkled ($r \to L^{-}$) zones   
\begin{equation}
L = W (\frac{\alpha^{*}(\tilde{K})}{\alpha})^{1/3} \ \ ; \ \ 
\sigma_{rr}(L) = \gamma (\frac{\alpha}{\alpha^{*}(\tilde{K})})^{1/3} \  .
\label{eq:lengh-wr}
\end{equation}  
As the confinement increases, $\alpha \gg \alpha^{*}(\tilde{K})$, we find that $L$ decreases indefinitely and the wrinkled annulus thus occupies most of the laminated area of the substrate \cite{Scaling-L-difference}.

Figure~6
%\ref{fig:axi-symmetric} 
shows the numerical solution for the shape and the compression-free stress field for various values of $\tilde{K}$, and compares them with the analytic solution for $\tilde{K} \to \infty$, Eqs.~(\ref{stressaxi20},\ref{eq-new2}) where the shape $\zeta(r) \to \zeta_{\rm sph}(r)$. 
%It is seen that substantial perturbation to the infinite-stiffness limit occur only for as $\tilde{K} \to \alpha^{-1} \ll 1$.   

%
%but with the {\emph{compression-free}} condition: $\sigma_{ii} \geq 0$ everywhere. For our laminated, axially-loaded film, this condition is naturally realized by an inner, purely tensile region $(I)$ at $0<r<L$, where both diagonal stresses are purely tensile: $\sigma_{rr}>0,\sigma_{\q\q} >0$, and an outer annulus $(O)$ at $L<r<W$, where the radial stress is still tensile $\sigma_{rr} >0$ whereas $\sigma_{\q\q}=0$        

\begin{figure*}
\centering
%\subfigure{\label{lame-a}\includegraphics{FFT-plots}}
\subfigure{\label{lame-a}
\includegraphics[width=15cm]
{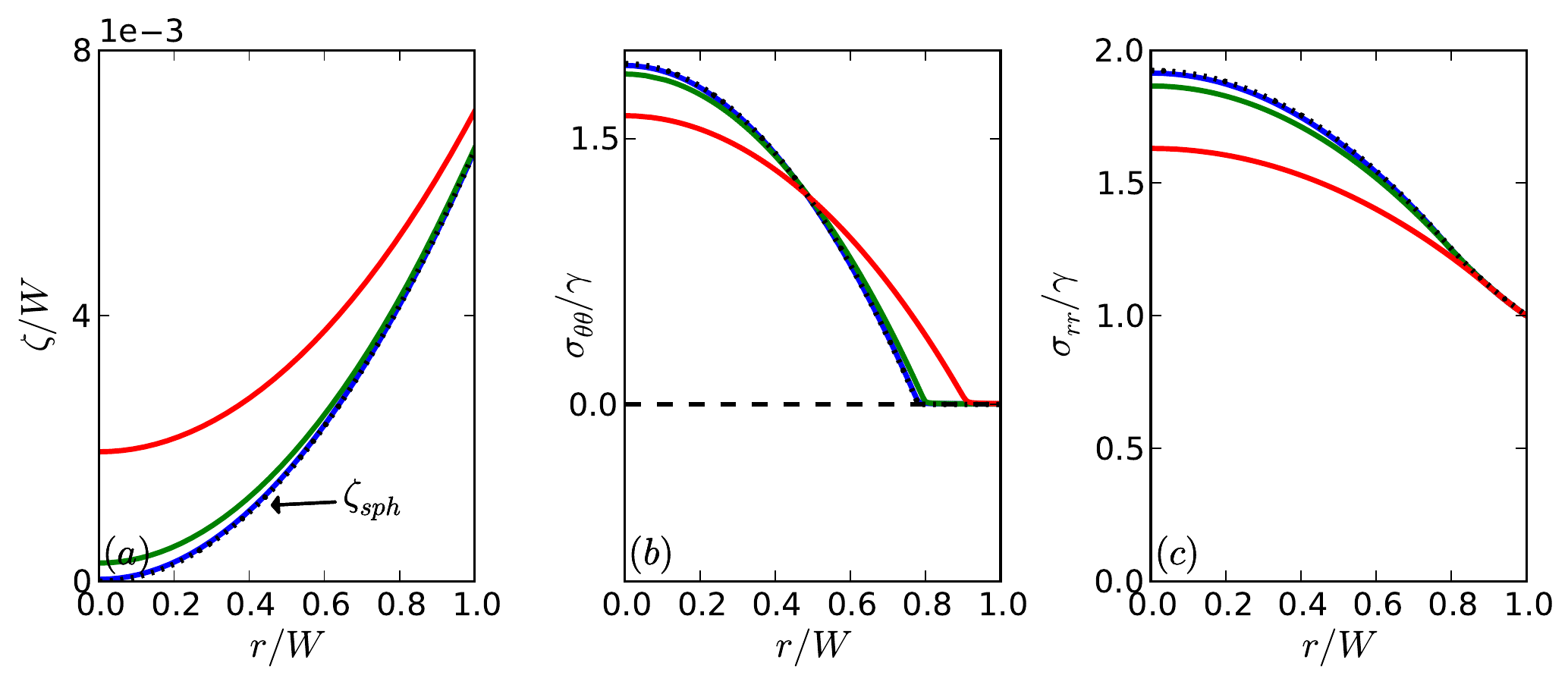}}
%\subfigure{\label{lame-b}}
%\subfigure{\label{lame-c}}
\caption{\label{fig:axi-symmetric}
Plots of the shape $\zeta(r)$ and the stresses $\sigma_{rr}(r),\sigma_{\q\q}(r)$ in the wrinkled configuration for a fixed 
%(large) 
value of the confinement $\alpha\approx 17$ and a few representative values of $\tilde{K}$:  $\tilde{K}=1$ (red),   $\tilde{K}=10$ (green), $\tilde{K}=100$ (blue), and $\tilde{K}=\infty$ (dotted black lines) in which case the (non-oscillatory) vertical displacement of plate is exactly $\zeta_{sph}$ and the stress is given by the analytic solution, Eqs.~(\ref{stressaxi20},\ref{eq-new2},\ref{eq:lengh-wr}).}  
%and $\tilde{K}$ close to $1/\alpha$. }
%Plots of the shape $\zeta(r)$ and the stresses $\sigma_{rr}(r),\sigma_{\q\q}(r)$ for a fixed (large) value of the confinement $\alpha$ and a few representative values of $\tilde{K} < 1$.}
\end{figure*}
% (Perhaps) second panel -- for fixed $\alpha \gg 1$, show how $L(\alpha)$ varies as $\tilde{K}$ increases, and how the change in exponents ($1/5$ to $1/3$) is correlated to the deformation of the substrate. 

%\subsection{The energy $U_{sub}$ due to bending stiffness}

\subsection{Asymptotic displacement and strain \label{sec:Asymdisp}}
The analysis in Sec.~\ref{sec:Compfree}
%Our derivation of the compression-free stress field 
suffices to compute the energy term $u^{\rm dom}$ in Eq.~(\ref{uwr}). In order to evaluate $u^{\rm sub}$ it is essential for us to discuss first a few constraints
on the displacement field, Eq. (\ref{disp-wrink}), which are imposed by the requirement that the stress approaches the compression-free field (Eqs.~\ref{stressaxi20},\ref{eq-new2}).

Our analysis % in Sec.~\ref{sec:compression-free} 
of the compression-free stress field assumed that in the wrinkled zone $L<r<W$, both hoop and shear stresses vanish: $\sigma_{\q\q} ,\sigma_{r\q} \to 0$ as $\epsilon \to 0$. Compatibility of these conditions with the Hookean stress-strain relations, Eq.~(\ref{eq:stresses}), imposes two conditions on the strain tensor in the limit $\epsilon \to 0$:  
%in the wrinkled region require that $\sigma_{r\theta} = \sigma_{\theta\theta}=0$. Using the constitutive law, Eqs. (\ref{eq:sigma_rr}-\ref{eq:sigma_rt}), the compression free condition can be expressed in terms of the strain as
\begin{align}
\varepsilon_{r\theta}& \to 0,\label{eq:compression-free-1}  \ , \\
\varepsilon_{\theta\theta} &\to -\Lambda\varepsilon_{rr}\label{eq:compression-free-2} \ , 
\end{align}
where the radial strain $\varepsilon_{rr}$ is readily obtained from Eq.~(\ref{eq-new2}) and Eq.~(\ref{eq:stresses}): 
\begin{equation}
\varepsilon_{rr}  =  \sigma_{rr}/(E_ft) \to \frac{\gamma}{(E_ft)}\frac{W}{r} \ . \label{eq:radial-strain}
\end{equation}
Similarly to the axisymmetric state (Eq.~\ref{geom-axi}), we find that the areal change of the wrinkled film $dA_{\rm fil}$, Eq.~(\ref{dA1}), approaches zero in the large confinement $\alpha \to \infty$. However, in contrast to the axisymmetric case, here this property truly indicates on the asymptotic isometry of the wrinkled state in this limit, since it stems from the simultaneous suppression of {\emph{every}} component of the strain tensor. %, rather than from their mutual cancellation.    

Considering now the geometric link between %connection between the 
strain and displacement, Eq.~(\ref{eq:strains}), 
we can characterize numerous components of the displacement field (\ref{disp-wrink}). 
%the strains (\ref{eq:compression-free-1}-\ref{eq:radial-strain}) imply numerous consequences, that reflect the distinct nature of wrinkling due to geometric ({\emph{i.e.}} curvature-imposed) confinement versus mechanical confinement \cite{PNAS11}:

%$\bullet$ 
{\emph{Radial strain:} For the axisymmetric component $\ru_r^{(0)}$ of the radial displacement we obtain, using Eqs. (\ref{eq:strain-radial-1},\ref{eq:radial-strain}): 
\begin{equation}
\ddr \ru_r^{(0)} + \frac{1}{2}(\ddr \zeta_{\rm sph})^2 \to  \frac{\gamma}{(E_ft)}\frac{W}{r} \ , \label{eq:radial-disp}
\end{equation}
%(where $\cdot'$ means derivative w.r.t. $r$). 
where $(\ddr \zeta_{\rm sph})^2 \sim (r/R)^2$. % for $r<W$. 
The geometric meaning of this equation is elucidated by considering the infinite confinement limit $\alpha = \delta_g/\delta_m \to \infty$, 
%(recall that the high bendability limit $\epsilon \to 0$), 
where the isometric mapping of radial lines on the curved sphere (namely, $\varepsilon_{rr} =0$) is obtained by radial displacement $\ru_r^{(0)}(r) \approx -\int_0^r \ddr (\zeta_{\rm sph})^2/2 \approx -r^3/6R^2$. Hence, in the large confinement limit the radial displacement $\ru_r^{(0)}$ is dominated by the geometric strain $\delta_g =(W/R)^2$, rather than 
by the mechanical strain $\frac{\gamma}{(E_ft)}\frac{W}{r} \sim \delta_m$. %\cite{Comment-Asymptotic-Areas-Ratio-3}.  
%, and is ``decoupled'' from the radial strain $\epsilon_{rr}$, Eq.~(\ref{eq:radial-strain}), which remains dominated by the mechanical strain $\delta_m=\gamma/Y$.    

%this equation shows that at l while the radial strain scales with the mecanical strain, $\epsilon_{rr} \sim $\delta_m$, the radial displacement scales,    

%$\bullet$ 
{\emph{Hoop strain:}} For the axisymmetric component of the hoop strain we obtain, using Eqs. (\ref{eq:hoopstrain},\ref{eq:compression-free-2}): 
\begin{equation}
\frac{\ru_r^{(0)}}{r} + \frac{m^2}{4r^2}f^2 \to \Lambda \frac{\gamma}{(E_ft)}\frac{W}{r},\label{eq:slaving-1} \ . \\
\end{equation}
%\frac{u_r^m}{r} &=0,\\
%\frac{2m}{r}u^{2m}_\theta - \frac{m^2}{4r^2}f^2&=0.\label{eq:slaving-2}
%\end{align}
The geometric meaning of this equation, whose analog in \cite{PNAS11,PRE12} was dubbed {\emph{slaving condition}}, is that the fraction of latitudinal length absorbed by the wrinkle undulations ($m^2f(r)^2/4r^2$) together with the shrinkage of lattitudes ($\ru_r^{(0)}/{r}$) must equal the appropriate hoop strain ($\varepsilon_{\theta\theta} \to -\Lambda\varepsilon_{rr}$, Eq.~\ref{eq:compression-free-2}), that is necessary for the collapse of hoop compression ($\sigma_{\q\q} \to 0$).  
%
%(for $\Lambda = 0$ this is often called ``inextensibility").  
%Eq.~(\ref{eq:slaving-1}) and the above discussion (following Eq.~\ref{eq:radial-disp}) highlight an important difference between wrinkle patterns on a planar background that are generated by radial tension \cite{PNAS11} versus imposing spherical curvature on the film. 
Considering again the large confinement limit, $\alpha \gg 1$, we note that in contrast to wrinkle patterns on a planar background 
%In the first case \
\cite{PNAS11}, where all three terms in Eq.~(\ref{eq:slaving-1}) are comparable, %and all scale with the mechanical tensile strain imposed on the film. In contrast, at 
the large confinement regime addressed by our study is characterized by balance of the two terms on the left side of Eq.~(\ref{eq:slaving-1}), %where both terms must scale 
which both scale with $\delta_g$, whereas the term on the right is much smaller, scaling with the 
%RHS scales with the 
mechanical strain $\delta_m \ll \delta_g$. Thus, similarly to
$\ru_r^{(0)}$, the product $m\cdot f$ is
%together with the 
%radial displacement, the amplitude and number of wrinkles are both 
determined in the large confinement limit by the geometric strain $\delta_g \sim (W/R)^2$, 
rather than by the mechanical strain $\delta_m \sim \gamma/(E_ft)$. 
%becoming ``decoupled'' from the radial strain $\epsilon_{rr} \sim \delta_m$.  
 % 
%Important difference from previous studies: whereas the radial strain scales with the mechanical strain (vanishing at large confinement), the product mf scales with the geometric strain. this ``decoupling" of radial strain and the wasted hoop strains is a distinctive feature of the wrinkles in this geometry. 

%The oscillating component of the hoop strain (namely, the part of Eq. (\ref{eq:hoopstrain}) that is $\propto \sin(2m\q)$) yields an additional equation:
%\begin{equation}
%\frac{2m}{r}\ru^{(2m)}_\theta - \frac{m^2}{4r^2}f^2=0 \ .\label{eq:slaving-2}
%\end{equation}
%This equation for the azimuthal displacement $\ru_{\q}^{(2m)}$ is necessary to eliminate a highly-energetic oscillating component of the hoop stress, but it will not be required for the evaluation of the energy $u_{sub}$. 

The oscillating component of the hoop strain $\varepsilon_{\q\q}$ (namely, the part of Eq. (\ref{eq:hoopstrain}) that is $\propto \sin(2m\q)$) yields an additional equation:
\begin{equation}
\frac{2m}{r}\ru^{(2m)}_\theta - \frac{m^2}{4r^2}f^2=0 \ .\label{eq:slaving-2}
\end{equation}
This equation for the azimuthal displacement $\ru_{\q}^{(2m)}$ is necessary to eliminate a highly-energetic oscillating component of the hoop stress, but it will not be required for the evaluation of the energy $u^{\rm sub}$.
     
%$\bullet$ 
{\emph{Shear strain:}}
Finally, the shear strain $\varepsilon_{r\q}$ has only an oscillating component $\propto \cos(m\theta)$, for which we obtain, using Eqs.~(\ref{eq:strain-shear-1},\ref{eq:compression-free-1}):
\begin{equation}
\frac{m}{r}\ru_r^{(m)} +\frac{m}{r}f\ddr \zeta_{\rm sph} \to 0 \ , \label{eq:shear-strain-20} \ . 
\end{equation}
Similarly to Eq.~(\ref{eq:slaving-2}), that determines the azimuthal displacement required to eliminate a finite amount of oscillating hoop stress,  %(in the high bendability limit $\epsilon \to 0$), 
Eq.~(\ref{eq:shear-strain-20}) determines an oscillating component of the radial displacement $\ru_r^{(m)}\cos(m\theta)$ that is required to eliminate a finite {\emph{shear stress}} \cite{Comment-Shear-1}. In other words, the out-of-plane undulations which relax the compressive hoop stress, must be accompanied by in-plane oscillations of the boundary of the same periodicity ($2\pi/m$) and of comparable amplitude $(\ru_r^{(m)} \sim f)$. A similar effect was noted in \cite{Kohn-Bella} in a problem of metric-generated cascades.
%We will elaborate below on the significance of this oscillatory component of the radial displacement 

%Eqs.~(\ref{disp-wrink},\ref{stressaxi2}, \ref{eq-new2}) for the stress field, together with Eqs.~(\ref{eq:radial-disp},\ref{eq:slaving-1}) for the wrinkle amplitude and the radial component of the displacement field allow us to evaluate the wrinkle energy $u^{\rm wr}$. This computation is the subject of the next subsection.  

% in the limit $\epsilon \to 0$.}. 

%$\bullet$ Whereas the expansion of $u_\q$ is required to eliminate $O(1)$ contribution to the hoop stress, the expansion of $u_r$ is required to eliminate $O(1)$ contribution to the shear stress. Mention similarity with the $u_\q$ evaluation. Both are required to minimize $u_{dom}$ but do not affect the evaluation of $u_{sub}$. 

%\subsection{``Substrate'' versus ``tensional" wrinkles}
%GIVE ONLY RESULT. DETAILS IN APPENDIX 
\subsection{``Wrinklogami": asymptotic isometry assisted by wrinkles \label{sec:wrinklogami}}
Our analysis of the compression-free stress (Sec.~\ref{sec:Compfree}) and the displacement field (Sec.~\ref{sec:Asymdisp}) allows us to evaluate the energy $u^{\rm wr}$ of the wrinkled state. 
% laminated sheet and the attached substrate. 
This analysis will reveal the nontrivial isometry attained by the wrinkle pattern in the doubly-asymptotic regime of high bendability and large confinement ($\epsilon^{-1},\alpha \gg 1$), and will enable us to identify the sector in the paramater space at which the laminated state becomes wrinkled.

\subsubsection*{Evaluating the wrinkle energy} 
%$u_{dom}$ and $u_{sub}$}
In the FT expansion, the wrinkle energy $u^{\rm wr}$ is decomposed into two components (Eq.~\ref{uwr}). As we explained in Sec.~\ref{sec:Principles-FT}, 
the energy $u^{\rm dom}$ is stored in the compression-free stress field %(\ref{general-elas-energy}) 
and the work done on the film by the adhesive substrate,  
%(Eq.~\ref{eq:define-Wsurf}), which must be evaluated, respectively, with the aid of our expressions for the compression-free stress (Eqs.~\ref{stressaxi20},\ref{eq-new2}), and the associated displacement field (Eq.~\ref{eq:radial-disp}). Consequenelty, this energy component 
and approaches a finite value $u^{\rm dom}(\alpha,\tilde{K})$ in the high-bendability limit $\epsilon \to 0$.   
%evaluated by Eq.~(\ref{general-elas-energy}), is the energy stored in the compression-free stress field (Eqs.~\ref{stressaxi20},\ref{eq-new2}), 
%which approximates the stress in the wrinkled sheet 
%as $\epsilon \to 0$. 
%Therefore, in our terminology the energy $u_{dom}$ 
%which has a finite value $u^{\rm dom}(\alpha,\tilde{K})$ in the high-bendability limit $\epsilon \to 0$. 
We will show below that $u^{\rm dom} (\alpha,\tilde{K})$ actually vanishes as $\alpha \to \infty$. In contrast, the energy $u^{\rm sub}$ is associated with the energetic costs of bending the sheet and deforming the substrate due to the azimuthal undulations of wrinkles. The FT expansion requires the energy $u^{\rm sub}$ to vanish as 
$\epsilon \to 0$ \cite{PNAS11}, but we will show that it may override $u^{\rm dom}$ in a sub-domain of the doubly-asymptotic regime $\epsilon^{-1} , \alpha \gg 1$.

%A basic difference between the energy components $u_{dom}$ and $u_{sub}$ pertains to their behavior in this doubly-asymptotic limit.  The term $u_{dom}$, evaluated by Eq.~(\ref{general-elas-energy}), is the energy stored in the compression-free stress field (Eqs.~\ref{stressaxi20},\ref{eq-new2}), which reaches a ``finite" value $u_{dom}(\alpha)$ as $\epsilon \to 0$. We will show below that $u_{dom} (\alpha)$ actually vanishes as $\alpha \to \infty$. In contrast, the term $u_{sub}$ is associated with the energetic costs of bending the sheet and deforming the substrate due to the azimuthal undulations of wrinkles. The FT expansion requires the energy $u_{sub}$ to vanish as $\epsilon \to 0$ \cite{PNAS11}, but we will show that it may override $u_{dom}$ in a sub-domain of the doubly-asymptotic regime $\epsilon  \ll 1, \alpha \gg 1$.   
     
\paragraph*{Evaluating $u^{\rm dom}$:} 
The energy $u^{\rm dom}$ (where normalization is, per our convention, by $(E_ft)W^2$) is the sum of the straining energy (Eq.~\ref{general-elas-energy}), evaluated for the compression-free stress field, and the work $w_{\rm surf} = -\gamma dA_{\rm sph} /(E_ft) W^2$ (Eq.~\ref{eq:define-Wsurf}).
In order to evaluate these contributions, we will consider the limit $\tilde{K} \to \infty$, where we can use the analytic expressions obtained above, Eqs.~(\ref{stressaxi20},\ref{eq-new2},\ref{eq:lengh-wr}) and Eq.~(\ref{eq:radial-disp}), to obtain a well-defined, $\tilde{K}$-independent expression that we denote as $u^{\rm dom}(\alpha)$. For sufficiently small values of $\tilde{K}^{-1}$, we will use this value to approximate $u^{\rm dom}(\alpha,\tilde{K})$.
%is therefore well approximated by this limit value.   

We evaluate the straining energy $u_{\rm strain}$ by dividing the integral in Eq.~(\ref{general-elas-energy}) to two parts: $\int_0^W = \int_0^L + \int_L^W$. For the first part, where the film is unwrinkled, %axisymmetric state, 
we use Eq.~(\ref{elasenergy-axi}), replacing $W \to L$, $\gamma \to \sigma_{rr}(L)$, and $\alpha \to \alpha^*$, and using Eqs.~(\ref{eq:lengh-wr}) for $L$ and %Eqs.~(\ref{eq-newnew1},\ref{eq-newnew2}) for
$\sigma_{rr}(L)$, substituting $\alpha^*=8$. For the second part, where the film is wrinkled, we substitute in the integral the only non-vanishing component of the compression-free stress field: $\sigma_{rr} = \gamma W/r$. We thus obtain: 
\begin{equation}
%u_{\rm strain} = \tfrac{\pi}{6} \delta_m^2 (7 - 6 \Lambda) + 
%\pi \delta_m^2 \log(\frac{\alpha}{\alpha^*})^{1/3} 
u_{\rm strain} = {\pi} \delta_m^2 \big( \tfrac{1}{6}(7 - 6 \Lambda) + 
 \log(\frac{\alpha}{\alpha^*})^{1/3} \big)
\label{eq:energy-strain}
\end{equation}    
For the work $w_{\rm surf}$, we use Eqs.(\ref{eq:define-Wsurf},\ref{eq:radial-disp}) and obtain: 
\begin{equation}
w_{\rm surf} = \frac{\bW_{\rm surf}}{(E_ft)W^2} = 
%{-\gamma dA_{sph}  = -2 \pi W \gamma [\ru_r(W)-\tfrac{W^3}{8R^2}] = 
\pi \delta_m^2 \big( \alpha  + 8(-4 + 3\Lambda + \log\frac{\alpha}{\alpha^*} \big)
%\tfrac{2}{3} \pi \delta_m^2 [-\tfrac{4}{3} + \Lambda -\log(\tfrac{\alpha}{\alpha^*})]
\label{elas-3-fft-Kinfty}
\end{equation}
%$\bullet$ In order to evaluate the elastic energy in the unwrinkled patch $0<r<L$, where the state is axisymmetric, we use Eq.~(\ref{elasenergy-axi}), replacing $W \to L$, $\gamma \to \sigma_{rr}(L)$, and $\alpha \to \alpha^*$, and using Eqs.~(\ref{eq:lengh-wr}) for $L$, and Eqs.~(\ref{eq-newnew1},\ref{eq-newnew2}) for $\sigma_{rr}(L)$ and $\alpha^*=8$. We obtain that the energy contribution from the unwrinkled patch is: \begin{equation} u_{elas}^{(r<L)} = \tfrac{\pi}{6} \delta_m^2 (7 - 6 \Lambda) \label{elas-1-fft-Kinfty} \end{equation}

%$\bullet$ In the wrinkled zone, the only non-vanishing stress component is $\sigma_{rr} = \gamma W/r$, and upon integration over the annulus from $r \in (L,W)$  we obtain the energy contribution: \begin{equation} u_{elas}^{(L<r<W)} = \pi \delta_m^2 \log(\frac{\alpha}{\alpha^*})^{1/3}  \ . \label{elas-2-fft-Kinfty} \end{equation}

%(Evaluating the radial displacement $\ru_r(r)$ and consequently of $\ru_r(W)$, requires one more matching condition, in addition to Eqs.~(\ref{eq-newnew1},\ref{eq-newnew2}), imposing continuity of $\ru_r(r)$ between the wrinkled and unwrinkled zones.)    

Considering the contributions from Eqs.~(\ref{eq:energy-strain},\ref{elas-3-fft-Kinfty}), we notice that in the limit of large confinement ($\alpha  = \phi/\delta_m \gg 1$) the energy $u^{\rm dom}$ is governed by the work of the adhesive force pulling on its edge. Namely, $w_{\rm surf} \gg u_{\rm strain}$, and hence: 
 \begin{equation} 
u^{\rm dom} \approx w_{\rm surf} \sim  \delta_m \phi \ .  %\delta_m^2 \log \alpha \ , 
\label{elasenergy-fft-1}
\end{equation}
The linear dependence of the energy component $u^{\rm dom}$ on the laminated fraction $\phi$
%which 
is depcited by the red thin curve in Fig.~3. 
%This  applies at the infinite-stiffness limit, where $K =\infty$. 
The neglibility of the energy $u_{\rm strain}$ %Eq.~(\ref{elasenergy-fft-1}) implies that the energy proportional to the 
%due to the film's strain 
%in comparison to the adhesive work $w_{\rm surf}$ 
in comparison to the work $w_{\rm surf}$,  
reflects the {\emph{asymptotic isometry}} attained by the wrinkle pattern
in the doubly-asymptotic limit of large confinement, $\alpha \gg 1$, and high bendability, $\epsilon^{-1} \gg 1$.
% (the last one enables the compression-free stress field). 
In Sec.~\ref{sec:asym-iso} we will elaborate further on the importance of this result. 
  
%On the one hand, the explicit energetic contribution due to strain is negligible (hence the word ``isometry"). On the other hand, the formation of such a nearly strainless state requires the sheet to uncover an area $dA_{\rm sph} \approx W^4/24 R^2$ of the spherical substrate, and is therefore associated with an energetic cost $w_{\rm surf}$ (Eq.~\ref{elas-3-fft-Kinfty}), in addition to the bending energy, which we discuss next.     
   
\paragraph*{Evaluating $u^{\rm sub}$:}
Let us turn now to the energy $u^{\rm sub}$. In Appendix~\ref{sec:wrinkling} we generalize the scaling analysis of \cite{PNAS13} for large confinement values, and show that in the low-deformability regime ($\tilde{K}^{-1} \ll 1$) the wrinkle number is determined by a balance between the azimuthal bending force $\sim B (m^4/r^4) f$ %, which favors small wrinkle number $m$, 
and the substrate restoring force $Kf$. %\cite{Comment-Weak-Confinement}. 
This balance means that the energy $u^{\rm sub}$ is govenred by the sum of two comparable contributions, of the energies $u_{\rm bend}$ and $u_{\rm Win}$, which can be evaluated, respectively, from Eqs.(\ref{bendingenergy1},\ref{WinklerEnergy2}). We evaluate these energies by noting that the product $m^2f^2$ is subjected to the slaving condition, Eq.~(\ref{eq:slaving-1}), which implies  
$ m^2 f^2 \sim r^4/R^2$ in the doubly-asymptotic limit ($\epsilon^{-1},\alpha \gg 1$). These considerations yield the following estimates of the wrinkle number $m$ and the energy $u^{\rm sub}$: 
\begin{subequations}
\label{usub}
\begin{gather}
m \sim (\frac{KW^4}{B})^{1/4} \sim (\frac{\tilde{K}}{\epsilon})^{1/4} 
\label{usub1} \\
u^{\rm sub} \sim \frac{B}{(E_ft)W^2} \int_0^W  \frac{m^4 f^2}{r^4} r dr  \sim \sqrt{\epsilon \tilde{K}} \phi^2  \sim \tilde{t} \sqrt{\tilde{K}}\phi \ , 
\label{usub2}
\end{gather}
\end{subequations}
where the bendability parameter is $\epsilon^{-1} = \epsilon_g^{-1}$, as defined in Eq.~(\ref{gener-bendability}). 
%where we used Eq.~(\ref{bendingenergy1},\ref{wrinkleform}) 
%and ~(\ref{eq:slaving-1}) to estimate the bending energy of the wrinkle pattern.
Note that the first expression for $u^{\rm sub}$ in Eq.~(\ref{usub2}) does not reveal the actual $\phi$-dependence of this energy when all ``pristine" parameters in Eqs.~(\ref{nondim1},\ref{nondim2}), except $\phi$, are held fixed.   
%
%The difference between the two expressions for $u^{\rm sub}$ in Eq.~(\ref{usub2}), is that 
%the first one has a nontrivial dependence on the coverage fraction when analyzing the energy of the wrinkled state upon variation of $\phi$ only, when all other parameters in Eqs.~(\ref{nondim1},\ref{nondim2}) are held fixed. 
Expressing $\epsilon = \epsilon_g$ (Eq.~\ref{gener-bendability}) through % the ``pristine" parameters 
$\phi$ and $\tilde{t}$, we obtained 
the last expression, which shows that in such an analysis, the energy $u^{\rm sub}$ increases linearly with $\phi$. This linear dependence is illustrated in Fig.~3 (red thick line) for various parameter regimes.

\subsubsection*{The energy of a laminated state}
The above evaluation of the energy terms $u^{\rm dom}$ and $u^{\rm sub}$ shows that in the high bendability, large confinement, low deformability regime ($\epsilon^{-1}, \alpha, \tilde{K} \gg 1$), the energetic cost of strain is negligible, and the wrinkle energy %of the laminated, wrinkled state 
is thus described by: 
\begin{equation}
u^{\rm wr} \approx w_{\rm surf} + 
u^{\rm sub} \ , 
%\big( u_{\rm bend} + u_{\rm Win} \big) \ , 
\label{eq:wrinkle-energy}
\end{equation}
where $u^{\rm sub} = u_{\rm bend} + u_{\rm Win}$ 
is governed %by the curved geometry (through ..) and 
by the bending modulus $B$ and the substrate stiffness $K$, and the work $w_{\rm surf} $ is proportional to the tension $\gamma$ exerted on the film's boundary. Remarkably, the energy $u_{\rm strain}$ does not appear in this expression, nor does the stretching modulus $Y=(E_ft)$. This fact is a direct consequence of the collapse of all components of the strain tensor (Eqs.~\ref{eq:compression-free-1}-\ref{eq:radial-strain}), and reflects the asymptotic isometry exhibited by the wrinkled state (Eq.~\ref{disp-wrink}) in the limit ($\epsilon^{-1}, \alpha \to \infty$) for some $\tilde{K} \gg 1$. 

%One may view Eq.~(\ref{eq:wrinkle-energy}) as a generalization of the well-known problem:  a long, rectangular film floating on a liquid bath, under uniaxial, uniform confinement, where a pattern of parallel wrinkles of wavelength $\lambda = (B/\rho_l g)^{1/4}$ emerges, with $\rho_l$ is the liquid density. Such a simple wrinkle pattern reflects a balance of bending energy   
%problemthe one-dimensional problem, where a long rectangular film floating on liquid is uni. The geometric strain there is exerted directly (rather than through the curved shape). The difference - the necessity to include the work term. 

The two parts of the energy of the wrinkled state, $w_{\rm surf}$ and $u^{\rm sub}$, are plotted in Fig.~3 as linear functions of the coverage fraction $\phi$, for some given value of the mechanical strain parameter $\delta_m$.  Here, $w_{\rm surf}$ is depcited by a single thin red line, whose slope is $\sim \delta_m$, and thick red lines are used to depict the behavior of $u^{\rm sub}$, for three ranges of the product $\tilde{t} \sqrt{\tilde{K}}$. For $\tilde{t} \sqrt{\tilde{K}} \ll \delta_m$, we can approximate the wrinkle energy by the thin red line ({\emph{i.e.}} $u^{\rm wr} \approx w_{\rm surf}$, regime III-B), whereas for $\tilde{t} \sqrt{\tilde{K}} \gg \delta_m $, the wrinkle energy can be approximated by the corresponding thick red line ({\emph{i.e.}} $u^{\rm wr} \approx u^{\rm sub}$, regimes II and III-A). Next, we use these evaluations of the energy $u^{\rm wr}$, to compare with the energy $u^{\rm axi}$ of the axisymmetric (unwrinkled) state. This comparison  allows us to evaluate the characteristic values at which the film becomes wrinkled ($\phi_{wr}$) and delaminates from the substrate ($\phi_{rig}$ or $\phi_{def}$) in each of these parameter regimes.          
   
Considering the energies $u^{\rm axi}$ of the unwrinkled state (Eq.~\ref{axi-trans}) and the wrinkle energy $u^{\rm wr}$ (Eq.~\ref{eq:wrinkle-energy}), we find that the mechanics of a laminated state is governed by the three dimensionless groups: bendability $\epsilon^{-1}$, confinement $\alpha$, and deformability $\tilde{K}^{-1}$. This is shown in Fig.~7a, which plots schematically the morphology and energy of the laminated state as $\epsilon^{-1}$ and $\alpha$ are varied, for a fixed value of $\tilde{K}$. If $\alpha < \alpha^*(\tilde{K}) \ \approx 8$, the laminated state is under pure tension, and the axisymmetric state is the stable laminated configuration. If $\epsilon^{-1} \ll \tilde{K}$, the substrate is too rigid and the wrinkle energy is too large, making the wrinkled state unfavorable in comparison to a compressed (unwrinkled) axisymmetric state. The parameter regime $\big( \alpha \gg \alpha^*(\tilde{K}), \epsilon^{-1} \gg \tilde{K} \big)$, where the wrinkle pattern is energetically favorable in comparison to the axisymmetric state, splits into two sub-domains: $\alpha \ll (\epsilon \tilde{K})^{-1/2}$, where the wrinkle energy is governed by the work of adhesion, such that $u^{\rm wr} \sim w_{\rm surf}$ 
(corresponding to regime III-B in Fig.~3); 
%(Eq.~\ref{elasenergy-fft-1}); 
and $\alpha \gg (\epsilon \tilde{K})^{-1/2}$, where the wrinkle energy is govened by bending and substrate deformation, such that $u^{\rm wr} \sim u_{\rm bend} + u_{\rm Win}$ (corresponding to regimes II and III-A in Fig.~3).  
%(Eq.~\ref{usub2}).   

%In the doubly- asymptotic limit, $\epsilon \ll 1,\alpha \gg 1$, the ratio %between the energy components 
%$u_{sub}/u_{dom}$, as well as the ratio between the total wrinkle energy $u_{wr} = u_{dom}+u_{sub}$ and the axisymmetric state $u_{axi}$, are completely determined by the three dimensionless groups: bendability $\epsilon^{-1}$, confinement $\alpha$, and deformability $\tilde{K}^{-1}$. Namely: 
%\begin{equation}\frac{u_{sub}}{u_{axi}} \sim \sqrt{\epsilon \tilde{K}} \alpha\ \ ; \ \ \frac{u_{wr}}{u_{axi}} \sim \max[\alpha^{-1},\sqrt{\epsilon \tilde{K}} ]\label{EnergyAsym}\end{equation}

%This feature is depcited in Fig.?, which describes the morphology of the laminated state in the bendability-confinement plane, and proves the claim made earlier in the paper (Sec.~\ref{sec:dimensionlessgroups}): despite the existence of four dimensionless control parameters, the morphology of the laminated state depends only on the 
%three ``relevant" groups: $\epsilon^{-1},\alpha$ and $\tilde{K}^{-1}$.   

%In the following subsection we discuss both parts of Eq.~(\ref{EnergyAsym}).    
\begin{figure*}
\centering
\includegraphics[width=79mm]{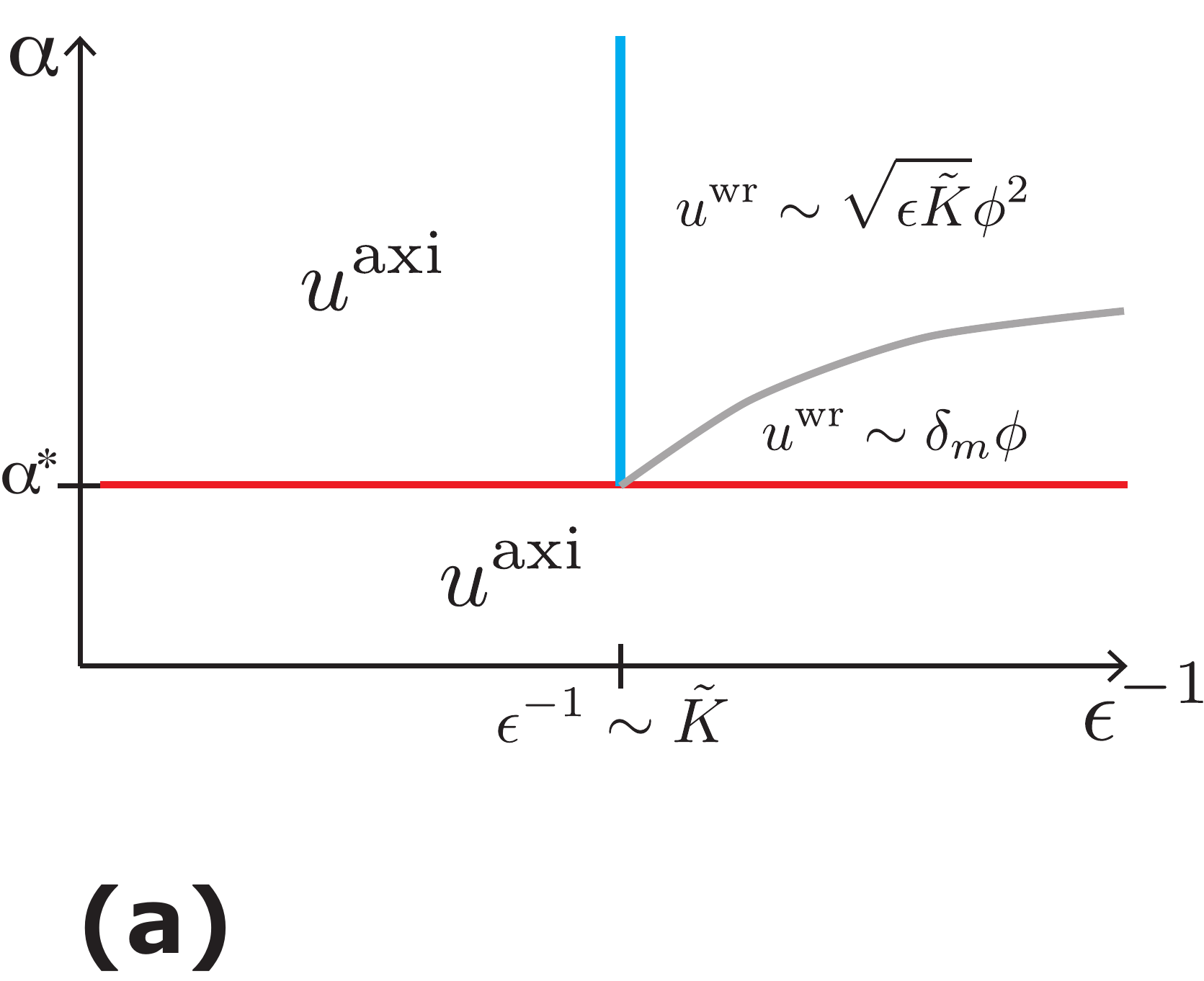}
\hspace{1cm}
\includegraphics[width=79mm]{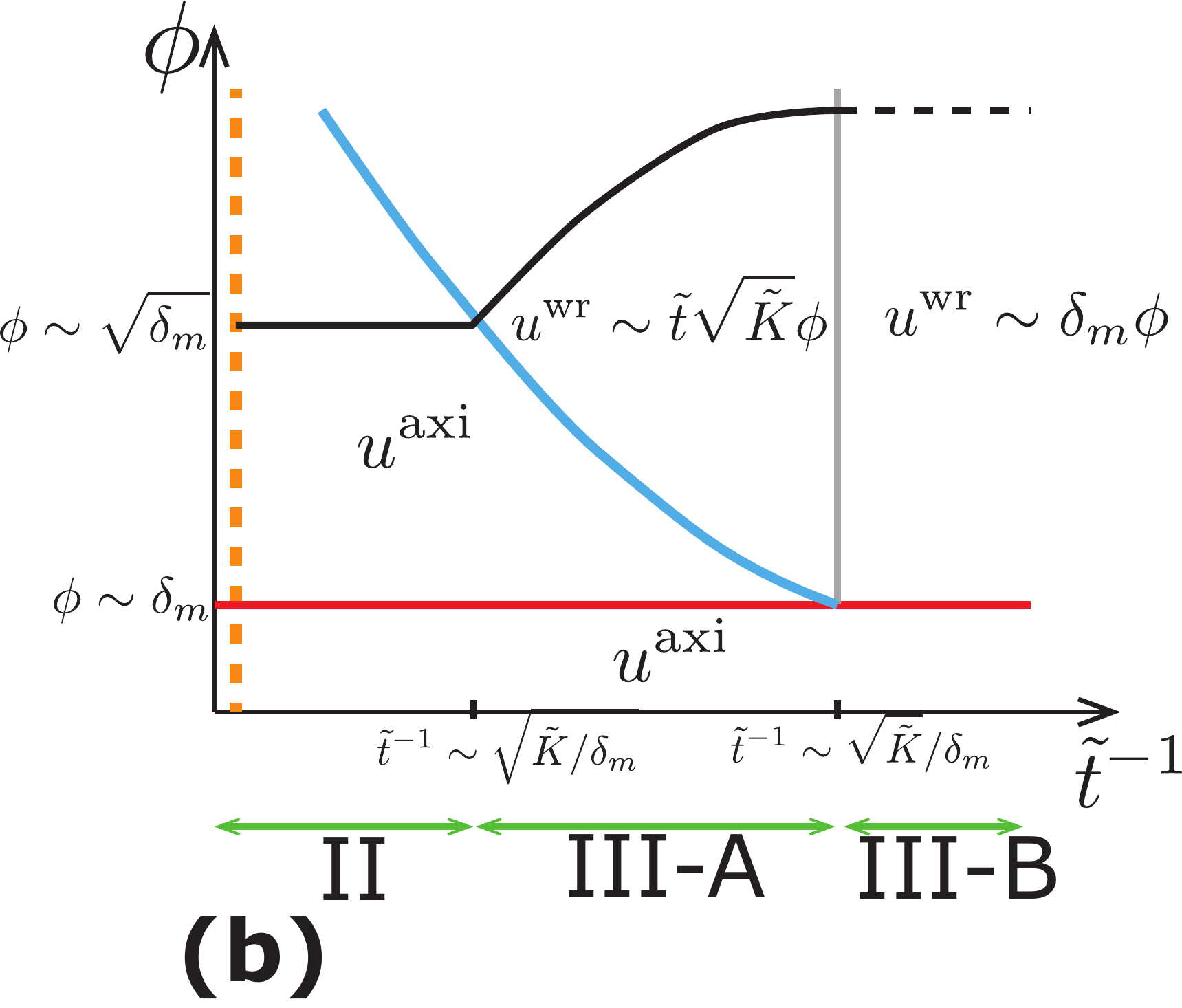}
%\includegraphics[width=79mm]{fig7a}
%\subfigure{\label{lame-a}\includegraphics{axisymmetric-plots}}
%\subfigure{\label{lame-a}
%\includegraphics[width=.5\columnwidth]{SF3.pdf}
%\subfigure{\label{lame-b}}
%\subfigure{\label{lame-c}}
\caption{\label{fig:energies-1}  
{\bf (a)} The behavior of the laminated states in the bendability-confinement plane ($\epsilon^{-1},\alpha$).
Here, we assume that the sheet is sufficiently thin such that the bendability is high ($\epsilon \ll 1$), and consider a fixed, small value of the deformability paramater $\tilde{K}^{-1} \ll 1$. The horizontal red line $\alpha  = \alpha_c(\tilde{K})$ corresponds to the critical confinement %($\alpha_c  \to 8$ for $\tilde{K} \to \infty$) 
below which the tensile (adhesive) force $\gamma$ exerted at the edge of the sheet is sufficiently strong such that the sheet is under pure tension. The vertical blue line $\epsilon^{-1}  \approx \tilde{K} \gg 1$ (where $\epsilon = \epsilon_g$) corresponds to the minimal bendability value %(correspondingly, maximal sheet's thickness) 
for which the energetic cost $u^{\rm sub}$ of bending and substrate deformation makes the wrinkled state energetically favorable in comparison to the axisymmetric (compressed, unwrinked) state. The gray curve separates between two parts of the parameter regime at which a wrinkle pattern is energetically favorable for the laminated state. Above this line, the energy $u^{\rm sub}$ (Eq.~\ref{usub2}) of bending and substrate deformation governs the energy $u^{\rm wr}$ of the wrinkled state. Below the gray line, bendability is sufficiently large %(or confinent is sufficiently small) 
and the wrinkle energy is governed by the energy $u^{\rm dom}$ (Eq.~\ref{elasenergy-fft-1}), associated with the work done by the tensile load at the film's edge. 
{\bf (b)} Re-plotting the diagram in panel a, where we replace the bendability and confinement parameters $\epsilon^{-1}$ and $\alpha$, 
%(that are useful for analyzing the wrinkled state), 
by the pristine dimensionless parameters of thickness $\tilde{t} = t/R$ and coverage fraction $\phi = (W/R)^2$, and consider fixed, small values of the deformability parameter $\tilde{K}^{-1} \ll 1$, and of the mechanical tension $\delta_m \ll 1$. 
%which are more natural for making comparison between the laminated and delamintaed states. 
%Again, we assume a fixed, small value of the deformability parameter $\tilde{K}^{-1} \ll 1$. 
The solid curves (red, blue, gray) correspond to the respective curves in panel a. The black curve marks the threshold $\phi_{def}$ above which the delamination is energetically favorable, and the vertical dashed orange line marks the maximal value of $\tilde{t}$, above which the film does not delaminate.   
The green double arrows show the parameter regimes that we call, respectively, II, III-A, and III-B (regime I, which corresponds to sufficiently soft substrate, {\emph{i.e.}} $\tilde{K} \ll 1$, is not shown in this figure).  
If $\sqrt{\delta_m / \tilde{K} }< \tilde{t} < \sqrt{\delta_m}$, the system is at regime II, where the energetically-favorable laminated state is axisymmetric (compressed, unwrinkled), and delamination becomes favorable for $\phi > \phi_{rig}$ (Eq.~\ref{max-phi-rigid}). If  $\tilde{t}<\sqrt{\delta_m / \tilde{K}}$, the system is at parameter regime III, where the film becomes wrinkled at $\phi >\phi_{wr}$ and delamination becomes energetically favorable for $\phi > \phi_{def}$, where $\phi_{wr},\phi_{def}$ are given by Eqs.~(\ref{eq:regime-III-A},\ref{eq:regime-III-B}). 
}           
%PLOT OF PHASE DIAGRAM OF EPSILON-1 AND ALPHA, OF THE LAMINATED STATE ONLY. SHOW THREE REGIMES, SEPARATED BY K-DEPENDENT CURVES, OF DOMINANCE OF UAXI, USUB, AND UDOM.   
\end{figure*}

% 
%The deep link between isometric embeddings and deformations of elastic sheets and shells has led in the past to novel .. Pogorelov proposed that the poking a shell, later - minimal ridges and d-cones, attributed to shape of a crumpled piece of paper.     
%
%tensionless deformations  
%
%, in a way that appears to be more energetically-efficient than a stress-focusing shape.

%A notable feature of the displacement field which underlies wrinkle patterns on a curved substrate is reflected in Eqs.~(\ref{disp-wrink}) and depicted in Fig.~S3: The out-of-plane undulations which relax the compressive hoop stress, must be accompanied by in-plane oscillations of the boundary of the same periodicity ($2\pi/m$) and of comparable amplitude $(\ru_r^{(m)} \sim f)$. A similar phenomena was noted in \cite{Kohn-Bella} in a different context of metric-generated cascades. 

%the surprising capability of a smooth wrinkle pattern to enbale a thin sheet on a spherical substrate, with no exerted tension, in a way that becomes indefintely close to isometric embedding in the asymptotic limit of indefinitely thin sheets.

%accommodate a spherical shape onto in a way that becomes {\emph{isometric embedding}} ({\emph{i.e.}} simultaneous cancellation of all strain components) of a film on a spherical topography in the limit of infinite confinement,  

 . 
\section{Pro-lamination by wrinkling \label{sec:Prolamination}}
The energy evaluations in the previous two sections allow us to determine the energetically favorable state: 
%of the system: 
laminated-unwrinkled ($u=u^{\rm axi}$, Eq.~\ref{axi-trans}), laminated-wrinkled ($u=u^{\rm wr}$, Eqs.~\ref{eq:wrinkle-energy},\ref{elasenergy-fft-1},\ref{usub2}), or delaminated ($u=\Gamma/E_ft$).      
%
%Evaluating the energies of the unwrinkled state ($u_{axi}, Eq.~\ref{axi-strans}) and the wrinkled state ($u_{wr}$, Eq.~\ref{uwr},\ref{eq:energy-scaling}), we can determine now the favorable state of the system    
In this section we will perform this energetic comparison, starting with the Winkler foundation, and then generalizing the results for an adhesive film on a compliant spherical substrate. Let us recall the simplifying assumption, $\Gamma \approx \gamma$ (Eq.~\ref{nondim3}), that we make in the current study.
%, where $\Gamma$ is the adhesion energy and the $\gamma$ is the substrate-vaopr surface tension that pulls on the film's edge. 
The general case ($\Gamma \neq \gamma$) will entail appearance of the ratio $\Gamma/\gamma$ in the various formulas, but should not affect the scaling laws derived in this section.

\subsection{Winkler foundation \label{sec:Winkler}} 
In order to facilitate the comparison of energies, it is useful to transform the coordinates $\epsilon^{-1}$ and $\alpha$ in Fig.~7a to $\tilde{t}^{-1}$ and $\phi$. The new diagram is depcited in Fig.~7b, where the red, blue, and gray curves are mapped from their counterparts in Fig.~7a by using the parameter transformation: $\epsilon \to (\tilde{t}/\phi)^2 \ ;  \ \alpha \to \phi/\delta_m$ (Eqs.~\ref{nondim1},\ref{gener-alpha},\ref{gener-bendability}).  
%Note that in drawing Fig.~7b we have to specify not only a small value of the deformability parameter ($\tilde{K} ^{-1} \ll 1$), but also a value of the mechanical tension $\delta_m \ll 1$. 
An additional curve (black) in Fig.~7b marks the maximal value of $\phi$, above which the energy of the laminated state becomes larger than the energetic cost of delamination $\gamma/(E_ft) =\delta_m$ (where we normalize energy, as usual, by $(E_ft) W^2$, and use the simplifying assumption, Eq.~\ref{nondim3}). For completness of our description, we added one more vertical curve (orange), that corresponds to the maximal value of the thickness parameter  $\tilde{t} \sim \sqrt{\delta_m}$, above which lamination is never favorable (see Sec.~\ref{sec:bendingenergy}).         
  
For a given pair of mechanical tension and deformability ($\delta_m,\tilde{K}$), Fig.~7b allows us to distinguish between the following parameter regimes: 

$\bullet$ Regime II, where $\sqrt{\delta_m/\tilde{K}} < \tilde{t} < \sqrt{\delta_m}$. In this regime, the laminated state of the film is always axisymmetric (unwrinkled), and delamination occurs at $\phi = \phi_{rig} =\sqrt{\delta_m}$. In dimensional units this parameter regime corresponds to: 
\begin{equation}
K> K_{rig} \ \ ; \ {\rm where} \  
K_{rig} \sim  \Gamma/t^2  \ .  
\label{eq:Krig}
\end{equation} 

Recalling Eq.~{\ref{regimeI}, we may express regime I, where the substrate is highly deformable (and is not included in Fig.~7b), through dimensional units, by identifying another characteristic stiffness: 
\begin{equation}
K < K_{soft} \ \ ; \ {\rm where} \  
K_{soft} \sim (E_ft)/R^2 \ . 
\label{eq:Ksoft}
\end{equation} 

The characteristic stiffness values $K_{soft}$ and $K_{rig}$, lead us to define the intermmediate parameter regime III: $K_{soft}<K<K_{rig}$, which splits into two parts, as shown in Fig.~7b:

$\bullet$ Regime III-A, where $\delta_m/\sqrt{\tilde{K}} < \tilde{t} < \sqrt{\delta_m/\tilde{K}} $. In dimensional units, regime III-A corresponds to: 
 \begin{equation}
\frac{\gamma^2}{E_f t^3} <K< K_{rig} \ . 
\label{eq:regime-III-A}
\end{equation} 
In this regime, the film becomes wrinkled at: 
\begin{subequations}
\label{eq:regime-III-A}
\begin{equation}
\phi_{wr} \sim \tilde{t} \sqrt{\tilde{K}} = \sqrt{t K/ E_f}  \ , 
% \  \ \ \ {\rm (regime III-A)} \  , 
\label{eq:regime-III-A-wr}
\end{equation} 
and delamination occurs at:  
\begin{equation}
\phi_{def} \sim \sqrt{\frac{\delta_m}{\tilde{K} \tilde{t}^2}} = \sqrt{\frac{\gamma }{K t^2}}  \ , 
\label{eq:regime-III-A-def}
\end{equation}     
\end{subequations}

$\bullet$ Regime III-B, where $\tilde{t} < \delta_m/\sqrt{\tilde{K}} $. In dimensional units, regime III-B corresponds to: 
\begin{equation}
K_{soft} < K < \frac{\gamma^2}{E_f t^3}  \ . 
\label{eq:regime-III-A}
\end{equation} 
In this regime, the film becomes wrinkled at: 
\begin{subequations}
\label{eq:regime-III-B}
\begin{equation}
\phi_{wr} \sim \delta_m = \gamma / E_f t  \ , 
% \  \ \ \ {\rm (regime III-A)} \  , 
\label{eq:regime-III-B-wr}
\end{equation} 
and delamination occurs at:  
\begin{equation}
\phi_{def} \sim O(1) \ . 
\label{eq:regime-III-B-def}
\end{equation}     
\end{subequations}
The last equation implies that, at least in the small-slope approximation used in this study ({\emph{i.e.}} $W/R \ll 1$), the film is sufficiently thin such that it can wrinkle so easily that delamination is not energetically-favorable even at arbitrarily large coverage fractions.

\vspace{0.3cm}

Importantly, both parts of the intermmediate-stiffness regime III are included in the low-deformability regime ($K > K_{soft}$), and therefore wrinkling involves no macroscale deformation of the substrate's shape. We call this phenomenon {\emph{pro-lamination}}, where the maximally laminated coverage fraction increases while the substrate retains its curved shape.   
Noticing that $K_{rig}/K_{soft} \sim (\gamma R^2/E_f t)$, we expect 
the parameter regime III to become particularly noticeable when the effective thickness $t/R$ decreases (Fig.~\ref{fig:fig1}a). In other words, pro-lamination should become a predominant phenomenon in the adhesion of ultrathin films on curved substrates.     

Our model system has four dimensionless groups: $\phi, \delta_m, \tilde{K},  \tilde{t}$ (Eqs.~(\ref{nondim1},\ref{nondim2}), assuming  $\gamma/\Gamma=1$). The schematic Fig.~7b is essentially a planar section, where the full 4d phase diagram of the model is projected onto a 2d hyper-plane spanned by the parameters $\phi$ and $\tilde{t}^{-1}$ (for fixed values of $\tilde{K}$ and $\delta_m$). The schematic phase diagram in Fig.~2a is another projection of the 4d parameter space onto a hyper-plane spanned by the dimensionless parameters $K R/E_f$ and $\Gamma/E_f R$, none of which depends on the thickness of the film or the coverage fraction $\phi$. The purpose of Fig.~2a is to describe the various scenarios that the system undergoes upon increasing $\phi$, for various values of the substrate stiffness and curvature, and the strength of adhesion.

\subsection{From Winkler foundation to elastic substrate \label{sec:fromWinkler}}

The response of an elastic substrate of Young modulus $E_s$ can be described through an {\emph{effective stiffness}} $K^{\rm eff} = E_s/\ell$, where $\ell$ is the characterstic lateral scale of a surface deformation \cite{Cerda03}. We start by assuming the existence of some $E_{s,soft}$, such that for $E_s >E_{s,soft}$ the system is in the low-deformability regime, and will address later the actual dependence of $E_{s,soft}$ on the parameters $t,R,E_f$.    

In the low-deformability regime, where the deformation of the spherical substrate is only at %associated only with 
the small wavelength and amplitude of the wrinkle pattern, the scale $\ell$ is the wrinkle wavelength, $\lambda \sim t(E_f/E_s)^{1/3}$, of a compressed film attached to compliant substrate \cite{Hutchinson98}. The effective stiffness is thus $K^{\rm eff} = E_s/\lambda = t^{-1} (E_s^4/E_f)^{1/3}$, and the deformability parameter becomes: $\tilde{K}^{\rm eff} = K^{\rm eff} R^2 / (E_f t) = (E_s/E_f)^{4/3}/ \tilde{t}^2$. Replacing $\tilde{K}$ by $\tilde{K}^{\rm eff}$      
allows us to evaluate the energy $u^{\rm wr}$ of the wrinkled state, by transforming the two parts of the energy, Eq.~(\ref{eq:wrinkle-energy}). The first part, $w_{\rm surf}$, which does not have an explicit dependence on the stiffness, is still given by Eq.~(\ref{elasenergy-fft-1}). The second part is evaluated by replacing $K \to K^{\rm eff}$ in Eq.~(\ref{usub2}), and we thus obtain: $u^{\rm sub} \sim (E_s/E_f)^{2/3} \phi$. 

In order to draw a schematic diagram analogous to Fig.~7b, which describes the energetically-favorable states in the low-deformability regime upon variation of $\phi$ and $\tilde{t}$, it is natural to consider some fixed, small values of the mechanical tension $\delta_m$ (similarly to Fig.~7b), and of the ratio $E_s/E_f$ (which replaces $\tilde{K}$). Recalling that the energy of the axisymmetric state, which does not depend explicitly on the stiffness, is still given by Eq.~(\ref{axi-trans}), we draw in Fig.~8 three diagrams, which correspond to regimes II, and to regimes III-A and III-B, where the pro-lamination effect is predicted \cite{footnote-diagrams}. As Fig.~8 shows, regimes II, III-A, and III-B are distinguished by the value of the ratio $E_s/E_f$.
%, which allows us to identify regime II (unwrinkled), and regimes III-A and III-B, where the formation of a wrinkle patter underlies pro-lamination. The parameter axes ($\tilde{t},\phi$) are similar to Fig.~7b, but the fixed parameters are different. Instead of a fixed $\tilde{K}$ in Fig.~7b, we assume in Fig.~8 a fixed value of the ratio $(E_s/E_f)$; insead of a fixed value of $\delta_m = \gamma/E_f t$ in Fig.7b, we find it more conveninent to assume in Fig.~8 a fixed value of $\gamma/E_f R$.     
\begin{figure*}
\centering
\includegraphics[width=169mm]{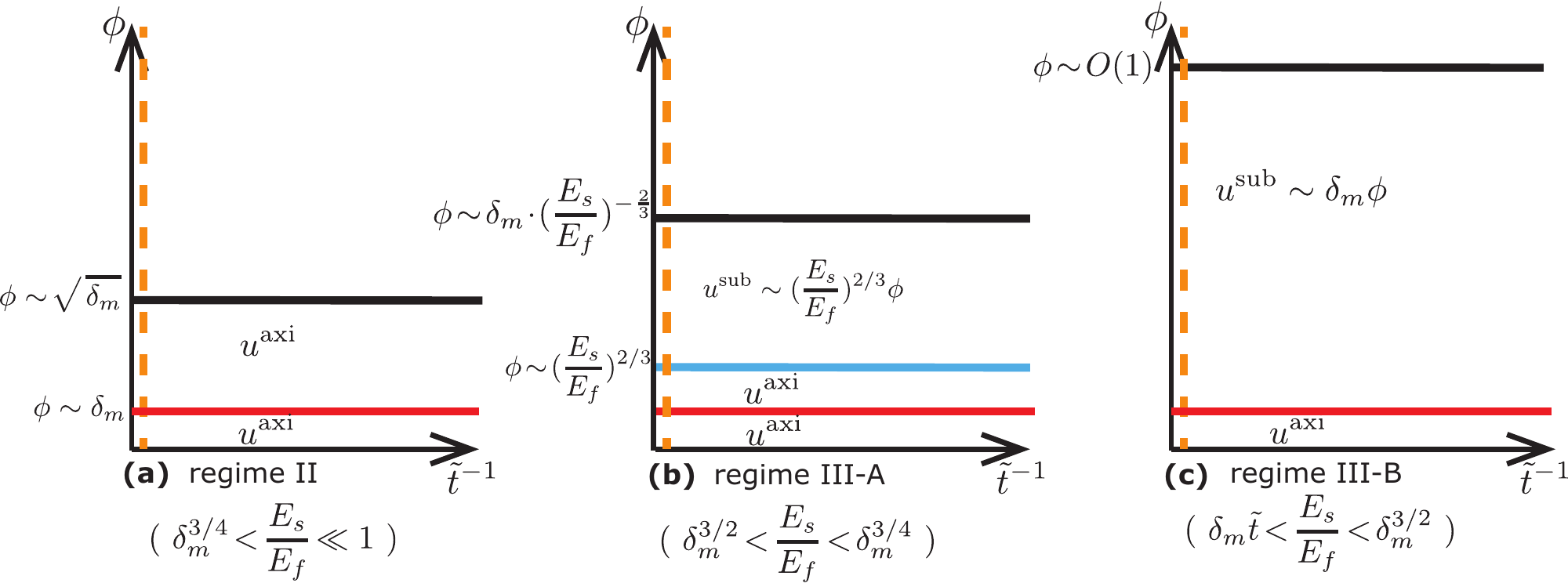}
%\subfigure{\label{lame-a}\includegraphics{axisymmetric-plots}}
%\subfigure{\label{lame-a}
%\includegraphics[width=.5\columnwidth]{SF3.pdf}
%\subfigure{\label{lame-b}}
%\subfigure{\label{lame-c}}
\caption{\label{fig:energies-2}  
Diagrams analogous to Fig.~7b, where the spherical substrate is assumed to be an isotropic solid of Young modulus $E_s$ (instead of stiffness $K$). The two parameter axes are $\tilde{t}^{-1}$ and $\phi$, as in Fig.~7b, and the parameters with small, fixed values are now $\delta_m = \gamma/E_f t$ (as in Fig.~7b) and   the ratio $(E_s/E_f)$ between the Young modulii of the substrate and the film.
%which are more natural for making comparison between the laminated and delamintaed states. 
%Again, we assume a fixed, small value of the deformability parameter $\tilde{K}^{-1} \ll 1$. 
Rather than a single diagram, we find it easier to plot here three separate diagrams \cite{footnote-diagrams}, for three characteritstic values of $E_s/E_f$, which correspond, respectively, to:   
% for each of the regimes \cite{footnote-diagrams}: 
{\bf (a)} regime II; {\bf (b)} regime III-A; and {\bf (c)} regime III-B. 
%The three diagrams differ by the fixed value of the ratio $E_s/E_f$. 
In each of the diagrams, the solid curves (red, blue, black, orange) correspond to the respective curves in Fig.~7b.
% The dashed vertical line separates regimes III-A and III-B. The borderline between regimes II and III-A is associated here with the values of the two horizontal lines (blue, black). 
%     
%The black curve marks the threshold $\phi_{def}$ above which the delamination is energetically favorable, and the vertical yellow curve marks the maximal value of $\tilde{t}$, above which the film does not delaminate.   
%The dashed vertical lines separate between the parameter regimes that we call, respectively, II, III-A, and III-B (regime I, which corresponds to sufficiently soft substrate, {\emph{i.e.}} $\tilde{K} \ll 1$, is not shown in this figure).  
Similarly to Fig.~7b, regime I, which corresponds to sufficiently soft substrate, $E_s <E_{s,soft}$, is not shown here. 
}           
%PLOT OF PHASE DIAGRAM OF EPSILON-1 AND ALPHA, OF THE LAMINATED STATE ONLY. SHOW THREE REGIMES, SEPARATED BY K-DEPENDENT CURVES, OF DOMINANCE OF UAXI, USUB, AND UDOM.   
\end{figure*}
Inspection of Fig.~8 allows us to characterize, similarly to Sec.~\ref{sec:Winkler}, the response to increasing values of $\phi$ in the low deformability regime: 

$\bullet$ Regime II, at which the film is unwrinkled, and delaminates from the substrate at $\phi = \phi_{rig} = \sqrt{\delta_m}$ (Eq.~\ref{max-phi-rigid}), corresponds to $(E_s/E_f)^{2/3} > \sqrt{\delta_m}$. In dimensional units, this leads us to identify regime II as: 
\begin{equation}
E_s> E_{s,rig} \ \ ; \ {\rm where} \ \
E_{s,rig} \sim  \frac{\gamma^{3/4}E_f^{1/4}}{t^{3/4}}  \ .  
\label{eq:Esrig}
\end{equation}   

$\bullet$ Regime III-A is defined by $\delta_m < (E_s/E_f)^{2/3} < \sqrt{\delta_m}$, which in dimensional units reads: 
 \begin{equation}
\frac{\gamma^{3/2}}{ E_f^{1/2}t^{3/2}} <E_s< E_{s,rig} \ . 
\label{eq:regime-III-A-Es}
\end{equation} 
In this regime, the film becomes wrinkled at: 
\begin{subequations}
\label{eq:regime-III-A-Es}
\begin{equation}
\phi_{wr} \sim (E_s/E_f)^{2/3}  \ , 
% \  \ \ \ {\rm (regime III-A)} \  , 
\label{eq:regime-III-A-wr-Es}
\end{equation} 
and delamination occurs at:  
\begin{equation}
\phi_{def} \sim \sqrt{\delta_m} (E_f/E_s)^{2/3}
 = \frac{\gamma^{1/2}E_f^{1/6}}{E_s^{2/3} t^{1/2}} \ , 
\label{eq:regime-III-A-def-Es}
\end{equation}     
\end{subequations}

$\bullet$ Regime III-B is defined by $(E_s/E_f)^{2/3} < {\delta_m}$ and $E_s > E_{s,soft}$, which in dimensional units reads: 
\begin{equation}
E_{s,soft} <E_s < \frac{\gamma^{3/2}}{ E_f^{1/2}t^{3/2}} \ ,  
\label{eq:regime-III-B-Es}
\end{equation} 
where $E_{s,soft}$ is defined below.  
Note that our discussion of the analogous regime III-B in Sec.~\ref{sec:Winkler} shows that the coverage fractions at which wrinkling and delamination occur do not depend explicitly on the stiffness parameter. Hence, as can be seen also from Fig.~8, we obtain expressions for $\phi_{wr}$ and $\phi_{def}$ which are identical to Eq.~\ref{eq:regime-III-B}.

$\bullet$ Regime I: Turning now to the high-deformability regime, we note that here the spherical substrate undergoes a signficant deformation beneath the attached film, and hence the chatacteristic lateral scale for the deformation is the film's size $W$, rather than the wrinkle wavelength $\lambda$. Hence, the effective stiffness is $\tilde{K}^{\rm eff}  \sim E_s/W$. Substituting this expression in Eq.~(\ref{eq:Ksoft}), we obtain: $E_{s,soft} \sim E_f t W/R^2 = E_f \tilde{t} \sqrt{\phi}$. Thus, in contrast to a Winkler substrate, the tendency of a spherical compliant substrate to deform under the laminated film depends on the coverage fraction $\phi$:    
%
%for a compliant substrate the low deformability regime, $E_s < E_{s,soft}$ depends on the coverage fraction $\phi$. 
The larger $\phi$ is, the larger should be $E_s$ in order for the substrate to retain its shape under a laminated, axisymmetrically deformed (unwrinkled) film. This difference between the Winkler's model and a compliant substrate underlies a small difference bewteen Figs.~2c and 2d: In Fig.~2c, one may consider $K_{soft}$ as a unique ($\phi$-independent) value of the substrate stiffness below which the substrate deforms appreciably; in Fig.~2d, the actual value of $E_s$ below which the substrate undergoes significant deformation, does depend on $\phi$. 

Similarly to our discussion in Sec.~\ref{sec:Winkler}, we want to find a value of $E_{s,soft}$, which separates between the parameter regimes I, where the spubstrate deforms appreciably before the emergence of wrinkles, and regime   
III-B, where the formation of wrinkles enables lamination of the film without macroscopic deformation of the substrate. Hence, we substitute $\phi_{wr} \sim \delta_m$ (which is the minimal value of $\phi$ for which the film is azimuthally compressed) in the expression $E_{s,soft}(\phi) \sim E_f \tilde{t} \sqrt{\phi}$, and thus identify the high deformability regime I as: 
\begin{equation}
E_s<  E_{s,soft} \ \ ; \ {\rm where} \  \
E_{s,soft} \sim  \frac{\gamma^{1/2}E_f^{1/2} t^{1/2}}{R}  \ .    
\label{eq:Essoft}
\end{equation} 
Similarly to our discussion of the Winkler's substrate, we note that Fig.~8 and Fig.~2b depict distinct projections of the full phase diagram of the model onto 2d hyper-planes in the 4d parameter space. In Fig.~8, the free parameters are $\phi$ and $\tilde{t}^{-1}$ (whereas $E_s/E_f$ and $\delta_m$ are assumed fixed values). In Fig.~2b, the plane is spanned by the dimensionless parameters $E_s/E_f$ and $\Gamma/E_f R$, which are both independent on the thickness $t$ and the coverage fraction $\phi$. Similarly to Fig.~2a, the purpose of Fig.~2b is to describe the various scenarios that the system undergoes upon increasing $\phi$, for various values of the substrate's Young modulus and curvature, and the strength of adhesion.

%This idea allows us to identify, similarly to Eqs.~(\ref{eq:Krig},\ref{eq:Ksoft}), two characteristic values of the Young modulus: $E_{s,soft} = \ell  \ K^{\rm eff}_{soft} \ ;  \ E_{s,rig} = \ell \ K^{\rm eff}_{rig}$, that define regime I (highly deformable substrate, $E < E_{s,soft}$), and regime II (highly rigid substrate, $E > E_{s,rig}$. The intermmediate regime III, where pro-lamination is expected, thus corresponds to $E_{s,soft}<E_s<E_{s,rig}$. 
     
%In order to find the characteristic modulii, $E_{s,soft}$ and $E_{s,rig}$, one must realize the existence of two distinct lateral scales for the deformation of the substrate. The first lateral scale is a macroscopic one, $\sim W$, which characterizes the flattening of a sufficiently soft substrate (regime I) beneath the film. Hence, the effective stiffness in regime I is $K^{\rm eff} = E_s/W$, and with the aid of Eq.~(\ref{eq:Ksoft}) we obtain: 
%\begin{equation} E_{s,soft} \sim \frac{E_f t W}{R^2} \ .  \label{eq:Essoft} \end{equation}
%The second lateral scale is $\lambda \sim t(E_f/E_s)^{1/3}$, which characterizes the wrinkle wavelength of a compressed film attached to a compliant substrate \cite{Hutchinson98}). In regime III, where the substrate retains its spherical shape, its deformation is characterized by the wrinkle wavelength scale $\lambda$, and therefore $K^{\rm eff} = E_s/\lambda \sim E_s^{2/3}E_f^{1/3}/t$. Finally,    

%%%%%%%%%%%%%%%%%%%%%%%%%%%
\section{Asymptotic isometry \label{sec:Asym-iso}}
%%%%%%%%%%%%%%%%%%%%%%%%%%%
Beyond its relevance for our problem, the stucture of the wrinkle energy, Eq.~(\ref{eq:wrinkle-energy}),
%Thus, the displacement field, Eq.~(\ref{disp-wrink}), 
reflects a surprising fact:
%the shape of a thin sheet may become arbitrarily close to a smooth, doubly-curved shape ({\emph{i.e.}} with nonzero Gaussian curvature) at negligible cost of strain and bend. More precisly, in the limit of vanishing thickness and vanishing tension exerted at boundary, the energies associated with strain and bending are both vanishing in comparison to the elastic energy associated with the smooth (unwrinkled) shape.  
It is possible to impose a doubly-curved shape on a solid film 
%by using small tension at the film's boundary, and avoiding a divergeing bending energy through a smooth wrinkle pattern
in a way that becomes asymotptically isometric to the undeformed film. 
%close to an isometric  embedding, 
(Namely, where all components of the strain tensor are being eliminated \cite{Comment-GaussTheorem}).  In contrast to the common usage of isometries in studies of elastic sheets, which refers to the limit of small thickness,  %(equivalently, large von Karman number), 
the asymptotic process underlying the ``wrinklogami" pattern is double - involving both small thickness of the sheet (quantified through the bendability, $\epsilon^{-1} \gg 1$), and a small exerted tension (quantified by the confinement $\alpha \gg 1$).  
In this section, we will expand on the meaning of the asymptoic isometry. We will highlight the generic nature of Eq.~(\ref{eq:wrinkle-energy}), and discuss its relevance for other physical systems.     
%rephrasing Eq.~(\ref{eq:wrinkle-energy}) in a general way that indicates on its relevance and implications for other physical systems.     

\subsection{When are thin sheets said to be isometric to their undeformed state ?}

One may distinguish between three classes of loading conditions that can be exerted on a thin elastic sheet: 
 
{\bf (A)} When a sheet is subjected to {\emph{purely tensile loads}}, the exerted work is transmitted primarily to an elastic energy stored in the stress field  (Eq.~\ref{general-elas-energy}). This applies not only at simple set-ups  
when a load induces a uniform tensile stress across the sheet ({\emph{e.g.}} pulling with equal force on all boundaries) , but also when the induced stress is non-homogenous and some zones in the sheet are wrinkled  due to compression. An example is the Lam\'e problem \cite{Timoshenko}, where an annular sheet is subjected to distinct radial tensile loads at its inner and outer edges, and part of the sheet develops radial wrinkles that relax the induced azimuthal compression. In such a case, the exerted work is transmitted to the tensile components of the compression-free stress field, and 
the energetic cost of the wrinkles (analogous to our $u^{\rm sub}$, Eq.~\ref{usub}), is a negligible fraction of that work \cite{PNAS11}.  

In problems of type (A), the main effect of the exerted loads is the deformation of the metric ({\emph{i.e.}} inducing strain), but the shape of the sheet remains close to its original, unstressed, planar shape. 

 {\bf (B)} When a sheet is confined in space, the exerted forces are {\emph{purely compressive}} and their associated work is transmitted solely to bending the sheet and deforming an attached substrate, 
or become partitioned between the energies associated with bending the sheet and straining the mid-plane in small ``stress focusing" zones. The first scenario occurs under uniaxial compression, where the sheet becomes buckled or wrinkled (due to an attached substrate), 
%, such as liquid bath whose stiffness stem from gravity, $K = \rho_{liq} g$ \cite{Milner89,Pocivavsek08,Huang10} or a compliant substrate \cite{Hutchinson98}), 
retaining everywhere a developable shape. The second scenario occurs when a sheet is confined into a ring \cite{Pomeau97,Cerda98} or a small box \cite{Witten07}, where the defomed shape is developable almost everywhere, except at narrow zones (the vertex of a ``d-cone" or along a ``minimal ridge") that contain strain. Similarly, when a sheet attached to a compliant substrate is subjected to bi-axial compression, the deformed shape is developable almost everywhere \cite{Audoly08,Kohn-Nguyen}. 

In most problems of type (B), the exerted forces barely affect the metric, but may have a significant effect on the shape of the sheet, which departs appreciably from a planar one. In numerous cases, a solution may be found by searching for a developable (or piecewise developable) shape -- an isometric transformation of the planar sheet to a shape that is compatible with the geometric constraints imposed on it \cite{Pomeau97,Cerda98,BenAmar08-econe, Sharon07}. The sheet thus gets arbitrariliy close to this isometric shape in the asymptotic limit $t \to 0$.

 {\bf (C)} The problem we address in this work represents another class of systems, where the sheet is subjected simultaneously to a geometric constraint by the spherical shape of the substrate that is imposed on the sheet, and to a tensile load at its boundary. Here, in cotrast to the first two classes (A,B), both the metric and the shape of the sheet are affected in a nontrivial manner. 
However, our anaysis in Sec.~\ref{sec:Asymdisp} (Eqs.~\ref{eq:compression-free-1}-\ref{eq:radial-strain}) shows that  
%Our anaysis in Sec.~? yields two, intimately related insights. 
%, which can be generalized to other problems in this class. 
%First, 
the deformation of the metric ({\emph{i.e.}} the strain) does not stem from the spherical shape itself, but rather from the mechanical strain $\delta_m = \gamma/Y$ exerted at the boundary. Hence, the sheet does approach a nontrivial isometry, where the shape is close to a spherical cap whose Gaussian curvature is nonzero,   
% ({\emph{i.e.}} a shape different than the original planar sheet), 
but notably -- the limit underlying this behavior is doubly asymptotic - being associated with small thickness of the sheet (quantified by the inverse bendability $\epsilon$) and small exerted tensile load (quantified by the ratio $\delta_m/\delta_g = \alpha^{-1}$). 
%The doubly asymptotic nature of the wrinkled isometry underlies a new type of mechanical response, and may provide a conceptual framework for understanding certain types of morphological transitions in confined sheets. 

%In the next subsection we will show how Eq.~(\ref{eq:wrinkle-energy}), which we call an {\emph{asymptotic isometry equation}}, reflects the asymptotic isometry of the wrinklogami state in our problem. We will also explain how equations with similar structure describe the energy of systems, such as the three above examples, where a nontrivial isometry is approached at the doubly asymptotic limit of small thickness and weak tensile load.      

%Below, we will briefly discuss these issues, but first -- let us 
%Before proceeding, it is useful to 
%mention a few other examples in this class of problems. 
%, where an elastic sheet develops a nontrivial shape under the influence of geometric confinemenet and weak tensile loads. 

% in this class which may also be characterized by a nontrivial isometry approached at the doubly asymptotic limit of small thickness and weak tensile load.    

\subsection{Asymptotic isometry equation \label{sec:asym-iso}}

With the above classification of loading types, Eq.~(\ref{eq:wrinkle-energy}) can be viewed as a specific example of a generic form for the energy in class ({C}). Here, a sheet of size $W$ is attached to a sphere of radius $R$, and becomes nearly strainless %under a given geometric constraint %(encapsulated by $\delta_g =\phi \sim  (W/R)^2$) 
in the singular, doubly asymptotic limit of small thickness ($\tilde{t} \sim t/W \to 0$) and weak tensile load ($\delta_m = \gamma/Y \to 0$).  
In this limit, the energy consists of two relevant terms. The first one is the work, which is linear in the tensile load ($\sim \delta_m$) and overrides the straining energy, which is quadratic in this parameter ($\sim \delta_m^2$). The second term is the energetic cost $u^{\rm sub}$ of bending the sheet and the deformation of a substrate.  
%(real one or an effective one \cite{Cerda03}). %associated with the formation of wrinkles (or, more generally, any ``micro-structure" required to avoid the would-be strain due to the geometric constraint). 
The work term is directly proportional to the tensile load, and is independent on the sheet's thickness; the second term vanishes as some power of the sheet's thickness. Since the two energetic components are determined by independent parameters, the work is essentialy  ``decoupled" from the elastic energy stored in the sheet.   
This type of energetic structure is strictly different from the analogous one in classes (A) and (B). In class (A), %the straining energy is not negligible -- 
the exerted work is transmitted primarily to straining the sheet; in class (B), %there is no need to consider explicitly the work term -- it simply balances the 
the work is transmitted to bending the sheet and deforming a substrate (if the shape is developable), or to the elastic energy stored in the stress-focusing zones of the sheet (in the case of a piecewise-developable shape). 
%substrate deformation and/or to elastic energy and any substrate deformation stored in wrinkles, cumples, or folds.    

The simplified nature of our problem in the low deformability regime, ($\tilde{K}^{-1} \ll 1$), is reflected in two intimately related facts. First, the  
wrinkles are superimposed on the original profile of the spherical substrate, and 
hence the work term, Eq.~(\ref{elasenergy-fft-1}), can be expressed using the radius $R$ of the undeformed substrate.  
Second, %the energetic cost of wrinkles $u^{\rm sub}$ is 
the optimal wrinkle pattern is determined by balancing the bending modulus $B$ and the actual substrate's stiffness $K$. As a consequence,
the energy component $u^{\rm sub}$, Eq.~(\ref{usub2}), 
may be viewed as a straightforward generalization of the energetic cost of wrinkles in a rectangular sheet attached to a substrate of stiffness $K$ under uniaxial compression \cite{Hutchinson98,Cerda03}. In order to understand the broad relevance of the asymptotic ismoetry equation, of which Eq.~(\ref{eq:wrinkle-energy}) is one example, it is useful to briefly consider some more complicated examples of systems in class (C).   

{\emph{(i)}} If the spherical substrate is sufficiently soft, such that the deformability parameter $\tilde{K}^{-1} \gg 1 $ (regime I in our classification in Sec.~\ref{sec:Winkler}), we expect not only the formation of wrinkles but also flattening of the substrate beneath the attached sheet.
%, and consequently a substantial deformation from a spherical profile. 
Such a behavior is demonstrated by a sheet floating on a liquid drop \cite{King12} (or even by an inflated mylar balloon \cite{Taylor}), where $\tilde{K} = 0$, and the curved shape is imposed by exerting a uniform Laplace pressure $P$, balanced by the surface tension $\gamma$ of the drop of radius $R = 2\gamma/P$ that pulls at the boundary of the sheet.  
When $\gamma$ and the bending modulus $B$ are sufficiently small, we expcet the wrinkled shape to approach an asymptotic isometry, similarly to our system. However, since the drop flattens beneath the sheet, the radius of curvature becomes $R_{\rm eff} \gg R$, and the radial profile $\zeta(r)$ on which the wrinkles are superimposed is significantly different from a spherical profile \cite{King12}. The energy may still be expressed in a similar manner to Eq.~(\ref{eq:wrinkle-energy}), but the actual computations of the work term (which is proportional to $\gamma/Y$) and the bending energy associated with wrinkles (which is proportional to some power of $\tilde{t}$) are more complicated \cite{Number2}.     

{\emph{(ii)}} Another example is the indentation of thin sheets that are placed on adhesive substrate or floating on a liquid bath \cite{Holmes10}. For a free-standing sheet, poking is an example of type (B) in the above classification, whereby the sheet attains a developable cone (``d-cone") shape, everywhere except at a small, stress-focusing vertex, whose size vanishes with the sheet's thickess \cite{Pomeau97,Cerda98}. Such a deformation is not possible if the sheet is required to remain floating on the liquid bath, and a nontrivial pattern of wrinkles emerges. Despite its complexity, such a wrinkle pattern may also become isometric to the undeformed sheet in the doubly asymptotic limit of weak tension (exerted by the liquid at the sheet's edge) and small thickness. The presence of multiple external forces (indetnation, liquid gravity, and surface tension) complicates the computation of the work and the bending energy \cite{Vella14}.       
 
{\emph{(iii)}} A third example is an elastic ribbon that is stretched and twisted around its main axis, where a plethora of patterns -- wrinkles, creases, and loops -- has been observed \cite{Green37,Chopin13}. In a recent theoretical study, it was noted that under a given twist
({\emph{i.e.}} a geometric constraint that imposes a helicoidal shape with nonzero Gaussian curvature), %which imposes geometric constraint), 
the ribbon approaches an isometry in the doubly asymptotic limit of vanishing tensile load and riboon's thickness \cite{Chopin14}. It was further argued that such an asymptotic ismoetry may be attained through wrinkles that cover the whole ribbon and are superimposed on a helicoidal shape, or through a ``creased helicoid" shape, in which the stress is focused in narrow zones, whose size vanishes in this limit. The energy of each of those asymptotically isometric patterns consists of a work (done by the stretching force) and %wrinkle-induced
bending energy, similarly to Eq.~(\ref{eq:wrinkle-energy}).       

\vspace{0.3cm} 

All of the above examples exhibit a more complex, morphologically richer behavior than our system. In addition to wrinklogami patterns, other patterns have been observed: In example {\emph{(i)}}, a ``wrinkle-to-crumple" transition has been found upon increasing the Laplace pressure in the drop, whereby the  stress appears to be localized in structures that resemble ridges and d-cones \cite{King12}; in example {\emph{(ii)}}, a somewhat different transition has been observed upon increasing the indentation depth \cite{Holmes10}, resembling a ``wrinkle-to-fold" transition in uniaxial compression of rectangular floating sheets \cite{Pocivavsek08,Diamant11}; in example {\emph{(iii)}}, various instabilities of the wrinkle pattern, which are also characterzied by stress-localizing ridges and loops, have been observed upon increasing the twist on the ribbon or decreasing the exerted tension \cite{Chopin13}. 

In the next subsection we will discuss the reason for the relative complexity of those systems in comparison to the simpler kind of asymptotic isometry that we studied here. 

\subsection{Various routes for asymptotic isometry}

%In order to understand the general structure of problems in class (C), it is useful to consider only the
In order to elucidate the various morphologies that are observed in problems of class (C), we will present in this section a formal discussion, assuming a film subjected to some geometric constraint ({\emph{e.g.}} twisting a ribbon by a given angle, indenting a film by a given amplitude, requiring a sheet to enclose a finite volume, {\emph{etc}}), and      
%Being subjected to geometric constraints of distinct types, 
%problems in class (C) are 
characterized by 
bending modulus $B \sim E_f t^3$ and stretching modulus $Y \sim E_f t$, a characteristic lateral scale $W$ ({\emph{i.e.}} the radius of a circular film or the width of a ribbon), and a tension $\gamma$ exerted on the film's boundary. 

$\bullet$ {\bf Route 1:} If it is possible to impose the geometric constraint without any tension ({\emph{i.e.}} at $\gamma=0)$ %({\emph{e.g.}} twisting a ribbon by a given angle, poking a film by a given amplitude) 
then the shape can be described as a small perturbation of a perfectly isometric map of a 2D film, similarly to problems in class (B). The small parameter in this perturbative expansion, depicted schematically by the long leg of the red curve in Fig.~9, is $BW^2/Y \sim (t/W)^2$. The isometric shape, around which the expansion is carried out, may often consist of isolated curves or points with infinite curvature \cite{Pomeau97,Cerda98,Korte10}, which are regularized by this perturbative expansion, yielding stress focusing zones \cite{Witten07}. 
%This type of behvaior, which is strictly different from the wrinklogami pattrens, is esssntially identical to problems in class (B).
Naturally, if the tension $\gamma$ is nonzero, but is sufficiently small, the film's shape may still be described by this expansion, by adding to the energy the tensile work, as is depicted in the short leg of the red curve in Fig.~9. In a formal language, such a route to isometry is an expansion around the ordered limit: $\lim_{BW^2/Y \to 0} \lim_{\gamma/Y \to 0}$.  
As we described in Sec.~\ref{sec:asym-iso},  an asymptotic isometry equation similar to Eq.~(\ref{eq:wrinkle-energy}) is valid also for such a state: the work term couples $\gamma$ to the displacement field of the isometric map,
% (at $\gamma=0$), 
and overrides the straining energy (which is proportional to $\gamma^2/Y$); the energy $u^{\rm sub}$ for such a state stems from the bending cost of the isometric map at $\gamma=0$ ({\emph{e.g.}} formation of stress-focusing zones, or a smoothly bent shape, such as for an unstretched, twisted ribbon).  
% (if the isometric map at $\gamma=0$ corresponds to a bent shape), associated with the isometric map, %(bent, unstrained zones), 
%or in stress-focusing zones.   
    
$\bullet$ {\bf Route 2:} The above paragraph describes an expansion around an actual isometric shape of a 2D film ({\emph{i.e.}} developable or piecewise-developable shape), in sharp contrast with the wrinklogami pattern, which we addressed in this paper. The first step in our derivation in Sec.~\ref{sec:Wrinkledstate}, depicted by the short leg of the blue curve in Fig.~9, starts with the singular limit of a film with a {\emph{finite}} modulus $Y$ and $B=0$, on which some tension $\gamma>0$ is exerted. The compression-free stress is attained by such a ``virtual" film through a wrinkling pattern with vanishing wavelength, and the energy $u^{\rm sub}$ is the energetic cost of wrinkles for a small $B>0$. At the second step, depicted by the long leg of the blue curve, we assume the tension is reduced, such that the energy stored in the tensile component in the compression-free stress field may become arbitrarily small. In a formal language, this asymptotic route to isometry entails an expansion around the ordered limit: $\lim_{\gamma/Y \to 0} \lim_{BW^2/Y \to 0}$, which is strictly different from the one underlying route 1. 

Thus, we recognize the existence of two distinct routes for isometric response of a thin film to a geometric constraint. Both routes lead to vanishing energetic cost in the singular limit of vanishing film's thickness and exerted tensile load.    
%
%each of them is valid which depends on the ratio between the tensile load and the film's thicnkess.  
The parameter regimes at which each of the two routes is valid are depicted by the green curve $\gamma^*(t)$ in Fig.~9:     
Route 1, where the film's shape may be approximated by an actual isometric, developable or piecewise-developable map, of a 2D sheet,  %(which often consists of stress-focusing zones), 
is valid at large ratios between the thickness and the tensile load (above the green curve). Route 2, where the film's approaches isometry in a non-developable fashion, through a wrinklogami pattern with a smooth distribution of stress, cannot be approximated by any isometric shape of a 2D sheet, and is valid at  a small thickness/tension ratios (below the green curve).   

We posit the existence of two parameter regimes at the doubly asymptotic limit ($t \to 0, \gamma \to 0$), at which strictly different types of deformations are expected, underlies phenomena such as ``morphological phase transitions" in problems of class (C). Such a transition, which should become sharper as $t \to 0$, is expected in the vicinity of a curve $\gamma^*(t)$, depicted in Fig.~9. The universal aspect of such a  transition, common to problems in class (C), is encapsulated in the asymptotic law: $\gamma^*(t) \to 0$ as $t \to 0$. The non-universal features of the transition, which may vary between systems, are in the exact function $\gamma^*(t)$, and furthremore, in the actual shapes associated with the two routes to isometry. For the indentation of a floating sheet, such a morphological transition could be the observed wrinkle-to-fold transition \cite{Holmes10}; for a floating film on a liquid drop, this mechanism may underlie a wrinkle-to-crumple transition \cite{King12}; for a stretched-twisted ribbon, such a scenario may explain some of the observed morphological transition between various helicoid-like shapes \cite{Chopin14}. In our system, such a morphological transition may not be dramatic, since the low deformability of the substrate forces the film to remain close to the substrate's spherical shape, hence any observed difference between the two routes is likely to be minor.
%\cite{Comment-why-wrinklogami-can-be-safely-used-for-energetic-estimates}.     

\begin{figure}
\centering
\includegraphics[width=79mm]{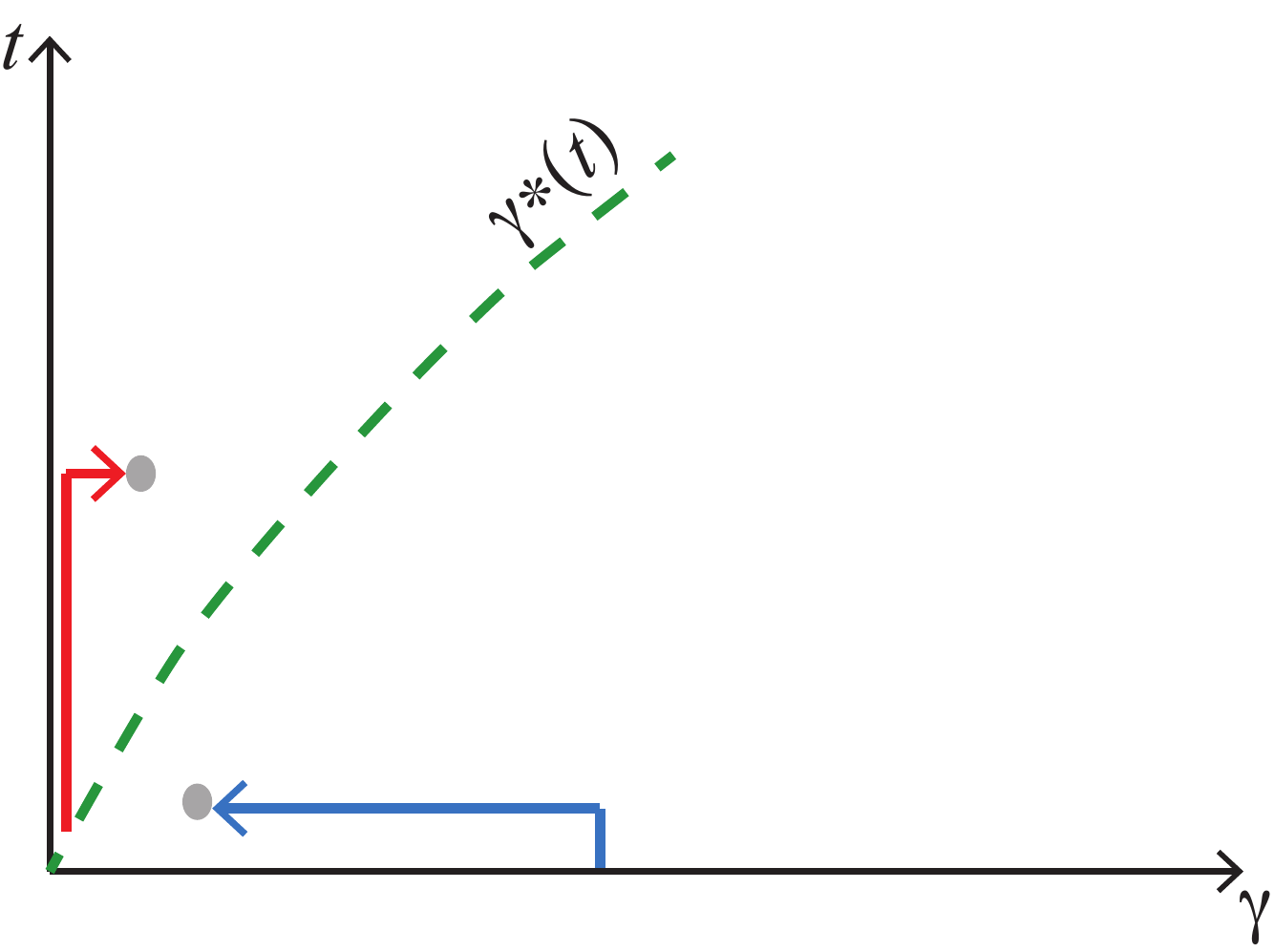}
%\subfigure{\label{lame-a}\includegraphics{axisymmetric-plots}}
%\subfigure{\label{lame-a}
%\includegraphics[width=.5\columnwidth]{SF3.pdf}
%\subfigure{\label{lame-b}}
%\subfigure{\label{lame-c}}
\caption{\label{fig:9}  
A schematic diagram, which depicts the two distinct types of asymptotic isometries in problems of class (C), in which a film of thickness $t$ is subjected to a fixed geometric constraint and a weak tensile load $\gamma$ on its boundary. The horizontal and vertical axes are measures for the thickness and the exerted tensile load, respectively, for a given geometric constraint. (The actual, dimensionless measures, are $B/YW^2$ and $\gamma/Y$). Below a curve $\gamma^*(t)$ we expect the emergence of a wrinklogami pattern, where the isometry is approached asymptotically by homogenous suppression of all stress components (blue trajectory: ``route 2 to isometry" in our classification). Above the curve $\gamma^*(t)$, we expect the shape to be approximated as a perturbation to real isometric transformation ({\emph{i.e.}} developable or piecewise-developable map) of a 2D film, which often involves the formation of stress focusing zones at ridges and vertices (red trajectory: ``route 1 to isometry" in our classification).} 
\end{figure}

\section{Summary \label{sec:summary}}
Our main goal in this paper was to develop a theoretical framework for understanding the behavior of thin adhesive films on curved substrates. The  elementary model we introduced here elucidates the dramatic difference between adhesion on rigid and deformable substrates, and led us to predict the prolamination effect: an adhesive film may remain laminated on a curved, slightly deformable substrate, by developing a wrinkle pattern (wrinklogami) that does not affect the macro-scale shape of the substrate.
A central feature, reverberated by Fig.~\ref{fig:fig1}, is the prevalence of the parameter regime III, at which the prolamination effect is expected, for very thin films. This prediction, which we expect to remain valid also for more complicated models of adhesion ({\emph{e.g.}} inhomogeneous substrate with non-spherical shape), highlights the broad potential usage of the prolamination effect for technologies that employ ultrathin polymer films as well as crystalline sheets, most notably graphene.     

Beyond its importance for adhesion phenomena, the ``wrinklogami" pattern that we found here is an example of {\emph{asymptotic isometry}}, whereby the film becomes strainless in the limit of small thickness and weak tensile load exerted on its edge. This asymptotic isometry concept, which characterizes thin sheets under sufficiently weak tensile loads, generalizes the standard usage of isometric maps, often used to describe the morphologies of elastic sheets under purely compressive loads. We proposed a general mechanism for morphological transitions between distinct types of asymptotic isometries, which may underlie various phenomena, such as wrinkle-to-fold transition in the indentation of floating films \cite{Holmes10}, and wrinkle-to-crumple transition in elastic sheets on liquid drops \cite{King12}. We hope to apply the asymptotic isometry equation, whose general structure we derived in this papaer, to study the universal and non-universal features of such morphological transitions.

\begin{acknowledgements}
We acknowledge support from UMass MRSEC on Polymers, NSF CAREER Award DMR-11-51780, and the hospitality of the Aspen center for physics, where part of this manuscript was written. We thank P. Buchak, G. Grason, R. Kohn, N. Menon,  H.-M. Nguyen, C. Santangelo, and D. Vella for many discussions. 
\end{acknowledgements}

\appendix
%\section{High and low deformability \label{sec:deformability}}
\section{The meniscus energy \label{sec:intro}}
%%%%%%%%%%%%%%%%%%%%%%%%%%%%%%%%%%%%%%
%In this section we describe the complete equations that govern the mechanical equilibrium of a film laminated on a curved, Winkler-type substrate, and the energies associated with this laminated state. 
%%%%%%%%%%%%%%%%%%%%%% 

%= U_{(\cdot)}/(E_ft)W^2$. %, that are used in the main text.  
%
%
%As was noted on the main text, we noramlize the energy by the stretching modulud $Y = E_ft$ and the area of the film $\pi W^2$.  The energy can be decomposed into parts that are associated with the  

%{\emph{(Surface energy:}} 

%\subsubsection{Susbstrate deformation energy \label{sec:subsutaredef}} 
The energy associated with the local, Winkler-type restoring force,
was given in Sec.~\ref{sec:Winkler-1}: 
\begin{gather}
\bU_{\rm Win}= \tfrac{K}{2} \int_0^{2\pi} \! d\q \int_0^W \! r dr  \  (\zeta-r^2/2R)^2  
\ + \  \bU_{\rm men}  \ . 
%
%\nonumber \\
%\bU_{subst}^{(2)} = \tfrac{K}{2} \int_{outside} d^2x \  [r(\bfx) - R]^2   \ ,
\label{WinklerEnergy2-app}
\end{gather}  
Here we will elaborate on the energetic contribution $\bU_{\rm men}$. This term is associated with the ``meniscus", at some zone $r>W$, across which the deformed substrate recovers its spherical shape: 
\begin{gather}
\bU_{\rm men} = \tfrac{K}{2} \int_{r>W} dx^2 \  [r(\bfx) - R]^2     \ , 
\label{WinklerEnergy22-app}
\end{gather}  
where the integral is over the surface of the substrate not covered by the film, and $r(\bfx)$ is the distance of a point $\bfx$ on the substrate's surface from the center of the undeformed sphere.    
We may evaluate $\bU_{\rm men}$ by noting that the substrate's surface recovers its spherical shape at a distance $\ell \approx \sqrt{\gamma/K}$ from the boundary $r=W$. 
This distance reflects an exponential decay of the meniscus shape, which stems from solving the equation $\gamma \ \delta \zeta'' - K\ \delta \zeta =0$ for $r>W$ subjected to some boundary value $\delta \zeta_{({r=W})}$ and required to vanish ($\delta \zeta (r) \to 0$) at $r \to \infty$. Here $\delta\zeta (r)$ is the deviation of the substrate's surface from its original, spherical shape $\zeta_{sph}(r)$. 
 %and its energy~(\ref{eq:meniscus}) are obtained by solving the eq  
This approximation, valid for $W/R \ll 1$ and $\gamma/KR^2 \ll 1$ (see below), is identical to the meniscus of a flat liquid interface with surface tension $\gamma$ and density $K = \rho g$, 
originating as  
%. This equation is the 
an Euler-Lagrange equation of %derived from 
the approximated energy density $\tfrac{1}{2}(\gamma (\delta \zeta')^2 + K\delta \zeta^2)$ of the substrate surface that is not covered by the film.
%Assuming $\gamma /K R^2 \ll 1$ (which can be realizd, even for the low deformability range $\tilde{K} \ll 1$, if the substrate's curvature $R^{-1}$ or the surface tension $\gamma$ are sufficiently small), 
The energetic cost of the meniscus can be thus approximated as:
%\footnote{The shape of the meniscus and its energy~(\ref{meniscus}) are obtained by solving the equation $\gamma \zeta'' - K\zeta =0$ for $r>W$ subjected to to some boundary value $\zeta(r=W)$ and required to vanish $\zeta \to 0$ at $r \to \infty$.  This approximation, valid for $W/R \ll 1$ and $\gamma/KR^2 \ll 1$, is identical to the meniscus of a flat liquid interface with surface tension $\gamma$ and density $K = \rho g$. This equation is the Euler-Lagrange equation derived from the approximated energy density $\tfrac{1}{2}(\gamma (\zeta')^2 + K\zeta^2)$ of the substrate surface that is not covered by the film.}:
\begin{equation}
\bU_{\rm men}  \approx \tfrac{1}{2} \sqrt{\gamma K} \cdot \delta \zeta(W)^2  \ . \label{eq:meniscus}
\end{equation} 
Similarly to our derivation of Eq.~(\ref{BC1}), the meniscus energy $\bU_{\rm men}$ gives rise %amounts 
to a {\emph{normal}} boundary force $-\delta \bU/\delta \zeta(W)$ exerted on the film at $r=W$: 
\begin{equation}
F_n(W) = -\sqrt{\gamma K} \delta \zeta(W)
 \label{BC2-app} \ . 
\end{equation}     
%While our numerical solutions (see Figs.~S1,S2) incorporate this BC, 
We note, however, that in the parameter regime on which we focus our study (where $\tilde{K}, \alpha,\tilde{t}^{-1}\gg 1$), this force is negligible with respect to the tangential boundary force, Eq.~(\ref{BC1}), and therefore one can safely ignore the meniscus effect by assuming that the film approaches smoothly the substrate surface, namely:
% \footnote{A detailed proof of this observation will appear elsewhere.}: 
\begin{gather}
\zeta(W) = \zeta_{sph}(W) \approx -W^2/2R  \nonumber \\ 
 \zeta'(W) = \zeta_{sph}'(W) \approx -W/R \ , 
\label{BC3-app} 
\end{gather}  
where we used, as usual, $W/R \ll 1$ to simplify the above equation. 

In fact, the negligibility of the meniscus effect (and consequently the use of the BCs~\ref{BC3-app}), may be valid also for a highly-deformable substrate ($\tilde{K}^{-1} \gg 1$), as long as the meniscus zone is sufficiently smaller than the radius $R$ of the spherical substrate. Using our dimensionless parameters, this implies that the BCs~(\ref{BC3-app}) are valid as long as: 
\begin{equation}
\tilde{K} \gg \delta_m \ .  
\end{equation}  
Since the mechanical strain $\delta_m = \gamma/(E_ft) \ll 1$, we expect that the meniscus has a negligible effect even at the high deformability regime ({\emph{i.e.}} when the substrate deforms appreciably beneath the film), as along as $\tilde{K}$ exceeds $\delta_m$. We will discuss elsewhere \cite{Number2} %we will discuss 
the qualitative change in the system's behavior when $\tilde{K}$ becomes smaller than this minimal value, a situation which is particularly relevant for the problem of a sheet on a liquid drop (where $\tilde{K} = 0$).

\section{The sub-dominant energy $u^{\rm sub}$ \label{sec:wrinkling}}
%$\bullet$ Follow steps in the original manuscript
%$\bullet$ Start by writing the perturbation to the shape (repeat prvious equation), and also the equations for $u_r$ and $u_\q$. This is actually a singular expansion in $\epsilon$ (refer to \cite{PRE12}). The full rigorous derivation of this series will appear elsewhere, but here we will provide heuristic arguments for its reasoning.    

%The compression-free stress field, which was addressed in the previous section,~.., as well as the fine features of the wrinkle pattern, reflceted by the number of wrinkles $m$ and the various terms in Eqs..., are derived at the singular limit $\epsilon \to 0$. In the following subsection we will show how the displacement field, Eqs. .. enable the simultaneous derivation of both energy components $u_{dom}$ in the previous section and $u_{sub}$. A rigorous justification of this expansion  

%\subsubsection{The slaving condition and the wrinkle amplitude}

%\vspace{0.5 cm}

%%%%%%%%%%%%%%%%%%%%%%%%%%%%%%%%%%%%%%%%%%%%%%
%%%%%%%%%%%%%%%%%%%%%%%%%%%%%%%%%%%%%%%%%%%%%%%%

%\subsubsection{``compressional'' versus ``tensional'' wrinkles \label{comp-tens-wr}}

%With the analysis of the two previous subsections, the stress and displacement fields underlying the wrinkle pattern are fully characterized for any given $m$, but the wrinkle number $m$ remains undetermined. As was noted already in \cite{Cerda03,PNAS11}, this last unknown is found by minimizing the sub-dominant energy $u_{sub}$ of the wrinkle pattern, and is hence crucial for evaluating this energy. 
In order to evaluate the wrinkle number $m$ and its associated energy cost $u^{\rm sub}$, we follow \cite{Cerda03,PNAS11}, and consider the normal force balance, Eq.~(\ref{eq:normal}), substituting for $\zeta(r,\q)$ the wrinkle shape, Eq.~(\ref{wrinkleform}).    
%
%Following \cite{Cerda03,PNAS11}, we consider the {\emph{leading oscillatory components}} ($\propto \cos(m\q)$) of the normal force balance, Eq.~(\ref{eq:normal}), substituting for $\zeta(r,\q)$ the wrinkle shape, Eq.~(\ref{disp-wrink-3}) 
%\footnote{
(Other oscillatory terms, such as $\sigma_{r\q}\ddr \zeta_{\q} \propto \sin(2m\q)$, have negligible energtic contribution %of higher order in $\epsilon$ 
with respect to the primary oscillatory terms $\propto \cos(m\q)$.).   
As a consequence of the divergence of the wrinkle number $m$ in the singular limit $\epsilon \to 0$, 
we note that this equation could be simplified, to leading order in the amplitude $f$: 
\begin{gather}
%\begin{equation}
B\frac{m^4}{r^4} f - (\sigma_{rr}^{(0)} \frac{d^2}{dr^2} f  + \sigma_{rr}^{(m)}  \frac{d^2}{dr^2} \zeta_{\rm sph}) + K f  \nonumber \\
=  - \frac{m^2}{r^2} \sigma_{\q\q} f \ , 
\label{eq:simplified-normal-balance}
\end{gather}
%\end{equation} 
where we used the superscript notations that were defined in Sec.~\ref{sec:Asymdisp}. The RHS is the destabilizing term (that derives from the energetic gain of compression release through out-of-plane buckling), and the LHS consists of the restoring forces ({\emph{i.e.} associated with energetic costs) that favor small wrinkle amplitude $f$. Note that the coupling of the radial stress to the radial component of the curvature yields two terms. The first term is associated with the coupling of the ``pre-tension" $\sigma_{rr}^{(0)}$, Eq.~(\ref{eq-new2}), with the excess radial curvature $ \frac{d^2}{dr^2} f $ along the wrinkle's direction \cite{Cerda03}. The second term, $\sigma_{rr}^{(m)} \frac{d^2}{dr^2} \zeta_{\rm sph}$,  stems from the coupling of the substrate's curvature $ (\zeta_{\rm sph}^{''})$ to the oscillatory, wrinkle-induced perturbation to the radial stress: $\sigma_{rr}^{(m)} = -(E_ft) \frac{d^2}{dr^2} \zeta_{\rm sph} \cdot f$, which we evaluated with the aid of Eqs.~(\ref{eq:shear-strain-20},\ref{eq:strain-radial-1},\ref{eq:stresses-radial}). We will expand elsewhere on the generic nature of such a restoring force, which is induced by a curvature that is imposed on a film.           
The last restoring force, $Kf$, stems from the actual stiffness of the spherical substrate. The energetic costs associated with these restoring forces, are, respectively: 
\begin{subequations}
\label{subenergies}
\begin{gather}
\frac{1}{4} \int_L^W rdr \ B \frac{m^4f^2}{r^4}  %\Rightarrow u_{bend} 
\sim \epsilon \phi^2 \ m^2  \label{suben1}  \ , \\
%\delta_m \delta_g \epsilon \ m^2  \label{suben1}  \ , \\
\frac{1}{4} \int_L^W rdr \ \sigma_{rr} (\ddr f)^2 %\Rightarrow u_{tens} 
\sim \delta_m \phi \ m^{-2}  \label{suben2}  \ , \\
\frac{1}{4} \int_L^W rdr \ (E_ft) (\ddr \zeta_{\rm sph})^2 f^2 %\Rightarrow u_{tens} 
\sim \phi^2 \ m^{-2}  \label{suben3}  \ , \\
\frac{1}{4} \int_L^W rdr \ K f^2  \Rightarrow u_{subst} \sim \phi^2 \tilde{K} \ m^{-2}  \label{suben4}  \ , 
\end{gather}
\end{subequations}
where we used the slaving condition, Eq.~(\ref{eq:slaving-1}) together with the approximation $\ru_r^{(0)} \approx -r^3/6R^2$ (which is valid in the large confinement regime, see Eq.~\ref{eq:radial-disp}) to eliminate any explicit dependence on the wrinkle profile $f(r)$, and Eq.~(\ref{eq-new2}) for the radial tension $\sigma_{rr}$. The radial derivative $\ddr f$ is estimated as $f/W$, recalling that at large confinement the wrinkles prevail the whole film
% that $L/W \sim \alpha^{-1/3}$ 
(see Eq.~\ref{eq:lengh-wr}), and hence %for large confinement ($\alpha \gg 1$) 
the characteristic radial scale for the variation of the wrinkle profile $f(r)$ is $W$. An additional factor $1/2$ originates from the azimuthal integration of $\cos^2(m\q)$, and all energies are normalized, per our convention, by $(E_ft)W^2$.
% Eq.~() for the radial tension $\sigma_{rr}$, and).  

%
%The energies associated with these forces are: $U_{bend}$, which is proprotional to bending modulus $B$ of the film; $U_{tens}$, which is proportional to the radial tension $\sigma_{rr} \approx \Gamma$; and $U_{subst}$, which is proportional to the stiffness $K$ of the substrate.  
%
%These forces are proportional, respectively, to the bending modulus $B$ of the film, the radial tension $\sigma_{rr} \approx \Gamma$, and to the stiffness $K$ of the substrate.
%A substrate of fixed $K$, known as Winkler foundation and assumed for simplicity in the current study, is characterized by a local restoring force (per area) $f = -K \Delta \zeta_{sph} ({\bf x})$ to deformations $\Delta \zeta_{sph} ({\bf x})$ of its shape at a surface point ${\bf x}$ \footnote{A Winkler foudnation with $K \approx E_s/H$ is obtained by coating a rigid sphere of radius $R$ with a narrow compliant layer of Young modulus $E_s$ and thickness $H$. From theoretical perspective, the local Winkler's response does capture the qualitative aspects of the more complicated, non-local response of homogenous solid substrates.}. 
%
%Anticipating the wrinkle number $m$ to diverge as the film becomes thinner, and recalling the slaving condition, Eq.~(\ref{slaving}) that implies inverse proportionality between $m$ and the wrinkle amplitude $f(r)$, one notices  that $U_{bend} \sim m^2$, whereas $U_{tens}$ and $U_{substs}$ both scale as $m^{-2}$. 
An inspection of the three energetic costs in Eqs.~(\ref{suben2}-\ref{suben4}) reveals that all of them are proportional to $m^{-2}$, so we need consider only the largest of them; the balance of this largest-among-three with the bending energy, Eq.~(\ref{suben1}), which is proportional to $m^{2}$, yields the energetically favorable wrinkle number \cite{Cerda03}. Since we consider the low-deformability regime, $\tilde{K}^{-1} \ll 1$, and adress the large confinement asymptotics $\alpha = \phi/\delta_m \to \infty$, the largest among the three terms (\ref{suben2}-\ref{suben4}) is the last one, which stems from the actual stiffness of the substrate. This type of energy balance is typical of ``compressional wrinkles" that are formed under uniaxial compression of a thin film on compliant substrate \cite{Hutchinson98}. 
We thus obtain the wrinkle number from the balance: 
\begin{equation}
\tilde{K} \phi^2 m^{-2} \sim \epsilon \phi^2 m^2  \ \ \Rightarrow \ \ m \sim (\frac{\tilde{K}}{\epsilon})^{1/4}  \ ,  \label{wrinkle-number} 
\end{equation}
and the associated energy: 
\begin{equation}
u^{\rm sub} \sim \sqrt{\tilde{K}\epsilon} \phi^2 \ .  
\end{equation}
For a given set of ``pristine" parameters $\tilde{t}, \tilde{K},\delta_m$, this energy is given as a function of $\phi$ by the expression: 
\begin{equation}
u^{\rm sub} \sim \tilde{t} \sqrt{\tilde{K}} \phi \ .  
\end{equation}

\bibliographystyle{plain}

\begin{thebibliography}{10}
\bibitem{Comment-Disp-Strain-Stress} 
Let us emphasize that Eq.~(\ref{eq:strains}) describes the geometric (strain-displacement) relation only for $|\nabla \ru| \ll 1$ (which necessarily implies $|\epsilon_{ij}| \ll 1$), while the Hookean (stress-strain) response only requires $|\epsilon_{ij}| \ll 1$ (which may be satisfied even in a situation where the gradient $|\nabla \ru|$ is not small and the geometric relation, Eq.~(\ref{eq:strains}), is not valid).
\bibitem{footnote1}
{The Winkler energy $U_{\rm Win}$ is appropriate  for an elastic layer of thickness $H$ and modulus $E$ above a rigid foundation when the characteristic length scale for deformations of the film $\gg H$. For such a system: $K\sim E/H$.}
\bibitem{Comment-BC-1}
Eq.~(\ref{dA2}) was obtained from the axisymmetric map $\ru_r(r)= -r^3/8R^2, \zeta(r) = \zeta_{sph}(r) = -r^2/2R$ that projects the disk 
onto the (rigid) spherical substrate by preserving its area 
(where we assume $W/R \ll 1$): $\epsilon_{ii} = \partial_r \ru_r  + (\tfrac{d\zeta}{dr})^2/2 + \ru_r/r = 0$.
Note that this map is obtained in the limit $\tilde{K} \to \infty, \alpha \to \infty$ of the axisymmetric laminated state (see Sec.~\ref{sec:deformability}).
\bibitem{Comment-deformability}
The definition (\ref{nondim2}) of the deformability parameter seems to be different from Eq.~23 of \cite{PNAS13}. However, the work reported in \cite{PNAS13} addressed the ``weak confinement" regime ($\alpha \gtrsim \alpha^*$), where $\gamma/W^2 \approx 8 Y/R^2$, such that the two definitions are nearly identical. The definition given in the current paper is more suitable for the general case, yielding naturally the identification of the  low-deformability parameter regime through $\tilde{K}^{-1} \gg 1$ (Eq.~\ref{regimeI}).
\bibitem{footnote2}
As we discussed above (after Eq.~(\ref{BC1}), we assume for the simplicity of the discussion $\Gamma \approx \gamma$, where $\gamma$ is the surface tension of the substrate. Our analysis is valid, up to a numerical pre-factor that does not affect the scaling behavior, for any finite ratio of $\gamma/\Gamma$.
\bibitem{footnote4}
{We consider here the most elementary mechanism for wrinkling (a simply periodic shape, Eq.~\ref{wrinkleform}), and for delamination (such that the whole film detaches from the substrate). The actual instabilities may be more complex (e.g. wrinkling cascades, and a periodic pattern of blisters, respectively). Nevertheless, we do not expect this to affect the scaling laws  %[Eqs. (\ref{phi-wr}-\ref{eq:elastic-soft-rig})] 
that are presented in Fig. 2.}
\bibitem{Comment-FT}
The FT expansion is around the singular limit at which the sheet approaches the compression-free stress field, where the small parameter is the inverse bendability $\epsilon$. Tension field theory may be considered as the leading order in this expansion.  
The FT expansion is strictly different from the standard post-buckling approach, valid in a narrow near-threshold (NT) parameter regime, which is essentially an expansion of FvK equations around the compressed, axisymmetric state, where the small parameter is the wrinkle amplitude (see \cite{PNAS11,PRE12,King12}).
\bibitem{Scaling-L-difference}
The scaling $L/W \sim  \alpha^{-1/3}$ is different from the result $L/W\sim \alpha^{-1/5}$ found for a wrinkled film on a liquid drop \cite{King12}. This reflects the difference between the soft liquid drop (akin to a highly-deformable Winkler substrate with $\tilde{K}^{-1} \gg 1$), and the low deformability case ($\tilde{K}^{-1} \ll 1$) addressed in the current study. A detailed discussion of this difference will appear elsewhere.
\bibitem{Comment-Shear-1}
Eq.~(\ref{eq:shear-strain-20}) shows that $\ru_r^{(m)}/f$ is finite, and therefore %remain finite as $\epsilon \to 0$. 
the oscillating component of the radial displacement  $\ru_r^{(m)} \to 0$ in the singular limit $\tilde{t} \to 0$ (similarly to $f$). Evaluation of the relevant contribution to the elastic energy $\sim (\ddr \ru_r^{(m)})^2$) shows that it is negligible compared to $u^{\rm sub}$.
\bibitem{footnote-diagrams} 
Since both parts of $u^{\rm wr}$ do not have explicit dependence on $\tilde{t}$, it is more transparent to draw in Fig.~8 three separate diagrams, rather than a single one, as in Fig.~7b.
\end{thebibliography}

\end{document}